\documentclass[nofootinbib]{revtex4}
\usepackage{graphicx}
\usepackage{dcolumn}
\usepackage{amsmath,amssymb,epsfig}
\usepackage{paralist}
\usepackage{comment} 
\usepackage{siunitx}
\usepackage{graphicx}
\usepackage{multirow}
\usepackage{color,soul}
\usepackage{suffix}
\usepackage{mathtools}

\usepackage[breaklinks=True,colorlinks=True, linkcolor=magenta,citecolor=blue,debug=True]{hyperref}

\usepackage{wasysym}

\def\beqn{\begin{eqnarray*}}
\def\eeqn{\end{eqnarray*}}

\newcommand{\be}{\begin{equation}}
\newcommand{\ee}{\end{equation}}
\newcommand{\ba}{\begin{eqnarray}}
\newcommand{\ea}{\end{eqnarray}}

\begin{document}

\title{Lunar Laser Ranging with High-Power CW Lasers}

\author{Slava G. Turyshev
}   

\affiliation{ 
Jet Propulsion Laboratory, California Institute of Technology,\\
4800 Oak Grove Drive, Pasadena, CA 91109-0899, USA
}%

\date{\today}

\begin{abstract}

We present a high-power continuous-wave (CW) lunar laser ranging (LLR) technique that has the potential to significantly improve Earth--Moon distance measurements. Using a 1~kW CW laser at 1064~nm and a 1~m-aperture telescope as an example, we develop a detailed link budget and analyze the prevailing noise sources to assess system performance when ranging to next-generation $\sim$10~cm corner-cube retroreflectors (CCRs). Unlike legacy arrays, these smaller CCRs are designed to yield lower intrinsic range errors, yet their reduced reflective area results in lower photon return rates, posing challenges for pulsed LLR systems. The photon-rich CW approach, by providing continuous high-power illumination, overcomes this limitation, reducing shot noise and enabling sustained millimeter-level ranging with a pathway to sub-0.1~mm precision. Furthermore, by alternating measurements between widely separated lunar reflectors, differential LLR mitigates common-mode station errors to achieve tens-of-micrometer precision, limited primarily by uncorrelated atmospheric turbulence. This scalable approach---integrating high-power CW lasers, narrowband filtering, and rapid atmospheric turbulence averaging---enables next-generation gravitational tests, precision lunar geodesy, and improved lunar reference frames in support of planetary exploration.

\end{abstract}

\maketitle

\section{Introduction}
\label{sec:introduction}

Lunar Laser Ranging (LLR) has served as a pivotal technique in gravitational physics and planetary geodesy for over five decades. Since the deployment of corner-cube retroreflector (CCR) arrays on the Moon during the Apollo era \cite{Faller_1970}, LLR has advanced to millimeter-level accuracy in measuring the Earth--Moon distance \citep{Dickey:1994,Williams:2004,Williams:2009,Williams2012}. These measurements have yielded stringent tests of general relativity \citep{Williams:2009}, refined our understanding of lunar rotation and tidal interactions, and provided insights into the Moon’s internal composition and long-term orbital evolution \citep{Williams:2008,Williams:2014,Briaud2023}.

Future science objectives, however, demand a further leap in LLR precision. Tests of general relativity at much improved levels of accuracy \citep{Turyshev:2008,Zhang-etal:2022}, probing deep lunar interior beyond the levels achieved by the GRAIL mission \cite{Konopliv-etal:2013,Zuber-etal:2013}, and searches for ultra-low-frequency ($\mu$Hz) gravitational waves \citep{Blas-Jenkins:2022,Turyshev-etal:2024} increasingly call for sub-millimeter and tens-of-micrometer capability. Furthermore, upcoming exploration initiatives under NASA's Artemis program will benefit from more accurate lunar reference frames, supporting mission navigation and surface operations.

Though pulsed-laser ranging (PLR) systems have underpinned high-precision LLR for many years, they confront intrinsic limitations that become more pronounced with next-generation CCRs. These new CCRs typically of $\sim 10\,\mathrm{cm}$ in diameter, reduce size-induced pulse spread as well as thermal and libration-related errors but also drastically reduce the total reflective area. Consequently, pulsed systems—relying on low duty cycles and high peak power—will experience limited photon return rates and face substantial challenges achieving higher signal-to-noise ratios (SNR).

High-power continuous-wave (CW) laser systems offer an alternative pathway to overcome these limitations. By emitting a steady photon flux at kilowatt-level average power, CW LLR can leverage multi-second coherent integrations to suppress shot noise, thereby improving ranging precision without relying on large reflector apertures. For instance, a \SI{1}{kW} CW laser at \SI{1064}{nm}, coupled to a \SI{1}{m} telescope, can potentially reach sub-millimeter single-range precision and approach tens-of-micrometer accuracy in differential measurements \citep{Shao:2018,Turyshev:2018}.

However, realizing this level of precision presents several technical and physical challenges. The beam footprint on the lunar surface must be controlled to mitigate losses due to atmospheric turbulence and beam divergence. Mechanical and thermal drifts within the telescope and optical mounts must be suppressed to the micrometer level or below, which necessitates stable pier designs, low-expansion materials, and stringent thermal regulation. Timing and frequency references must maintain fractional frequency stabilities of $10^{-11}\text{--}10^{-13}$ to eliminate long-term drifts that could mask sub-millimeter signals. Additionally, larger photon fluxes pose elevated demands on background noise suppression and detector throughput, particularly during high lunar-phase brightness or partial daytime operations.

This paper analyzes the feasibility and design considerations of high-power CW LLR under realistic conditions. The discussion includes an examination of the potential photon return rates when targeting $10\,\mathrm{cm}$ CCRs, the ensuing shot-noise behavior, and the key noise sources including atmospheric turbulence and mechanical instabilities. Strategies for mitigating these limitations—such as narrowband optical filtering, turbulence averaging, and real-time station calibration—are addressed. In particular, we consider differential LLR, wherein alternating measurements between retroreflectors separated by $\sim1000\,\mathrm{km}$ suppress common-mode station errors, allowing differential precisions in the tens-of-micrometer range under favorable turbulence (e.g., $r_0\approx20\,\mathrm{cm}$ at $1064\,\mathrm{nm}$).

This “photon-rich” approach, uniting kilowatt-class CW lasers with advanced filtering and stable infrastructure, stands to improve lunar geodesy and enable more stringent gravitational tests. Sub-millimeter and tens-of-micrometer measurements would refine the Moon’s reference frame, probe its deeper interior, and support detection of gravitational phenomena at very low frequencies. 

The paper is organized as follows. In Section~\ref{sec:lllr-classic}, we review classical LLR methods, summarizing current precision levels and typical error budgets. Section~\ref{sec:highpower_cw_expanded} then introduces the principles of high-power CW ranging and how continuous illumination can increase the photon return. A detailed link budget is established in Section~\ref{sec:flux_model}, and Section~\ref{sec:eb} addresses the prevailing noise sources—thermal and mechanical drifts, atmospheric turbulence, and shot noise—together with potential mitigation techniques. Section~\ref{sec:differential_cw_llr_1000km} extends the analysis to differential ranging, evaluating the possibility of tens-of-micrometer precision by comparing signals from widely spaced lunar CCRs. Finally, Section~\ref{sec:conclusion} summarizes the main conclusions and offers a perspective on future steps to achieve these precision levels.

\section{Lunar Laser Ranging}
\label{sec:lllr-classic}

LLR experiments measure the round-trip travel time, \(\Delta t\), of a laser pulse emitted from an Earth-based station to a retroreflector on the Moon's surface and back. This single observable constrains the instantaneous Earth--Moon distance and also serves as a powerful tool for testing fundamental physics and geodesy 
\citep{Murphy_etal_2008,Samain1998,Williams:2004,Williams:2009,Murphy:2013}. 
The following sections provide a concise analytical framework, discuss measurement precision, and detail the main sources of error.

\subsection{Experimental Setup}

Lunar laser ranging experiments rely on five key components:  
(1) a high-power Earth-based laser transmitter and telescope,  
(2) retroreflector arrays deployed on the lunar surface,  
(3) high-precision photon detection and timing systems,  
(4) advanced data processing and modeling infrastructure, and  
(5) long-term data archival for scientific analyses. The setup is detailed below:

\begin{enumerate}

\item \textit{Laser Transmitter and Telescope:}  
LLR stations typically use pulsed Nd:YAG lasers operating at 1064\,nm or frequency-doubled 532\,nm wavelengths, with pulse widths ranging from a few picoseconds to several nanoseconds \citep{Samain1998,Williams2012}. Telescopes with apertures between 0.5\,m and 3.5\,m are used to transmit the outgoing beam and collect the faint return signal. Beam divergence is minimized to near-diffraction-limited levels ($\sim1''$, or $\sim5\,\mu{\rm rad}$), ensuring efficient illumination of the lunar retroreflectors. Precise optical alignment is critical to maximize photon flux and minimize transmission losses \citep{Degnan1993, Murphy_etal_2008}.

\item \textit{Lunar Retroreflectors:}  
The lunar surface hosts five retroreflector arrays deployed during the Apollo missions (11, 14, and 15) and on the Lunokhod 1 and 2 rovers \citep{Dickey:1994, Murphy_etal_2010}. The Apollo arrays, made of fused silica CCRs, feature 3.8\,cm diameters. Apollo 15 contains 300 CCRs, yielding a total reflecting area of $\sim0.34\,{\rm m}^2$, while Apollo 11 and 14 each house 100 CCRs with a reflecting area of $\sim0.11\,{\rm m}^2$. In contrast, the Lunokhod arrays consist of 14 triangular CCRs with 11\,cm edge lengths, providing a total effective retro-reflecting area of $\sim0.0443\,{\rm m}^2$ per array. These arrays experience signal degradation due to thermal gradients, dust accumulation, and libration-induced tilts, which distort the optical centroid and reduce photon return efficiency \citep{Murphy_etal_2010}.

\item \textit{Detection and Timing:}  
Single-photon avalanche diodes (SPADs) or photomultiplier tubes (PMTs) are used to detect returning photons with timing resolutions of 20--50\,ps. Photon arrival times are referenced against ultra-stable frequency standards, such as hydrogen masers, providing fractional stabilities of $10^{-13}$ to $10^{-14}$ over integration periods \citep{Battat_etal_2009, Lisdat2016}. Timing systems are calibrated using short-distance metrology targets, achieving nanosecond-level precision critical for millimeter-scale range measurements.

\item \textit{Data Processing and Modeling:}  
The primary observable in LLR is the \emph{round-trip travel time} of laser pulses, converted to the Earth--Moon distance as $\rho = \frac{1}{2} c \Delta t$, where $\Delta t$ is the measured time interval and $c$ is the speed of light. Corrections for instrumental delays, atmospheric refraction, tidal loading, and Earth orientation are applied to refine the measurement. State-of-the-art lunar ephemerides models, such as DE440, further reduce systematic uncertainties to the millimeter level \citep{Park-etal:2021, Williams_etal_2014}.

\item \textit{Normal Points and Long-Term Archival:}  
Individual photon detections exhibit scatter at the centimeter level. To reduce statistical noise, thousands of photon returns are averaged into "normal points" over short time intervals (e.g., a few minutes) \citep{Murphy_etal_2012}. These normal points are archived and used to constrain orbital dynamics, station coordinates, gravitational theories, and geophysical properties in global analyses.

\end{enumerate}

LLR stations require sub-arcsecond pointing accuracy, nanosecond-level timing precision, and rigorous calibration to achieve millimeter-scale measurement accuracy. By combining precise data with advanced geophysical and orbital models, LLR provides unique insights into gravitational physics, lunar geodesy, and Earth-Moon system dynamics under optimal conditions.

\subsection{Observation Equation}

The LLR experiment measures the round-trip travel time of a laser pulse sent from Earth to the Moon, reflected by a retroreflector, and returned to the station. Key variables are defined as follows:
\begin{itemize}
    \item $\mathbf{r}_{\tt E}(t)$: Position vector of the Earth-based laser station in a geocentric or inertial reference frame at the laser emission time $t$.
    \item $\mathbf{r}_{\tt M}(t')$: Position vector of the lunar retroreflector at the reception (return) time $t'$.
    \item $c$: Speed of light in vacuum.
    \item $\Delta t = t' - t$: Total two-way round-trip travel time of the laser pulse.
\end{itemize}

The one-way light travel time, $\tau_{\mathrm{one\,way}}$, is approximated as:
\[
    \tau_{\mathrm{one\,way}} \;=\; \frac{1}{c} 
    \bigl\lVert \mathbf{r}_{\tt M}(t_{\mathrm{moon}}) - \mathbf{r}_{\tt E}(t_{\mathrm{earth}}) \bigr\rVert,
\]
where $t_{\mathrm{moon}}$ and $t_{\mathrm{earth}}$ differ by the finite light travel time. The total round-trip time $\Delta t$ includes both paths and corrections for relativistic and atmospheric effects:
\[
    \Delta t 
    \;=\; 2 \,\tau_{\mathrm{one\,way}} 
    \;+\; \epsilon_{\mathrm{relativity}}
    \;+\; \epsilon_{\mathrm{atmos}}.
\]
Here, $\epsilon_{\mathrm{relativity}}$ accounts for general relativistic corrections, including the Shapiro delay and gravitational redshift, while $\epsilon_{\mathrm{atmos}}$ captures refractive delays caused by the Earth’s troposphere and ionosphere \citep{Mendes2004, Williams:2004}.

The observed Earth-Moon distance $\rho$ is calculated as:
\begin{equation}
\rho = \tfrac12\,c\,\Delta t,
    \label{eq:basic_range}
\end{equation}
but achieving mm-level accuracy requires incorporating additional physical effects. The modeled range $\rho_{\mathrm{model}}$ is given:
\begin{equation}
    \rho_{\mathrm{model}} 
    \;=\; \bigl\lVert \mathbf{r}_{\tt M}(t') - \mathbf{r}_{\tt E}(t) \bigr\rVert 
    \;+\; \delta_{\mathrm{relativity}}
    \;+\; \delta_{\mathrm{atmos}}
    \;+\; \delta_{\mathrm{tides}}
    \;+\; \delta_{\mathrm{Moon}}
    \;+\; \delta_{\mathrm{instrum}}
    \;+\;\cdots
    \label{eq:range_model}
\end{equation}
Key corrections include:
\begin{itemize}
    \item \textit{Relativistic Effects ($\delta_{\mathrm{relativity}}$)}: The Shapiro delay contributes delays up to $\sim25\,{\rm ns}$ (equivalent to $\sim3.7\,{\rm m}$) for the Sun and $\sim0.02\,{\rm m}$ for the Earth \citep{Williams:2004,Turyshev-etal:2013}.
    \item \textit{Atmospheric Refraction ($\delta_{\mathrm{atmos}}$)}: Tropospheric delay corrections reduce uncertainties to $<1\,{\rm mm}$ using advanced models such as Mendes-Pavlis \citep{Mendes2004}.
    \item \textit{Tidal Effects ($\delta_{\mathrm{tides}}$)}: Solid Earth tides introduce range variations up to $0.5\,{\rm m}$, while ocean tidal loading contributes displacements on the millimeter scale \citep{fes2014}.
    \item \textit{Lunar Libration and Orientation $(\delta_{\mathrm{Moon}}$)}: Libration-induced tilts in the Apollo 15 retroreflector array introduce timing spreads of up to $300\,{\rm ps}$, corresponding to $45\,{\rm mm}$ \citep{Turyshev-etal:2013,Murphy:2013}.
    \item \textit{Instrumental Biases $(\delta_{\mathrm{instrum}})$}: Station timing offsets and detector delays are regularly calibrated to minimize residual errors to sub-millimeter levels \citep{Battat_etal_2009}.
\end{itemize}

The observed range $\rho_{\mathrm{obs}}$ is iteratively compared with the modeled range $\rho_{\mathrm{model}}$, and relevant parameters—such as station coordinates, orbital elements, and tidal effects—are refined using least-squares or Kalman filter techniques:
\[
    \rho_{\mathrm{obs}} \;=\; \tfrac12 c \Delta t_{\mathrm{obs}}.
\]

By iteratively fitting these parameters to extensive LLR datasets, this framework refines Earth-Moon ephemerides and geophysical models, enabling tests of gravitational theories and achieving mm-scale measurement precision \citep{Williams2012, Park-etal:2021}.

\subsection{Error Budget}

The short-term (per-session) measurement precision in LLR experiments is primarily determined by the timing uncertainty, $\sigma_{\Delta t}$. The relationship between the timing error and range error, $\sigma_{\rho}$, is:
\[
\sigma_{\rho} \;=\;  {\textstyle\frac{1}{2}} c\,\sigma_{\Delta t}.
\]
For example, a time uncertainty of 1\,ps corresponds to a 0.15\,mm (round-trip) range uncertainty. Achieving millimeter-level ranging thus requires controlling timing jitter to a few picoseconds, in addition to minimizing systematic biases.

While precise timing is critical, overall accuracy is also influenced by systematic and modeling uncertainties, including station geodesy, atmospheric refraction, and lunar orbital dynamics. Modern LLR systems, such as APOLLO (Apache Point), Grasse, Matera, and McDonald Observatory, typically achieve 2--5\,mm \emph{per-normal-point} precision, with single-photon returns exhibiting greater scatter (10--20\,mm). Combining thousands of photon detections into a single normal point reduces statistical noise significantly \citep{Murphy_etal_2012, Battat_etal_2009}.

Table~\ref{tab:error_budget} summarizes the most significant error contributions to present-day LLR measurements. The listed uncertainties correspond to representative two-way (round-trip) values, though specific magnitudes depend on station hardware, environmental conditions, and analysis methods.

\begin{table}[ht!]
\centering
\caption{Updated LLR Error Budget (Round Trip).}
\label{tab:error_budget}
\begin{tabular}{p{0.27\textwidth} c p{0.609\textwidth}}
\hline
{Error Source} & {Magnitude} & {Comments} \\
\hline\hline
Atmospheric Delay          
& 1--3 mm        
& Residual tropospheric delay after modeling (e.g., Mendes--Pavlis) using in-situ meteorological data. Unmodeled delays are elevation-dependent and reduce to mm-level uncertainties at high elevation angles \citep{Mendes2004}. \\

Clock and Oscillator Stability 
& $<1$ mm        
& Hydrogen masers or Cs fountains provide sub-ps timing stability (fractional stability of $10^{-14}$ to $10^{-15}$ over relevant timescales) \citep{Murphy:2013}. \\

Detector Timing Jitter and Walk 
& 1--3 mm      
& Timing jitter of 20--50\,ps from SPADs or PMTs is minimized via threshold-walk corrections and advanced electronics \citep{Battat_etal_2009}. \\

Laser Pulse Shape and Centroid 
& 1--2 mm    
& Pulse widths of 50--200\,ps require precise centroiding to ensure consistency in laser pulse timing and energy stability. \\

Photon Statistics and Background 
& 2--5 mm     
& Single-photon scatter is typically 10--20\,mm due to low photon return rates; averaging thousands of photon returns in a normal point reduces this to 2--5\,mm \citep{Murphy:2013}. \\

Station Coordinates (Geodesy) 
& 1--2 mm    
& Accurate station positions are determined through GNSS, SLR, or VLBI, with uncertainties directly propagating to LLR measurements \citep{Vokrouhlicky_etal_2019}. \\

Earth Orientation \& Tidal Models 
& 1--5 mm
& Errors in Earth orientation parameters (EOP), including UT1 and polar motion, as well as tidal loading corrections, contribute millimeter-level uncertainties \citep{Williams_etal_2014, altamimi2016}. \\

Lunar Ephemerides \& Librations 
& 1--5 mm
& Errors in the Moon's orbital and libration models (e.g., DE440, INPOP) limit accuracy but are continuously refined through precise observations \citep{Murphy:2013}. \\

Retroreflector Array Geometry 
& 3--7 mm    
& Libration-induced tilts, thermal gradients, and multi-cube array structure cause photon pulse spreading (~6 cm per pulse, ~3 cm one-way), averaging to 3--7 mm per normal point. Apollo 15 arrays are more sensitive due to their larger size \citep{Murphy_etal_2012, Murphy:2013}. \\

Systematic Calibration Offsets 
& 1--2 mm   
& Timing chain delays, including cable lengths and electronic biases, are minimized through regular calibration using local metrology targets \citep{Battat_etal_2009}. \\
\hline
\end{tabular}
\end{table}

The error sources in Table~\ref{tab:error_budget} are defined as follows:

\begin{enumerate}[1)]
    \item \textit{Atmospheric Delay}: The neutral atmosphere introduces a path delay of up to \(2\,\text{m}\) near the horizon, which diminishes to 20--30\,cm at high elevation angles. Modern models \citep{Mendes2004} reduce residual uncertainties to 1--3 mm.
    
    \item \textit{Clock and Detector Stability}: Hydrogen masers provide fractional timing stability of \(10^{-14}\) to \(10^{-15}\), corresponding to sub-ps timing precision. Detector jitter, typically 20--50\,ps, is mitigated by advanced hardware and threshold-walk corrections \citep{Lisdat2016, Battat_etal_2009}.
    
    \item \textit{Photon Statistics}: Due to low photon return rates, single-shot range scatter is 10--20 mm per pulse, but normal points, formed by averaging thousands of returns, reduce statistical uncertainty to 2--5 mm per normal point \citep{Murphy:2013}.
    
    \item \textit{Earth and Lunar Models}: Residual errors in EOPs (e.g., UT1, polar motion), lunar ephemerides (e.g., DE440) contribute among the largest sources of systematic uncertainty, adding 1--5 mm depending on model accuracy \citep{Williams_etal_2014}.
    
    \item \textit{Retroreflector Array Geometry}: Apollo 15 exhibits a ~6 cm pulse broadening per pulse (~3 cm one-way) due to its multi-cube design and libration effects. This error averages to 3--7 mm per normal point, making it a dominant contributor to systematic uncertainty \citep{Murphy_etal_2012, Murphy:2013}.
    
    \item \textit{Calibration Offsets}: Regular metrology and calibration against local targets minimize timing chain biases and systematic offsets to the 1--2 mm range \citep{Murphy:2013}.
\end{enumerate}

By addressing these error sources through advanced modeling, calibration, and instrumentation, LLR consistently achieves millimeter-level precision, enabling stringent tests of gravitational physics and improved geophysical models.

\subsection{LLR mm-Level Capabilities and Outlook}
\label{sec:llr-capabilities-outlook}

LLR currently achieves sub-centimeter precision in measuring the Earth--Moon distance, relying on the two-way light-time relationship (\ref{eq:basic_range}). Advanced LLR stations achieve \emph{per-normal-point} accuracies of 2--5\,mm \citep{Murphy_etal_2012}, with single-session normal points occasionally reaching 1--2\,mm under optimal conditions, including favorable atmospheric seeing and precise station calibration \citep{Battat_etal_2009}. Sustained accuracies of 2--3\,mm have been reported globally over several years, enabled by refinements in optical alignment, atmospheric modeling, and timing electronics \citep{Samain1998, Williams_etal_2014}.

Several technological and methodological advancements are now propelling LLR toward achieving—and potentially surpassing—the \si{\milli\metre} precision threshold:
{}
\begin{itemize}
  \item \emph{High-Power and High-Repetition-Rate Lasers:}  
  Advances in laser technology, including kW-class average-power lasers and pulsed systems operating at kHz-to-MHz repetition rates, enable sub-\SI{100}{ps} or even sub-\SI{10}{ps} pulse durations. These improvements significantly enhance photon flux on smaller CCRs, reducing shot noise and enabling finer temporal resolution \citep{Degnan1993, Sakamoto2014}.

  \item \emph{Detector Enhancements:}  
  State-of-the-art single-photon detectors, such as low-jitter SPADs and superconducting nanowire single-photon detectors (SNSPDs), now achieve timing jitters as low as \SI{10}{ps}, improving both the timing centroid precision and the SNR of photon returns \citep{Marsili2013, Murphy_etal_2008}.

  \item \emph{Refined Atmospheric and Tidal Modeling:}  
  Improved tropospheric delay models (e.g., Mendes--Pavlis), tidal loading corrections, and lunar interior modeling have reduced systematic errors to the millimeter level \citep{Mendes2004, Williams_etal_2014}. Incorporating modern lunar ephemerides such as DE440 and INPOP21a, along with advanced filtering techniques like Kalman filters, further minimizes residuals in range measurements \citep{Park-etal:2021, Fienga_etal_2021}.

  \item \emph{Station Geodesy and Timekeeping:}  
  Combined geodetic techniques using GNSS, SLR, and VLBI now constrain station positions to 1--2\,mm \citep{Vokrouhlicky_etal_2019}. Ultra-stable clocks, such as hydrogen masers and optical lattice clocks, provide picosecond timing accuracy with fractional stabilities of $10^{-14}$ to $10^{-15}$ over relevant timescales, critical for achieving sub-mm performance \citep{Lisdat2016}.
\end{itemize}

As advancements in hardware and analysis converge, LLR continues to serve as a cornerstone of high-precision gravitational science. Current applications include stringent tests of the equivalence principle, constraints on the variability of the gravitational constant, and investigations of relativistic precession effects \citep{Williams:2004, Turyshev2007}. Furthermore, LLR offers unique insights into the Moon's internal structure, Earth’s rotational dynamics, and tidal dissipation processes, significantly advancing our understanding of planetary and geophysical phenomena \citep{Dickey:1994, Williams_etal_2014}.

\subsection{Next-Generation CCRs and Challenges for LLR}
\label{sec:photon-flux-scaling}

Next-generation CCRs, with diameters of 10--17\,cm \citep{Currie_etal_2011,Turyshev-etal:2013}, address limitations of Apollo and Lunokhod arrays but introduce challenges due to their smaller aperture sizes. While Apollo 15’s \(0.34 \, \text{m}^2\) reflector suffers from photon pulse spreading primarily due to its multi-cube array design---further modulated by lunar librations---resulting in \(\pm7\,\text{cm}\) range errors, smaller CCRs (\(0.00785 \, \text{m}^2\) for 10\,cm) reduce this by ensuring uniform photon path lengths. Their compact design also improves thermal stability under lunar temperature variations (90--390\,K) and allows for robotic deployment with reduced mass (\(\leq 3.5\,\text{kg}\)). However, smaller apertures challenge photon collection, requiring increase in laser power, beam divergence control, and detection efficiency to achieve mm- and sub-mm-level precision.

Using photon flux measurements reported by APOLLO and OCA values \citep{Samain1998,Murphy_etal_2012,Murphy:2013}, the photon flux for the new 10\,cm CCR is estimated by scaling the observed flux according to the ratio of geometric areas. Table~\ref{tab:photon-flux-scaling} demonstrates the severe limitations imposed by the smaller aperture size of the new CCRs.  Note that APOLLO operates from a 3.5 m telescope and employs a laser averaging 2.3\,W at 532 nm, generating 100 ps pulses at a 20 Hz repetition rate and 115 mJ per pulse \cite{Murphy:2013}. OCA operates a 1.5 m telescope equipped with a dual-cavity Nd:YAG laser at a wavelength of 532 nm, generating 200 ps pulses with 300 mJ per pulse for lunar measurements, and a repetition rate of 10 Hz \cite{Samain2009}.

\begin{table}[ht]
    \centering
    \caption{Effective geometric areas, photon flux registered by APOLLO and OCA for existing retroreflector arrays \cite{Murphy_etal_2012,Murphy:2013,Courde-etal:2017}, and photon flux (in photons/s) anticipated from a 10\,cm next-generation CCR.}
    \label{tab:photon-flux-scaling}
    \begin{tabular}{lccc|cc}
        \hline
        Reflector &  Array Area & APOLLO & OCA  & \multicolumn{2}{c}{Scaled to 10 cm CCR (photons/s)} \\
                  & (m\(^2\))      & (photons/s)  & (photons/s)  & APOLLO & OCA \\
        \hline
        Apollo 11    & 0.11   & 19.13 & 0.28 & 1.37 & 0.02 \\
        Apollo 14    & 0.11   & 30.42 & 0.44 & 2.17 & 0.03 \\
        Apollo 15    & 0.34   & 62.92 & 0.94 & 1.45 & 0.02 \\
        Lunokhod 1   & 0.04 & 8.28  & 0.15 & 0.89 & 0.02 \\
        Lunokhod 2   & 0.04 & 5.20  & 0.10 & 0.56 & 0.01 \\
        \hline
    \end{tabular}
\end{table}

Table~\ref{tab:photon-flux-scaling} highlights several critical points. Apollo 15 achieves the highest photon flux among the arrays, with \(62.92 \, \text{photons/s}\) registered by APOLLO, but the scaled flux for a 10\,cm CCR drops to \(1.45 \, \text{photons/s}\). Apollo 14, despite having the same area as Apollo 11, exhibits stronger performance due to its higher photon return efficiency. Lunokhod 1 and  2 produce much lower flux values due to their smaller geometric areas and degraded performance.

The drastic reduction in photon flux for the new CCRs underscores the challenges of maintaining mm-level ranging precision with such small reflectors. For example, the scaled flux values for OCA, already lower than APOLLO due to differences in system sensitivity, further emphasize the limitations of next-gen CCRs. Lunokhod 2 shows a photon flux of just \(0.01 \, \text{photons/s}\) for a 10\,cm CCR. These results highlight the need for advancements in laser power, beam divergence control, and photon detection efficiency to compensate for the reduced aperture size of the new CCRs.

Despite these challenges, achieving sub-mm and tens-of-\(\mu\mathrm{m}\) precision offers transformative opportunities for fundamental physics and planetary science. Addressing the limitations of smaller CCRs and stability requirements will enable breakthroughs in gravitational research, lunar science, and geodesy. Continued advancements in laser systems, detection technologies, and environmental modeling will keep LLR at the forefront of precision science, enhancing our understanding of the Earth--Moon system and testing general relativity.

\subsection{Outlook for LLR}
\label{sec:submm-prospects-PLR}

Building on decades of progress that have brought LLR from the cm-scale to the mm-scale, the community is now striving for sub-mm and even tens-of-\(\mu\text{m}\) precision. 
Reaching this milestone requires transformative improvements in timing stability, laser pulse characteristics, atmospheric path corrections, geophysical modeling, and lunar retroreflector performance. Below, we discuss the key technical challenges and necessary innovations to push pulsed LLR systems toward the 0.1-mm frontier, focusing on ranging to compact (\(\sim10\,\text{cm}\)-diameter) CCRs on the Moon.

\subsubsection{Timing and Laser Pulse Requirements}
\label{sec:PLR_femtosecond}

State-of-the-art LLR systems rely on pulsed Nd:YAG lasers at 532 nm or 1064 nm, with pulse widths of 50–100\,ps \citep{Murphy:2013, Battat_etal_2009}. However, achieving 0.1 mm precision requires that the total two-way timing uncertainty be reduced to  \(\sim 0.67 \)\,ps. This means that modern LLR systems must reduce pulse widths to $<$ 10 ps, ensuring single-shot timing uncertainties do not exceed 3 mm. Achieving 0.1 mm precision requires reducing this by a factor of 30, which means:
{}
\begin{itemize}
    \item \emph{Ultra-narrow pulses:} Sub-10 ps pulses, preferably \(\sim 5\) ps, will reduce range ambiguity while maintaining high peak power. Even a 10 ps pulse width limits range precision to \(\sim 1.5\) mm, necessitating statistical averaging over large photon counts \citep{Degnan1993}.
    \item \emph{Ultra-stable timing references:} Optical frequency combs and hydrogen masers must provide fractional stability of \(10^{-14}\)–\(10^{-15}\) over $10^3$ s to suppress oscillator-induced range drifts to the sub-0.1 mm level \citep{Lisdat2016, Campbell2017}.
    \item \emph{Low-jitter detection:} Current SPAD and PMT detectors have timing jitters of 20–50 ps \citep{Murphy:2013}. Advanced SNSPDs (Superconducting Nanowire Single-Photon Detectors) with jitter below 5 ps can improve resolution by an order of magnitude \citep{Marsili2013}.
\end{itemize}

Even with perfect timing, photon statistics impose a fundamental limit. Aiming for 0.1 mm accuracy per normal point requires averaging at least \( N = ( {\sigma_{\text{single shot}}}/{\sigma_{\text{target}}} )^2 \approx 900 \), meaning that stations must collect at least 900 high-SNR photons per normal point, requiring higher pulse energy, efficient beam transmission, and advanced detectors.

\subsubsection{Atmospheric Path Corrections}

The Earth’s troposphere remains a dominant error source, limiting LLR precision to a few mm under current correction models such as Mendes-Pavlis \citep{Mendes2004}. Achieving 0.1 mm precision demands:

\begin{itemize}
    \item \emph{Real-time atmospheric profiling:} Tropospheric delay fluctuations must be corrected to better than 100\,$\mu$m, requiring real-time monitoring of temperature, pressure, and humidity gradients with advanced models \citep{Samain1998}.
    \item \emph{Multi-wavelength ranging:} Simultaneous measurements at 532 nm and 1064 nm can separate dispersive and non-dispersive effects, reducing residual atmospheric uncertainty by a factor of 10 \citep{Degnan1993}.
    \item \emph{Active atmospheric sensing:} Ground-based water vapor radiometers can somewhat improve real-time delay corrections. Current uncertainties of \(\sim 3\) mm must be reduced by an order of magnitude \citep{Samain2009}.
\end{itemize}

With these advancements, atmospheric-induced errors can be limited to the 0.1 mm level, provided that real-time meteorological data are continuously incorporated into improved path delay models.

\subsubsection{Geodesy, Lunar Modeling, and Retroreflector Constraints}

Reaching 0.1 mm precision also requires improvements in station stability, lunar ephemerides,  CCR performance:

\begin{itemize}
\item \emph{Station tie stability:} Local ties between co-located instruments must maintain alignment within \(100\,\mu\)m to limit geodetic reference frame errors. Current \(\sim1\) mm station coordinate uncertainties must be reduced by a factor of 10 using advanced GNSS, VLBI, and SLR, ensuring their radial contribution remains \(\leq 50\,\mu\)m, which is necessary for achieving 0.1 mm LLR precision \citep{Williams:2014}.

    \item \emph{Lunar libration modeling:} Current lunar ephemerides exhibit residuals of \(\sim 5\) mm \citep{Park-etal:2021}. Improved lunar gravity models and high-cadence LLR observations must constrain libration effects to better than 0.1 mm.

\item \emph{Next-generation CCRs:} The multi-cube Apollo and Lunokhod arrays introduce pulse spreading of up to \(\pm 30\) mm due to their extended structure. Smaller, thermally stable 10 cm CCRs mitigate this but reduce photon return. Achieving 0.1 mm ranging requires higher laser power and optimized beam collimation \citep{Currie_etal_2011, Turyshev-etal:2013}.

\end{itemize}

While some proposals advocate for active transponders, their deployment faces challenges related to power, thermal control, and oscillator stability. They remain a promising alternative to high-precision passive retroreflectors \citep{Turyshev-etal:2010}.

\subsubsection{Summary for Pulsed LLR}

As we discussed above, achieving 0.1-mm LLR precision requires short-pulse lasers, ultra-stable frequency references, low-jitter detectors, multi-wavelength ranging, and real-time tropospheric modeling. Combined with thermally optimized CCRs or active transponders and refined lunar ephemerides, these advancements will enable high-precision tests of gravitational physics, including EP violations and constraints on the variability of the gravitational constant \citep{Williams:2004, Turyshev:2008, Murphy:2013}, while improving models of the lunar interior, tidal dissipation, and Earth–Moon dynamics \citep{Williams2012, Park-etal:2021}. These developments will further establish LLR as a cornerstone of high-precision geodesy and fundamental physics.

High-power continuous-wave lasers offer the most straightforward solution to overcome the photon collection and precision ranging challenges posed by smaller next-generation CCRs, as shown in Table~\ref{tab:photon-flux-scaling}. Combined with advancements in detection technologies and environmental corrections, these systems are essential for advancing beyond the current mm-level precision. These critical innovations will be discussed in the next section.

\section{LLR with High-Power Continuous-Wave Lasers}
\label{sec:highpower_cw_expanded}

High-power CW laser ranging to the Moon departs significantly from traditional pulsed-laser LLR in its method of encoding and extracting the \emph{round-trip travel time} (\(\Delta t\)). Instead of transmitting short bursts at high peak power, a CW system continuously illuminates the lunar retroreflectors with kW-level average power. This uninterrupted photon stream can be modulated in amplitude or frequency, and the returned signal is measured coherently. Under ideal conditions of phase stability, large-aperture telescopes, and multi-frequency or multi-chirp strategies, the technique can—in principle—push Earth--Moon ranging toward sub-\(10\,\mu\mathrm{m}\) accuracy. 

This section details how range observables are constructed from CW modulations, how \emph{multi-second} Doppler shifts are handled, and how the SNR governs final precision. Expressions are provided for both amplitude modulation (AM) and frequency-modulated continuous wave (FM) approaches, emphasizing their applicability to LLR.

\subsection{Fundamental Geometry, Doppler, and Precision Requirements}
\label{sec:fundamental_geometry}

In an LLR experiment, the \emph{primary observable} is the total round-trip time \(\Delta t\) of photons traveling from the ground station to a reflector on the Moon and back. If \(r_{\tt EM}\) denotes the Earth--Moon distance and \(c\) is the speed of light, then
\begin{equation}
    \Delta t \;=\; \frac{2r_{\tt EM}}{c}.
    \label{eq:LLR_geom}
\end{equation}
Numerically, \(r_{\tt EM} \approx 3.84\times10^8\,\mathrm{m}\) and thus \(\Delta t \approx 2.56\,\mathrm{s}\).  If the desired \emph{one-way} distance precision is \(\delta r_{\tt EM} \approx 30\,\mu\mathrm{m}\), the corresponding fractional stability requirement is
\begin{equation}
    \frac{\delta r_{\tt EM}}{r_{\tt EM}}
    \;\approx\;
    7.80\times10^{-14},
    \quad
    \text{which implies}
    \quad
    \delta(\Delta t)
    \;\lesssim\;
    0.1\,\mathrm{ps}.
    \label{eq:LLR_fractional_stability}
\end{equation}

During this \(2.56\)-s round trip, the lunar  velocity \(v_{\tt EM}\sim1023\,\mathrm{m/s}\), changes the Earth-Moon range  by \(\Delta r_{\tt EM}\approx v_{\tt EM}\Delta t\simeq 2569.56\)\,m. This leads to a Doppler velocity ratio \(v_{\tt EM}/c \approx 3.41\times10^{-6}\) and entails that any CW modulation frequency must be stabilized or compensated in real time to avoid cm- or m-level biases in a \(\sim30\,\mu\mathrm{m}\)-precision experiment. These combined constraints on timing stability and Doppler compensation largely shape CW LLR system requirements.

Note that, in addition to orbital velocity, a comprehensive CW LLR model must account for all time-dependent contributions as shown in \eqref{eq:range_model}. Consequently, besides the traditional Earth–Moon range, it is important to introduce and discuss the range-rate observable for LLR. With current technology, this observable can yield highly precise measurements that may significantly enhance the scientific return of advanced CW LLR experiments \cite{Williams:2018}.

\subsection{Amplitude-Modulated (AM) Continuous Wave}
\label{sec:AM_observable}

\subsubsection{Multi-Frequency AM Model: Range and Range-Rate}

In a CW LLR system, a high-power (e.g., 1\,kW) laser beam is amplitude modulated at a radio frequency \(f_m\). The transmitted signal is modeled as
\[
s_{\mathrm{tx}}(t)=A(t)\cos\Bigl(2\pi f_m\,t\Bigr),
\]
where \(A(t)\) is the known modulation envelope and \(t\) denotes the local time at the station.

After reflection from a lunar CCR, the CW beam returns with a round-trip delay \(\tau\) (with \(\tau \approx \Delta t \approx 2.56\,\mathrm{s}\)) and is attenuated by a factor \(\alpha\). Neglecting minor scintillation and amplitude fluctuations, the received signal is given by
\[
s_{\mathrm{rx}}(t)=\alpha\,A(t-\tau)\cos\Bigl[2\pi f_m\,(t-\tau)\Bigr].
\]

To estimate the propagation delay \(\tau\), we perform a cross-correlation between the received signal and a replica of the transmitted envelope. The correlation function is defined as
\[
C(\tau')=\int_{T} s_{\mathrm{rx}}(t)\,A(t-\tau')\cos\Bigl[2\pi f_m\,(t-\tau')\Bigr]\,dt,
\]
where the integration is carried out over a time interval \(T\) covering the signal duration. Under ideal conditions, \(C(\tau')\) attains its maximum at \(\tau' = \tau\), thereby yielding an accurate estimate of the round-trip delay.

The propagation delay \(\tau\) introduces a phase lag in the modulated signal, defined by
\[
\phi\equiv2\pi f_m\,\tau.
\]
Since the delay is related to the one-way range \(R\) via
\[
\tau=\frac{2R}{c},
\]
with \(c\) denoting the speed of light, the nominal round-trip phase lag becomes
\begin{equation}
\phi=2\pi f_m\,\frac{2R}{c}=\frac{4\pi f_m R}{c}.
\label{eq:phase_lag_AM}
\end{equation}
In practice, the phase \(\phi\) is measured by comparing the received signal with a stable local oscillator at the modulation frequency. However, since phase is inherently measured modulo \(2\pi\), the absolute phase is ambiguous by an integer multiple of \(2\pi\); that is, the measured phase is of the form \(\phi+2\pi N\) with \(N\in\mathbb{Z}\). This ambiguity directly leads to an uncertainty in the absolute range, which is given by
\begin{equation}
R=\frac{c}{4\pi f_m}\bigl(\phi+2\pi N\bigr).
\label{eq:AM_distance_equation}
\end{equation}
Multiple modulation frequencies are employed to resolve this integer ambiguity and determine the correct value of \(N\).

When the target (e.g., the Moon) is in motion, the delay \(\tau\) varies with time. Differentiating the range expression with respect to time yields the range-rate:
\[
\dot{R}=\frac{d}{dt}\left(\frac{c\,\tau}{2}\right)=\frac{c}{2}\,\frac{d\tau}{dt}.
\]
Alternatively, since \(\phi=2\pi f_m\,\tau\), differentiation with respect to time gives
\[
\dot{\phi}=2\pi f_m\,\dot{\tau}\quad\Longrightarrow\quad \dot{\tau}=\frac{\dot{\phi}}{2\pi f_m},
\]
so that the range-rate can also be expressed as
\begin{equation}
\dot{R}=\frac{c}{2\omega_m}\,\dot{\phi}=\frac{c\,\dot{\phi}}{4\pi f_m},
\label{eq:AM_RR_equation}
\end{equation}
where \(\dot{\phi}\) is the time derivative of the phase.

It is important to note that while the absolute phase measurement is ambiguous modulo \(2\pi\) (and thus the absolute range has an ambiguity due to the unknown integer \(N\)), the range-rate \(\dot{R}\) is derived from the time derivative of the phase. When the phase is properly unwrapped over time—tracking its continuous evolution—the constant \(2\pi N\) term does not contribute to \(\dot{\phi}\). Therefore, provided that phase unwrapping is performed correctly, the range-rate measurement is unambiguous.

\subsubsection{Range and Range-Rate Uncertainty for AM LLR}
\label{sec:AM-range-uncertainty}

In an AM LLR, one measures the round-trip phase shift $\phi$ of an RF intensity modulation at frequency $f_m$.  During  the total round-trip time is roughly $2r_{\tt EM}/c\simeq2.56\,\mathrm{s}$, the RF modulation accumulates a significant phase shift, from which $R$ is inferred.  Specifically, from (\ref{eq:AM_distance_equation})
a small phase uncertainty $\delta\phi$ maps into \cite{Rife1974,kay1993,vantrees2001}
\begin{equation}
  \delta R_{\mathrm{AM}} 
  \;=\;
  \frac{c}{4\pi f_m}\delta\phi.
  \label{eq:deltaR_AM_phase}
\end{equation}
For instance, at \(f_m = 1\,\mathrm{GHz}\) the factor \(c/(4\pi f_m)\) is  \(\approx 2.39\,\mathrm{cm}\); hence, a phase error \(\delta\phi\) of \(1\times10^{-3}\,\mathrm{rad}\) yields a range uncertainty of 23.9\,$\mu$m.

In the ideal photon-counting regime, where Poisson fluctuations dominate, the fundamental phase error is given by
\begin{equation}
  \delta\phi \approx \frac{1}{\mathrm{SNR}},
  \label{eq:phase-SNR}
\end{equation}
with the signal-to-noise ratio defined as
\begin{equation}
  \mathrm{SNR} = \frac{N_{\mathrm{sig}}}{\sqrt{N_{\mathrm{sig}}+N_{\mathrm{noise}}}}
  = \mathrm{SNR}_{1\,\mathrm{s}}\sqrt{\frac{T_{\mathrm{int}}}{1\,{\rm s}}},
  \label{eq:SNR_poisson}
\end{equation}
where \(N_{\mathrm{sig}} = \dot{N}_{\mathrm{sig}}\,T_{\mathrm{int}}\) and \(N_{\mathrm{noise}} = \dot{N}_{\mathrm{noise}}\,T_{\mathrm{int}}\) are the mean numbers of signal and noise photons detected over an integration time \(T_{\mathrm{int}}\). Substituting \eqref{eq:phase-SNR} into \eqref{eq:deltaR_AM_phase} yields the shot-noise limited range uncertainty:
\begin{equation}
    \delta R_{\mathrm{AM,\,shot}} \simeq \frac{c}{4\pi f_m}\,\frac{1}{\mathrm{SNR}_{1\,\mathrm{s}}\sqrt{T_{\mathrm{int}}/1\,\mathrm{s}}}.
    \label{eq:range-shot}
\end{equation}

Similarly, the range-rate is obtained from the time derivative of the range \eqref{eq:AM_RR_equation}.
Assuming that the uncertainty in the phase derivative is limited by the same photon-counting statistics, one can approximate
\begin{equation}
\delta\dot{\phi} \approx \frac{1}{T_{\mathrm{int}}\,\mathrm{SNR}}=\frac{1}{T_{\mathrm{int}}\,\mathrm{SNR}},
\label{eq:delta_phi_dot}
\end{equation}
so that the uncertainty in the range-rate becomes
\begin{equation}
    \delta\dot{R}_{\mathrm{AM,\,shot}} = \frac{c}{4\pi f_m}\,\delta\dot{\phi} \approx \frac{c}{4\pi f_m}\,\frac{1}{T_{\mathrm{int}}\,\mathrm{SNR}}=
 \frac{c}{4\pi f_m }\,   \frac{1}{T_{\mathrm{int}}\,\mathrm{SNR}_{\mathrm{s}}\sqrt{T_{\mathrm{int}}/1\,\mathrm{s}}}.
    \label{eq:range_rate_uncertainty}
\end{equation}

Thus, the shot-noise–limited uncertainties in range and range-rate are determined by \eqref{eq:range-shot} and \eqref{eq:range_rate_uncertainty}, respectively. Achieving sub-0.1 mm precision in range and high accuracy in range-rate requires high signal photon rates (from kW-class transmit power and large receiving apertures), sufficiently long integration times, and low noise levels. Multiple modulation frequencies are employed to resolve the \(2\pi\) ambiguity in the absolute range, while continuous phase unwrapping yields an unambiguous range-rate measurement.

\subsection{Frequency-Modulated (FM) Continuous Wave}
\label{sec:FM_observable}

\subsubsection{Linear Chirp FM Model:  Range and Range-Rate}
\label{sec:FM_model}

In an FM LLR system the transmitted signal is frequency modulated with a linear chirp and is modeled as
\[
s_{\text{tx}}(t)=A\cos\Bigl[2\pi (f_m t+{\textstyle\frac{1}{2}}\beta t^2\big)\Bigr],
\]
where \(A\) is the constant amplitude, \(f_m\) is the RF modulation frequency, and \(\beta\) (in Hz/s) is the chirp rate.  

After reflection from a lunar retroreflector, the received signal is delayed by \(\tau\) and attenuated by a factor \(\alpha\), so that
\[
s_{\text{rx}}(t)=\alpha\,A\cos\Bigl[2\pi f_m (t-\tau)+\pi\beta (t-\tau)^2\Bigr].
\]

A heterodyne detection scheme is used in which the received signal is mixed with a local oscillator that replicates the transmitted signal. The resulting beat signal has a phase given by the difference
\[
\Delta\phi(t)=\Bigl[2\pi f_m t+\pi\beta t^2\Bigr]-\Bigl[2\pi f_m(t-\tau)+\pi\beta (t-\tau)^2\Bigr].
\]

Expanding the delayed phase and neglecting the small quadratic term \(\pi\beta\tau^2\), one obtains
\[
\Delta\phi(t)\approx 2\pi f_m\tau+2\pi\beta t\,\tau.
\]
Differentiating with respect to time and dividing by \(2\pi\) yields the instantaneous beat frequency
\begin{equation}
f_b(t)=\frac{1}{2\pi}\frac{d}{dt}\Delta\phi(t)\approx \beta\,\tau \quad\Longrightarrow\quad
\tau=\frac{f_b}{\beta}\,.
\label{eq:freq_b-FM}
\end{equation}
Since the one-way range is related to the delay by $R={\textstyle\frac{1}{2}} {c\,\tau} $ one has
\begin{equation}
R= \frac{c}{2\beta}\,f_b\,.
\label{eq:FM_beat_freq_chirp}
\end{equation}
 
If the target is moving, the delay (and hence the beat frequency) becomes time-dependent, and differentiating the range with respect to time yields the range-rate
\begin{equation}
\dot{R}=\frac{c}{2}\,\frac{d\tau}{dt}=\frac{c}{2\beta}\,\frac{d f_b}{dt}\,.
\label{eq:range-rate-FM}
\end{equation}

Range ambiguity is inherent to FM LLR because the measured phase is wrapped modulo \(2\pi\), which can lead to an ambiguity in the absolute range. One approach to resolve this is to use a long chirp that covers the entire Earth–Moon distance. Alternatively, employing several shorter chirps with different chirp rates, \(\beta_i\), allows one to apply multi-slope processing and phase unwrapping techniques. This strategy effectively combines the large unambiguous range of a gentle chirp with the fine resolution of a steep chirp. Importantly, while the absolute range measurement is affected by phase wrapping, the range-rate is derived from the time derivative of the phase. With proper phase unwrapping, any constant \(2\pi\) offset is eliminated upon differentiation, resulting in an unambiguous range-rate measurement.

Range resolution improves with increased chirp bandwidth (since \(\beta = B/T_{\mathrm{chirp}}\)) or by reducing the sweep time \(T_{\mathrm{chirp}}\). Therefore, precise calibration of \(f_m\) and \(\beta\), high-resolution measurement of the beat frequency \(f_b(t)\), and stable phase references are essential for accurate determination of both range and range-rate.

\subsubsection{Range and Range-Rate Uncertainty for FM LLR}
\label{sec:FM-range-uncertainty}

In an FM CW LLR experiment, distance is inferred from the beat frequency \(f_b\) generated by heterodyning a transmitted linear chirp with the weak return signal from the Moon. Suppose the optical frequency is swept linearly at rate \(\beta\) (in Hz/s) over a chirp period \(T_{\mathrm{chirp}}\) (yielding a bandwidth \(B=\beta\,T_{\mathrm{chirp}}\)). With a round‐trip time \(\Delta t\) ($\sim$\,2.56\,s), the observed beat frequency scales as 
$
f_b \approx \beta\,\Delta t,
$
which directly relates to the range (\ref{eq:freq_b-FM}). For an ideal linear chirp, as seen from \eqref{eq:FM_beat_freq_chirp}, a small beat‐frequency uncertainty \(\delta f_b\) maps into a range uncertainty
\[
\delta R_{\mathrm{FM}} = \frac{c}{2\beta}\,\delta f_b.
\]

In the shot-noise–limited regime, coherent detection theory \citep{Rife1974,kay1993,vantrees2001} indicates that the fractional beat-frequency error scales as
$
\delta f_b \approx {1}/(T_{\mathrm{chirp}}\cdot {\mathrm{SNR})},
$
where the signal-to-noise ratio is defined as in \eqref{eq:SNR_poisson} and $T_{\rm int}\geq T_{\rm chirp}$. Substituting this into the expression for \(\delta R_{\mathrm{FM}}\) yields the shot-noise–limited range uncertainty:
\[
\delta R_{\mathrm{FM, shot}} \approx \frac{c}{2\beta\,T_{\mathrm{chirp}}}\,\frac{1}{\mathrm{SNR}} 
\approx \frac{c}{2B}\,\frac{1}{\mathrm{SNR}_{1\,\mathrm{s}}\sqrt{T_{\mathrm{int}}/1\,\mathrm{s}}}\,.
\]

Similarly, the range-rate is determined from the time derivative of the range. From (\ref{eq:range-rate-FM}), if the uncertainty in the beat-frequency derivative is \(\delta({d f_b}/{dt})\), then the corresponding range-rate uncertainty is
\[
\delta\dot{R} = \frac{c}{2\beta}\,\delta\Big(\frac{d f_b}{dt}\Big).
\]
Under shot-noise–limited conditions, one may approximate
$
\delta\left({d f_b}/{dt}\right) \approx {1}/({T_{\mathrm{chirp}}^2\cdot\mathrm{SNR}})
$, where $T_{\rm int}\geq T_{\rm chirp}$, 
so that
\[
\delta\dot{R} \approx \frac{c}{2\beta}\,\frac{1}{T_{\mathrm{chirp}}^2\,\mathrm{SNR}} 
\approx \frac{c}{2B}\,\frac{1}{T_{\mathrm{chirp}}\,\mathrm{SNR}_{1\,{\rm s}}\,\sqrt{T_{\mathrm{int}}/1\,\mathrm{s}}}\,.
\]

It is important to note that although the absolute phase measurement is ambiguous modulo \(2\pi\) (leading to an integer ambiguity in \(R\)), proper phase unwrapping ensures that the phase derivative is continuous and unambiguous. Consequently, the range-rate \(\dot{R}\) is free from the \(2\pi\) ambiguity.

Achieving sub–0.1 mm shot-noise precision in range—and correspondingly high accuracy in range-rate—requires high signal photon rates (from kW-class transmit power and large receiving apertures), sufficiently long integration times \(T_{\mathrm{int}}\), and low noise levels. In addition, precise calibration of the chirp rate \(\beta\) over \(T_{\mathrm{chirp}}\) and stable phase references are essential.

\subsection{Phase Unwrapping: Multi-Frequency / Multi-Chirp Protocol}
\label{sec:multi_freq_multi_chirp}

As discussed in Sections~\ref{sec:AM_observable}--\ref{sec:FM_observable}, single-frequency amplitude modulation (AM) or a single linear chirp (FM) cannot directly span the \(\sim384\,000\,\mathrm{km}\) Earth–Moon baseline without ambiguity. Each modulation frequency (\(f_m\)) or chirp slope (\(\beta\)) restricts the unambiguous range to
\begin{equation}
    d_{\mathrm{max}} 
    \;=\; \frac{c}{2f_m} 
    \quad \text{(for AM)} 
    \qquad \text{or} \qquad
    d_{\mathrm{max}} 
    \;\approx\; \frac{c}{2\beta} f_b
    \quad \text{(for FM)},
    \label{eq:multi_freq_unambig}
\end{equation}
as given in \eqref{eq:AM_distance_equation} and \eqref{eq:FM_beat_freq_chirp}. For instance, a single GHz-class tone in AM yields \(d_{\mathrm{max}} \approx 15\,\mathrm{cm}\), far smaller than the Earth–Moon distance. Moreover, the Moon's radial velocity (\(\lesssim 1\,\mathrm{km/s}\)) can impart Doppler offsets ranging from tens of kilohertz to megahertz, depending on the carrier or modulation frequency.

A well-established solution is to employ multiple modulation tones (AM) or multiple chirp sweeps with different slopes (FM), thereby narrowing down the absolute distance from a coarse scale to a fine scale.

\subsubsection{Coarse-to-Fine Frequency (AM) or Slope (FM) Approach}

A typical multi-frequency AM protocol may use three or more discrete RF tones \(f_1, f_2, f_3, \dots\) organized as follows:
\begin{enumerate}
  \item \textit{Coarse Tone} (\(f_1 \sim 10\text{--}50\,\mathrm{MHz}\)):  
  Provides a synthetic wavelength of \(c/(2f_1) \approx 3\text{--}15\,\mathrm{m}\). Measuring the round-trip phase at \(f_1\) localizes the range \(r_{\tt EM}\) to a bin of roughly a few meters (modulo the coarse-wavelength spacing).

  \item \textit{Intermediate Tone} (\(f_2 \sim 100\text{--}500\,\mathrm{MHz}\)):  
  Refines the distance to the centimeter or millimeter level since each full \(2\pi\) cycle now corresponds to a shorter distance of \(c/(2f_2)\). Phase unwrapping between \(f_1\) and \(f_2\) yields a unique solution over the full Earth–Moon baseline.

  \item \textit{Fine Tone} (\(f_3 \sim 1\text{--}10\,\mathrm{GHz}\)):  
  Finally, a high-frequency tone locks in sub-mm or even tens-of-\(\mu\)m precision once the coarse and intermediate phases have identified the correct ambiguity bin. Equation~\eqref{eq:phase_lag_AM} (for AM) or Equation~\eqref{eq:FM_beat_freq_chirp} (for FM) then translates the measured phase or beat frequency into a fine distance \(r_{\tt EM,\,fine}\).
\end{enumerate}

Mathematically, if \(\phi_i\) denotes the measured round-trip phase at tone \(f_i\), then
\[
\phi_i = 2\pi f_i \frac{2r_{\tt EM}}{c} + \phi_{\mathrm{Dopp}}(t) + \delta_{\mathrm{noise}},
\]
where \(\phi_{\mathrm{Dopp}}(t)\) accounts for Doppler shifts due to the Moon's velocity, and \(\delta_{\mathrm{noise}}\) aggregates detector shot noise, turbulence-induced wavefront fluctuations, and oscillator jitter. Each \(\phi_i\) is first mapped into a set of possible distances separated by \(c/(2f_i)\), and the combination of all \(\{\phi_i\}\) yields a unique solution for \(r_{\tt EM}\) over the full \(\sim384\,000\,\mathrm{km}\).

\subsubsection{Multi-Chirp or Multi-Slope FM Variant}

An FM LLR station may replace discrete RF tones with several chirp sweeps characterized by slopes \(\beta_1, \beta_2, \dots\). Each chirp sweep produces a beat frequency that encodes the round-trip delay to the Moon. From (\ref{eq:freq_b-FM}), for a chirp with slope \(\beta_i\), the beat frequency is given by
\[
f_{b,i} = \beta_i\,\tau = \beta_i \frac{2r_{\tt EM}}{c}.
\]
A typical multi-chirp FM protocol includes:
\begin{enumerate}
  \item \textit{Gentle Chirp} (\(\beta_1\)):  
        Provides a large unambiguous range of \(c/(2\beta_1)\), spanning thousands of kilometers, and localizes \(r_{\tt EM}\) to a bin of roughly a few meters.
  \item \textit{Intermediate Slope} (\(\beta_2\)):  
        Refines the range to the centimeter or millimeter level since each full \(2\pi\) cycle corresponds to a shorter distance \(c/(2\beta_2)\). Phase unwrapping between \(\beta_1\) and \(\beta_2\) yields a unique solution.
  \item \textit{Steep Slope} (\(\beta_3\)):  
        Achieves sub-mm precision once the correct ambiguity bin is identified; the measured beat frequency is then converted into a fine distance \(r_{\tt EM,\,fine}\) via Equation~\eqref{eq:phase_lag_AM} (for AM) or Equation~\eqref{eq:FM_beat_freq_chirp} (for FM).
\end{enumerate}
Here, the chirp rate is defined as \(\beta_i = B/T_{\mathrm{chirp}}\), where \(B\) is the chirp bandwidth and \(T_{\mathrm{chirp}}\) is the duration of the chirp. High linearity in \(\beta_i\) (on the order of \(10^{-6}\)) ensures that systematic errors remain below the centimeter level during phase unwrapping. Doppler shifts appear as time-varying offsets in \(f_b(t)\); knowledge of the Moon's velocity profile enables real-time or post-processed correction of the Doppler-induced phase offset \(\phi_{\mathrm{Dopp}}\).

\subsubsection{Range-Rate as a New LLR Observable}

Although absolute phase measurements are inherently ambiguous modulo \(2\pi\) — leading to an integer ambiguity in the absolute range \(r_{\tt EM}\) — the range-rate is determined from the time derivative of the phase. With proper phase unwrapping over time, any constant \(2\pi\) offset is eliminated upon differentiation, resulting in a continuous and unambiguous phase derivative. Consequently, the range-rate, which is obtained from the derivative of the beat frequency, is inherently free from the \(2\pi\) ambiguity. This characteristic allows the range-rate to serve as an independent, highly precise observable in lunar laser ranging, thereby complementing traditional range measurements and providing additional constraints on the Earth–Moon system dynamics. In addition to improving absolute range estimates, incorporating range-rate measurements can enhance many LLR-derived science parameters \cite{Williams:2018}.
 
 \subsection{Systematic Error Contributions for CW  LLR}
\label{sec:systematicsAMFM}

While the expressions in Sections~\ref{sec:AM-range-uncertainty} and \ref{sec:FM-range-uncertainty} detail how shot noise sets a fundamental limit in amplitude‐modulated (AM) or frequency‐modulated (FM) LLR, realistic experiments face additional systematic effects.  In many scenarios, these effects dominate over the ideal \emph{photon‐counting} (or \emph{beat‐frequency}) precision:

\begin{itemize}
\item \textit{Atmospheric turbulence} $(\delta R_{\mathrm{atm}})$:
  Without multi‐wavelength compensation, turbulent cells in the troposphere can shift the measured path by tens to hundreds of micrometers, especially over multi‐second integrations.

\item \textit{Mechanical drifts \& thermal expansion} $(\delta R_{\mathrm{mech}})$:
  Telescopes, mirrors, or optical benches can move by micrometers due to temperature changes, vibration, or dome seeing, introducing quasi‐static offsets in either the AM phase or the FM beat frequency. (If the optical bench is made of steel, and its length is 1 m, then a $0.1^\circ$\,C change can cause $\sim1\,\mu\mathrm{m}$ expansion.)

\item \textit{Oscillator/frequency reference} $(\delta R_{\mathrm{osc}})$:
  In AM, drifts in the modulation frequency $f_m$ can masquerade as range changes unless stabilized.  In FM, slope ($\beta$) or local‐oscillator fluctuations similarly corrupt the beat frequency.  Calibration to an ultra‐stable reference (e.g., hydrogen maser, frequency comb) is crucial.

\item \textit{Electronic offsets and calibration} $(\delta R_{\mathrm{elec}})$:
  Cable delays, digitizer timing, or partial instrument calibration each introduce biases from a few micrometers to millimeters.  Regular station calibration and consistent local references can reduce these offsets.

\item \textit{Doppler offset} $(\delta R_{\mathrm{Dopp}})$:
  The Moon’s $\pm1\,\mathrm{km/s}$ radial velocity shifts either the returned AM phase or the FM beat by tens/hundreds of kHz at GHz scale.  Ephemeris‐based precompensation or post‐fit velocity corrections are essential to avoid cm‐level systematic errors.
\end{itemize}

To quantify their combined impact, one typically sums the relevant terms in a root‐sum‐square (RSS) budget:
\begin{equation}
  \delta R_{\mathrm{total}}^2
  \;=\;
  \delta R_{\mathrm{AM/FM, shot}}^2
  \;+\;
  \delta R_{\mathrm{atm}}^2
  \;+\;
  \delta R_{\mathrm{mech}}^2
  \;+\;
  \delta R_{\mathrm{osc}}^2
  \;+\;
  \delta R_{\mathrm{elec}}^2
  \;+\;
  \delta R_{\mathrm{Dopp}}^2
  \;+\;\dots
  \label{eq:total-range-error-FM}
\end{equation}
If unmitigated, turbulence, thermal drifts, or oscillator instabilities can exceed the random  limit by an order of magnitude or more.  Nonetheless, lowering the shot‐noise floor to a few tens of $\mu$m is indispensable for next‐generation sub‐mm LLR; once the \emph{photon‐counting or beat‐frequency} precision is no longer limiting, improvements in turbulence control, thermal stabilization, and Doppler calibration can push routine Earth--Moon distance measurements toward the tens‐of‐$\mu$m domain.  As a result, high‐power (kW‐class) CW lasers, large‐aperture telescopes, and sophisticated station hardware provide a robust path to sub‐mm or even tens‐of‐$\mu$m performance for both AM and FM LLR.

 \subsection{Outlook for High-Power CW LLR}
\label{sec:submm-prospects}

High-power CW lasers offer a compelling approach for advancing LLR into the sub-mm domain, even when dealing with the weak return signals of next-generation small CCRs. By modulating the beam in amplitude or frequency at kW power levels, these systems can achieve tens-of-$\mu$m precision. However, four critical challenges must be addressed:

\begin{enumerate}[1.]
    \item \textit{Distance Unambiguous Resolution:}  
    Multi-frequency schemes (in AM) or multi-chirp schemes (in FM) must be designed to span the \(\sim 4 \times 10^8 \, \mathrm{m}\) Earth--Moon baseline without ambiguity. This requires precise control over phase measurements and beat-frequency calculations to avoid integer-cycle errors in the final range.
      
    \item \textit{Reference Stability:}  
    Achieving sub-ps timing stability over the 2.56\,s round-trip delay requires oscillator fractional drifts on the order of \(\sim10^{-13}\). Hydrogen masers or optical frequency combs can meet this requirement, but calibration offsets or frequency drifts must be continuously monitored to prevent systematic errors that translate into millimeter or larger range biases.

    \item \textit{Real-Time Doppler Monitoring:}  
    The Moon’s \(\pm 1 \, \mathrm{km/s}\) radial velocity induces Doppler shifts that vary over the 2.56\,s round-trip, potentially shifting the measured phase or beat frequency by tens of kHz. Both hardware-level corrections and post-processing adjustments—based on accurate ephemerides—are necessary to track these shifts with high accuracy, thereby maintaining a \(\sim30 \, \mu\mathrm{m}\) range error floor.

    \item \textit{Mechanical and Thermal Stability:}  
    Even small thermal expansions or mechanical drifts can introduce range errors at the tens-of-\(\mu\)m scale. The use of low-expansion materials and rigorous environmental control is essential to ensure consistent performance.
\end{enumerate}

Under optimal conditions, high-power CW LLR can exceed current few-mm precision, achieving sub-100\,\(\mu\mathrm{m}\) accuracy. Shot-noise estimates indicate that \(\sim30 \, \mu\mathrm{m}\) precision is attainable with \(10^5\)--\(10^6\) detected photons over integration times of 10s to 100s of seconds, provided that mechanical, thermal, and systematic errors are effectively mitigated.

\begin{table}[h!]
\centering
\caption{Comparison of AM and FM methods for CW LLR.}
\label{tab:AM_vs_FM}
\begin{tabular}{|l|c|c|}
\hline
{Characteristic}        & {Amplitude Modulation (AM)}     & {Frequency Modulation (FM)}      \\ \hline\hline
Implementation Complexity      & Moderate                              & Advanced (reflecting high precision requirements)       \\ \hline
Modulation Type                & Intensity (Envelope)                   & Frequency (Phase Chirp)                 \\ \hline
Unambiguous Range              & \(c/(2f_m)\)                         & \(c/(2B)\) (extendable via multi-chirp) \\ \hline
Noise Sensitivity              & Sensitive to amplitude noise          & Robust via coherent detection            \\ \hline
Reference Stability            & Moderate RF stability                 & Requires ultra-stable RF, e.g., \(10^{-13}\) fractional stability \\ \hline
RF Hardware Requirements       & Standard RF electronics               & High-performance RF synthesis and ADCs   \\ \hline
Computational Complexity       & Moderate                              & High (real-time frequency tracking)       \\ \hline
Cost and Scalability           & More cost-effective                   & Higher investment, justified by ultimate precision \\ \hline
\end{tabular}
\end{table}

Both AM and FM provide viable approaches for CW LLR, each with distinct benefits. While AM modulates the intensity (envelope) of the optical carrier and can be implemented using standard RF electronics, FM directly modulates the optical frequency via phase chirping, offering enhanced noise immunity and finer range resolution. Table~\ref{tab:AM_vs_FM} summarizes the key differences.

In summary, both  techniques have reached a mature state with well-established hardware and signal processing methodologies. In practical LLR systems today, AM offers a robust and cost-effective means of achieving absolute range measurements with moderate complexity, while FM delivers enhanced phase sensitivity and finer resolution. The choice between them is already available and proven; indeed, many current systems successfully implement one or the other. In some cases, a hybrid approach—combining the robust absolute range capability of AM with the fine-scale precision of FM—may provide an optimal solution without waiting for future technological advances.

\section{Link Budget for High Power LLR}
\label{sec:flux_model}

Achieving sub-mm and potentially tens-of-$\mu$m accuracy in LLR requires not only advanced timing and phase measurement systems but also sufficient photon flux to minimize shot noise. Historically, pulsed-laser LLR stations using short ($\sim10\,\mathrm{ns}$) pulses with direct time-of-flight (TOF) detection have achieved single-shot precision at or near the millimeter level \cite{Degnan1993,Samain1998}. Recent advances in high-power CW laser technology and multi-frequency phase metrology indicate that CW-based architectures are strong contenders for achieving sub-mm accuracy.

\subsection{Representative System Parameters}
\label{sec:param_table}

We assume a 1\,m telescope, a \SI{10}{cm} CCR, and good seeing conditions (Fried parameter \(r_0 = 20\,\mathrm{cm}\) at 1064\,nm). Atmospheric turbulence and telescope diffraction define the outbound beam divergence, with wavefront errors uncorrected beyond basic optics. Table~\ref{tab:key-parameters-improved} outlines the baseline parameters used to estimate photon return rates.  These parameters apply to both CW and high-average-power pulsed laser systems at \(\lambda \approx 1064\,\mathrm{nm}\), paired with 1--2\,m-class telescopes and high-efficiency single-photon detectors (e.g., superconducting nanowire SPDs). In addition, we also included stellar aberration loss of $0.75$ for a CCR without dihedral offset angles which is typically 0.64--0.86 \cite{Murphy:2013,Murphy:2006}. For consistency, we assume a single \SI{10}{cm} CCR, consistent with modern lunar CCR designs \citep{Currie_etal_2011,Turyshev-etal:2013}.

\begin{table}[htb]
\centering
\caption{Baseline parameters for a \SI{1}{kW} CW beam at 1064\,nm, reflecting from a \SI{10}{cm} lunar CCR and collected by a \SI{1}{m} telescope. The approximate total system efficiency, \(\eta_{\text{eff}} \approx 0.27\), includes atmospheric transmission (both out and back), telescope throughput, CCR efficiency, and detector quantum efficiency.}
\label{tab:key-parameters-improved}
\begin{tabular}{lcl}
\hline
Parameter & Value & Comments \\
\hline\hline
\multicolumn{3}{l}{\textit{Laser Parameters}}\\
Wavelength \((\lambda)\) & \SI{1064}{nm}
  & NIR window, minimal atmospheric absorption \\
Laser Power \((P_{\text{laser}})\) & \SI{1}{kW}
  & CW or high-average-power pulsed \\
Beam Quality \((M^2)\) & 1.1
  & Near-diffraction-limited \\
\hline
\multicolumn{3}{l}{\textit{Telescope Parameters}}\\
Aperture \((d_{\text{tel}})\) & \SI{1}{m}
  & Controls diffraction and photon collection \\
Atmospheric Seeing \((r_0)\) & \SI{20}{cm}
  & Improved turbulence conditions at 1064\,nm \\
Beam Divergence \((\theta_{\mathrm{div}})\) & \(5.48\,\mu\mathrm{rad}\)
  & Combined diffraction + turbulence \\
\hline
\multicolumn{3}{l}{\textit{Retroreflector Parameters}}\\
CCR Diameter \((d_{\rm CCR})\) & \SI{10}{cm}
  & Representative lunar corner-cube size \\
CCR Efficiency \((\eta_{\text{CCR}})\) & 0.7
  & May reach 0.8-0.9 in newer designs \\
Stellar Aberration loss \((\eta_{\text{SA}})\) & 0.75
  & For dihedral angle offset  $(0''.0''.0'')$ \\
\hline
\multicolumn{3}{l}{\textit{Atmospheric/Optical Efficiencies}}\\
Atmospheric Transmission \((\eta_{\text{atm}})\) & 0.85
  & Applies each way (out and back) \\
Telescope Throughput \((\eta_{\text{tel}})\) & 0.8
  & Mirror and coating losses \\
Detector QE \((\eta_{\text{det}})\) & 0.8
  & e.g., superconducting nanowires \\
\hline
\end{tabular}
\end{table}

The total round-trip efficiency is conveniently expressed as
\begin{equation}
\eta_{\mathrm{eff}}
=
\eta_{\text{tel}}^2
\times
\eta_{\text{atm}}^2
\times
\eta_{\text{CCR}}
\times \eta_{\text{SA}}
\times
\eta_{\text{det}}
\approx 0.194.
\end{equation}
Improved coatings, retroreflector design, ultra-low-loss optics could increase \(\eta_{\mathrm{eff}}\) to 0.35--0.40, boosting photon return.

\subsection{Photon Flux Equations for CW Lasers}
\label{sec:flux_equations}

\subsubsection{Laser Emission Rate and Single-Photon Energy}

A single photon at \(\lambda = \SI{1064}{nm}\) carries
\begin{equation}
E_{\mathrm{photon}}
=
\frac{h c}{\lambda}
\;\approx\;
1.87\times10^{-19}\,\mathrm{J},
\label{eq:photon_energy}
\end{equation}
where \(h\) is Planck’s constant and \(c\) is the speed of light. Consequently, a \SI{1}{kW} beam emits
\begin{equation}
\frac{P_{\text{laser}}}{E_{\mathrm{photon}}}
\,=\,
\frac{1000\,\mathrm{J/s}}{1.87\times10^{-19}\,\mathrm{J}}
\,\approx\,
5.35\times10^{21}\,\mathrm{photons\,s}^{-1}.
\label{eq:emission_rate}
\end{equation}

\subsubsection{Outgoing and Return Footprints}
\label{sec:beam_divergence}

The total beam divergence accounts for both diffraction and wavefront distortions from atmospheric turbulence:
\begin{equation}
\theta_{\mathrm{div}}
=
\sqrt{
  \biggl(\frac{1.22 \lambda}{d_{\text{tel}}}\biggr)^2
  \;+\;
  \biggl(\frac{\lambda}{r_0}\biggr)^2
}
\,\approx\,
5.48\times10^{-6}\,\mathrm{rad}. 
\label{eq:theta-nonAO}
\end{equation}

For an Earth--Moon center-to-center distance $r_{\tt EM} \approx 3.84\times10^8\,\mathrm{m}$, the beam spot on the Moon attains kilometer-scale dimensions. The return beam divergence from CCR far-field diffraction is:
\begin{equation}
\theta_{\mathrm{return}}
\approx
\frac{1.22\lambda}{d_{\rm CCR}}
\,\approx\,
12.98\times10^{-6}\,\mathrm{rad},
\end{equation}
yielding a tens-of-kilometers spot on Earth’s surface (see Table~\ref{tab:spot_areas_flux}).

\subsubsection{Link-Budget Formula}

Combining the beam geometry, two-way transmission, and collection aperture yields an approximate flux at the detector \cite{Degnan1993,Samain1998}:
\begin{equation}
\dot N_{\mathrm{photons}}
=
\frac{P_{\text{laser}}}{E_{\mathrm{photon}}}
\,\times\,
\eta_{\mathrm{eff}}
\,\times\,
\frac{A_{\mathrm{CCR}}}{A_{\mathrm{spot,Moon}}}
\,\times\,
\frac{A_{\mathrm{tel}}}{A_{\mathrm{spot,Earth}}},
\label{eq:flux_eq_cw}
\end{equation}
where
\begin{gather}
A_{\mathrm{CCR}}
=
\pi\bigl(\tfrac{1}{2} d_{\mathrm{CCR}}\bigr)^2,
\quad
A_{\mathrm{tel}}
=
\pi\bigl(\tfrac{1}{2} d_{\text{tel}}\bigr)^2,
\quad
A_{\mathrm{spot,Moon}}
=
\pi\bigl(r_{\tt EM}\,\theta_{\mathrm{div}}\bigr)^2,
\quad
A_{\mathrm{spot,Earth}}
=
\pi\bigl(r_{\tt EM}\,\theta_{\mathrm{return}}\bigr)^2.
\end{gather}

Table~\ref{tab:spot_areas_flux} summarizes the estimates of each term in the link equation  \eqref{eq:flux_eq_cw}. It consolidates the spot-size and flux estimates, assuming stable beam quality ($M^2 \approx 1.1$) and constant atmospheric transmission. The results are striking: under a typical transmission scenario (e.g., Table~\ref{tab:key-parameters-improved}), a station operating a 1\,kW CW laser from a 1-m telescope targeting a 10\,cm CCR on the Moon is expected to achieve a return flux of \(\dot{N}_{\mathrm{photons}} = 5884\)~photons/s—a factor of at least \(2.7 \times 10^3\) greater than the flux anticipated from modern LLR stations (see Table~\ref{tab:photon-flux-scaling}).

\begin{table}[t]
\centering
\caption{Estimated beam footprints on the Moon and Earth with \(r_0 = 20\,\mathrm{cm}\), assuming a \SI{1}{kW} CW laser at 1064\,nm, a \SI{10}{cm} lunar CCR, and an overall efficiency \(\eta_{\mathrm{eff}} \approx 0.194\).}
\label{tab:spot_areas_flux}
\begin{tabular}{lcc}
\hline
{Quantity} & {Value} \\
\hline\hline
\multicolumn{2}{l}{\textit{Transmit beam divergence and spot on the Moon}} \\
Beam divergence \((\theta_{\mathrm{div}})\)
 & \(5.48\times10^{-6}\,\mathrm{rad}\)\\
Spot radius on Moon
 & \(\sim2.1\,\mathrm{km}\) \\
Spot area on Moon \((A_{\mathrm{spot,Moon}})\)
 & \(\sim1.39\times10^{7}\,\mathrm{m}^{2}\)\\
\hline
\multicolumn{2}{l}{\textit{Return beam divergence and spot on Earth}}\\
Return beam divergence \((\theta_{\mathrm{return}})\)
 & \(\sim12.98\times10^{-6}\,\mathrm{rad}\) \\
Spot radius on Earth 
 & \(\sim5.0\,\mathrm{km}\) \\
Spot area on Earth \((A_{\mathrm{spot,Earth}})\)
 & \(\sim7.8\times10^{7}\,\mathrm{m}^{2}\)\\
\hline
\multicolumn{2}{l}{\textit{Component Areas}}\\
CCR Area \((A_{\mathrm{CCR}})\)
 & \(\sim0.00785\,\mathrm{m}^{2}\) \\
Telescope Area \((A_{\mathrm{tel}})\)
 & \(\sim0.785\,\mathrm{m}^{2}\) \\
\hline
\multicolumn{2}{l}{\textit{Estimated photon flux}}\\
Photon rate \(\dot{N}_{\mathrm{photons}}\)
 & \(\sim5.88\times10^{3}\,\mathrm{s}^{-1}\)\\
\hline
\end{tabular}
\end{table}

\subsection{Comparison to Existing Pulsed Stations Ranging to New-Gen 10\,cm CCR}
\label{sec:comparison_scaled_ccr}

Many existing LLR stations, such as APOLLO (Apache Point) and Grasse (OCA), are optimized for larger multi-corner Apollo or Lunokhod arrays, which provide significantly greater reflective areas than a single \SI{10}{cm} corner cube. Table~\ref{tab:ccr_scaling} summarizes approximate photon-return rates (photons/s) under improved seeing conditions for three well-known pulsed-laser facilities, comparing their \emph{actual} multi-corner arrays to a hypothetical scenario where each station targets a single 10\,cm CCR. The table estimates returns based on the ratio of reflective areas and beam divergence angles, suggesting that APOLLO would detect only \(\sim2.17\) photons/s from a single 10\,cm cube (Table~\ref{tab:spot_areas_flux}).

The table also includes a notional LLR facility  operating a \SI{1}{kW} CW beam at 1064\,nm (Table~\ref{tab:key-parameters-improved}). Targeting the same 10\,cm reflector, the CW illumination delivers $\sim$ 5,884 photons per second (see Table~\ref{tab:spot_areas_flux}), significantly reducing shot noise. This enables sub-mm precision, provided other systematics—such as thermal drifts, oscillator instabilities, and atmospheric turbulence—are well-controlled (Appendices~\ref{sec:calibration_maintenance} and \ref{sec:hardware_components}). Note that, APOLLO operate at 532\,nm (OCA can do both: 532\,nm and 1064\,nm), while the notional CW LLR station operates at 1064\,nm, thus projecting twice the photon flux in the outgoing beam due to the higher energy of individual photons at longer wavelengths.

\begin{table}[htb]
  \centering
  \caption{Reported photon returns per second at two pulsed-laser LLR stations for their standard multi-corner arrays and for a single 10\,cm CCR, plus a \SI{1}{kW} CW concept on that same 10\,cm CCR. As in Table~\ref{tab:photon-flux-scaling}, the reported photon returns are from  \citep{Murphy_etal_2012,Murphy:2013} for APOLLO and \cite{Courde-etal:2017} for Grasse (OCA).}
  \label{tab:ccr_scaling}
  \begin{tabular}{lccc}
  \hline
  & {Aperture (m)} & {Apollo 15} & 
Single 10\,cm CCR\\
  {Station}
   & {}
   & {(photons/s)}
   & {(photons/s)}\\
  \hline
  APOLLO & 3.5 
    & 62.92
    & 2.17 \\
 OCA & 1.5 
    & 0.94
    & 0.03 \\
1\,kW CW & 1.0 
    & $\sim3.68\times 10^4$
    & \(\sim5.88\times10^3\)\\
  \hline
  \end{tabular}
\end{table}

In practice, the specific numbers in Table~\ref{tab:ccr_scaling} can vary due to aperture size, laser efficiency, and atmospheric conditions. Nonetheless, today’s pulsed-laser stations typically collect tens of photons per second on large Apollo 15 arrays but would detect only a fraction of that on a single 10\,cm CCR. By contrast, a \SI{1}{kW} CW laser from a 1-meter telescope at TMO compensates for the reduced retroreflector area, achieving significantly higher photon-return rates than pulsed LLR  with legacy multi-CCR arrays, paving the way for sub-mm or tens-of-\(\mu\mathrm{m}\)-level LLR measurements.

APOLLO currently operates a pulsed \(532\,\mathrm{nm}\) laser and a 3.5\,m telescope, firing \(\sim\!100\,\mathrm{mJ}\) pulses at \(20\)--\(30\,\mathrm{Hz}\) with an average power of \(\sim2.3\,\mathrm{W}\). Each pulse contains \(\sim3 \times 10^{18}\) photons, with only \(\sim 2\) photon per pulse detected under favorable seeing. Replacing APOLLO’s pulsed source with a \SI{1}{kW} CW laser at 1064\,nm (using the same aperture and high detector efficiency) could increase the return flux by orders of magnitude, reaching $\sim3.1\times 10^5$ of photons per second. This comparison underscores why kW-class CW lasers, paired with large-aperture telescopes and advanced phase metrology, may push LLR uncertainties well below the millimeter level, perhaps into the tens-of-\(\mu\mathrm{m}\) domain.

\subsection{Practical Caveats and Model Validation}

Equation~\eqref{eq:flux_eq_cw} encapsulates the core physics of the link budget but relies on simplifying assumptions. Several practical factors influence the achievable photon return flux:

\begin{itemize}
\item \emph{Retroreflector reflectivity:} The nominal $\eta_{\text{CCR}}=0.6$ factor for hollow CCRs can degrade over time due to dust accumulation, micrometeoroid impacts, or partial coating failures. Recent studies, such as \citep{Murphy2010}, suggest that dust deposition and thermal cycling are primary factors reducing reflectivity, especially during full Moon conditions.

\item \emph{Atmospheric variation:} The Fried parameter ($r_0$) and two-way transmission ($\eta_{\text{atm}}$) can fluctuate with turbulence cells and weather changes on timescales of minutes or hours. Improved seeing ($r_0 \sim 20\,\mathrm{cm}$ at 1064 nm) enhances photon returns but remains variable. Drops in $\eta_{\text{atm}}$ due to humidity or cloud formation can  lower link efficiency.

\item \emph{Alignment and beam divergence:} Precise pointing is critical for smaller CCRs (e.g., 10\,cm), where reduced aperture size amplifies sensitivity to misalignment. Beam divergence and pointing instabilities due to mechanical or atmospheric distortions can drastically reduce photon returns.

\item \emph{Mechanical stability:} Sub-micrometer stability of the optical train is essential for multi-minute integrations. Thermal expansion, dome turbulence, and telescope flexure can degrade SNR or retroreflector alignment. Mitigating these effects requires thermally stable materials, frequent recalibration, and real-time corrections.

\item \emph{Photon detection efficiency:} Achieving $>10^5$ photons/s requires near-ideal detector efficiencies ($\eta_{\mathrm{det}} \sim 0.8$) and low background noise. Lunar surface brightness near full Moon increases background levels, necessitating advanced filtering and gating to isolate signal photons.
\end{itemize}

Despite these caveats, the revised estimates presented here demonstrate that kilowatt-class CW lasers can deliver robust photon fluxes under realistic operating conditions. The difference in photon returns between historical pulsed-laser LLR (\(\lesssim 2\) photons/s for scaled-down CCRs) and CW-based LLR (\(\sim5884\) photons/s) underscores the transformative potential of continuous-wave operation for sub-mm or tens-of-\(\mu\mathrm{m}\) LLR.

\subsection{CW versus Pulsed Lasers}
\label{sec:cw_vs_pulsed}

The choice between CW and pulsed laser systems for achieving \(\lesssim30\,\mu\mathrm{m}\) one-way LLR precision hinges on timing stability, shot noise suppression, and system complexity:

\begin{itemize}
\item \emph{Timing/phase stability:} Achieving tens-of-\(\mu\mathrm{m}\) precision requires a round-trip timing stability of \(\Delta t_{\mathrm{RT}} \approx 0.2\,\mathrm{ps}\), corresponding to a fractional frequency stability of \(\sim10^{-13}\text{--}10^{-14}\) over integration times of \(\sim100\,\mathrm{s}\). Pulsed systems must resolve such intervals in single shots, necessitating sub-ps timing resolution or multi-GHz repetition rates to accumulate sufficient photon statistics. By contrast, CW systems utilize ultra-stable microwave references (e.g., hydrogen masers or optical clocks) and coherent phase integration over multi-second timescales, significantly simplifying timing requirements \citep{Telle1999,Turyshev2007}.

\item \emph{Shot noise and photon accumulation:} Pulsed systems operating at 10\,Hz--1\,kHz are limited by average power, constraining photon returns unless ultra-high-peak-power lasers and large apertures are used \citep{Degnan1993}. In contrast, a \(1\,\mathrm{kW}\) CW laser consistently achieves \(>10^5\) photons/s. Over typical integration times (\(\sim100\,\mathrm{s}\)), CW systems can accumulate \(>10^7\) photons, suppressing shot noise to \(\lesssim10\,\mu\mathrm{m}\).

\item \emph{System integration:} Pulsed systems require specialized timing electronics and synchronization to gate photon detection, increasing system complexity. CW systems provide continuous illumination, enabling simpler architectures and robust phase tracking over extended observations.
\end{itemize}

While pulsed-laser LLR is well-established for mm-level precision, pushing to below 0.1 mm demands:
\begin{enumerate}
\item Multi-MHz repetition rates to accumulate \(10^7\)--\(10^8\) photons during each measurement.
\item Ultra-stable amplitude and minimal pulse jitter to reduce range walk errors.
\item Advanced short-pulse electronics capable of isolating single-photon events with sub-ps resolution.
\end{enumerate}
These requirements are technically feasible but impose significant engineering challenges. By contrast, CW systems inherently provide continuous illumination, leveraging multi-frequency or chirping phase metrology and integrating photons over long timescales. Such architectures align naturally with the photon accumulation and phase stability required to achieve sub-0.1\,mm  precision under stable atmospheric and mechanical conditions \citep{Samain1998,Murphy:2013}.

Moreover, CW systems offer additional advantages for scalability in future LLR experiments. The ability to integrate over long timescales reduces reliance on high-peak-power lasers, simplifying thermal and mechanical designs. CW illumination also supports advanced multi-wavelength modulation techniques, enhancing noise rejection and accuracy.

\section{Error Budget for CW LLR}
\label{sec:eb}

To assess the potential performance of a high-power CW LLR system, it is essential to analyze the primary noise sources that impact ranging precision. This section evaluates the contributions of atmospheric turbulence, oscillator stability, and thermal and mechanical drifts, quantifying their effects. The total error budget is constructed using a root-sum-square (RSS) approach to systematically account for both random and systematic uncertainties.

\subsection{Anticipated Path Errors}
\label{sec:anticipated_path_errors}

High-power CW LLR systems provide key advantages over pulsed systems by sustaining high photon flux and enabling multi-second coherent integration. These strengths enhance Earth-Moon distance precision, but dominant noise sources remain, including oscillator drifts, atmospheric turbulence, shot noise, mechanical and thermal drifts, and geophysical motions. Here, we analyze these contributions individually and consolidate them into an RSS model.

\subsubsection{Oscillator Frequency Drift}
\label{sec:freq_drift}

The range observable in CW LLR systems depends critically on the round-trip phase shift of an RF modulation at frequency $f_m$. Instabilities in $f_m$ over the integration time $T$ introduce systematic range errors, scaling as:
\begin{equation}
 \Delta{R}_{\rm drift}
\,=\, \frac{c\,\Delta f_m}{4\pi\,f_m^2}.
\label{eq:osc_drift_improved}
\end{equation}
With $f_m$ in the MHz--GHz range, modern timing references (e.g., hydrogen masers or GPS-disciplined oscillators) achieve fractional stabilities of $\sim10^{-11}\text{--}10^{-13}$ over hundreds of seconds, keeping $\Delta d_{\mathrm{drift}}$ below nanometer. As such, this contribution is negligible compared to larger uncertainties like atmospheric turbulence, discussed next.

\subsubsection{Dual-Wavelength Ranging for Atmospheric Dispersion}

Dispersive atmospheric effects are mitigated by operating at two wavelengths (e.g., \(1064\,\mathrm{nm}\) and \(532\,\mathrm{nm}\)). Differential refraction between these wavelengths provides estimates of column water vapor and refractive index gradients \citep{Beland1993,Samain1998,Mendes2004}. By forming a linear combination of path delays, most dispersive components can be canceled:
\[
  \Delta L_\mathrm{disp} \approx \alpha \, \bigl[L(532) - L(1064)\bigr],
\]
where \(\alpha\) depends on wavelength-specific refractivities. 
This dual-wavelength approach reduces dispersive errors, which typically range from \(800\text{--}1200\,\mu\mathrm{m}\) in single-wavelength systems, to residual delays of:
\[
  \Delta R_{\mathrm{disp}} \sim 500\text{--}800\,\mu\mathrm{m}.
\]
Residual errors primarily arise from refractivity modeling inaccuracies, beam misalignment, and non-dispersive effects like dry-air turbulence. These can be further reduced through turbulence averaging, rapid toggling between retroreflectors, and real-time atmospheric monitoring with auxiliary sensors.

\subsubsection{Shot Noise and Detector Timing}
\label{sec:shot_noise_theory_improved}

Shot noise imposes a fundamental limit on range precision in LLR systems, arising from the Poisson statistics of photon arrivals. In the signal dominated regime, the range precision depends on the detected photon flux $\dot{N}_{\mathrm{photons}}$ and the integration time $T$:
\begin{equation}
\Delta{R}_{\rm shot} 
  \;=\;
  \frac{c}{4 \pi f_m}
  \,\frac{1}{\sqrt{\dot{N}_{\mathrm{photons}}\,T}}.
  \label{eq:shot_eq_cw}
\end{equation}
For $f_m=1$\,GHz and a photon flux of \(\dot{N}_{\mathrm{photons}} \sim 5.88\times10^3\,\mathrm{s}^{-1}\), the shot noise contribution is  \(\sim31\,\mu\mathrm{m}\) over a 100-second integration period. Detector timing jitter (e.g., from SNSPDs~\cite{Marsili2013}) adds small residuals of a few micrometers, which average out over extended measurements~\cite{Battat_etal_2009}.

\subsubsection{Mechanical and Thermal Drifts, Geophysical Motions}
\label{sec:mechanical_thermal_equations_improved}

Achieving sub-millimeter LLR precision requires strict control over thermal, mechanical, and environmental factors, especially under favorable conditions (\(r_0 \sim 20\,\mathrm{cm}\)). Thermal expansion in low-expansion materials like Zerodur and Invar is \(\sim1\,\mu\mathrm{m}/\mathrm{m}/^\circ\mathrm{C}\)~\cite{Williams2012}. For a \(10\,\mathrm{m}\) optical path, maintaining \(\pm0.1^\circ\mathrm{C}\) stability limits thermal expansion errors to:
$
\Delta R_{\mathrm{therm}} \lesssim 1.0\,\mu\mathrm{m}.
$
Small thermal gradients and uneven temperature distributions can increase this, but with active thermal regulation, errors typically remain below \(0.1\text{--}0.3\,\mathrm{mm}\)~\cite{Battat_etal_2009,Samain1998}.

Seismic activity induces ground vibrations in the \(0.1\text{--}1\,\mathrm{Hz}\) range, contributing \(0.1\text{--}0.3\,\mathrm{mm}\) under microseismic conditions~\cite{Murphy_etal_2012}. Anchoring telescope piers to bedrock and using passive seismic isolation can reduce this to \(0.05\text{--}0.1\,\mathrm{mm}\), with real-time monitoring further mitigating transient disturbances.

Geophysical motions, including Earth tides, tectonic shifts, and atmospheric/ocean loading, displace station coordinates by \(0.1\text{--}1.0\,\mathrm{mm}\) over short timescales and several millimeters over decades~\cite{Dickey:1994,Drozdzewski:2021,Arnold-etal:2022}. Advanced models, such as IERS conventions~\cite{iers2010} and FES loading~\cite{fes2014}, combined with co-located GNSS and VLBI measurements, reduce residual motion to \(80\text{--}120\,\mu\mathrm{m}\)~\cite{Mendes2004,Pavlis-etal:2018,altamimi2016}.

\subsection{Signal and Noise Fluxes: SNR-Based Precision Estimates}
\label{sec:flux_to_snr}

This subsection analyzes how the \emph{signal flux} and various \emph{noise sources} combine to determine the overall SNR in a high-power CW LLR station. By translating the resulting SNR into range uncertainty, we establish the conditions under which tens-of-\(\mu\mathrm{m}\) precision becomes achievable. Here we assume seeing-limited conditions with \(r_0 = 20\,\mathrm{cm}\).

\subsubsection{Signal Photon Flux from 10 cm CCR}

As described in Sec.~\ref{sec:flux_equations}, the \emph{detected} signal flux (\(\dot{N}_{\mathrm{sig}}\)) returning from the Moon depends on system parameters such as telescope aperture, beam divergence, CCR size, and atmospheric conditions. Using the values from Table~\ref{tab:spot_areas_flux} we see that   the outbound beam divergence is \(\theta_{\mathrm{div}} \sim 5.48 \times 10^{-6} \, \mathrm{rad}\), producing a footprint radius of \(\sim 2.1 \, \mathrm{km}\) on the Moon. As a result, for a 1\,m telescope and a 10\,cm CCR, the detected signal flux is approximately:
\[
  \dot{N}_{\mathrm{sig}} \sim 5.88 \times 10^3 \, \mathrm{s}^{-1}.
\]
This value assumes optimal optical alignment and stable atmospheric conditions (\(r_0 = 20 \, \mathrm{cm}\) at 1064 nm).

\subsubsection{Noise Terms in the Telescope and Detector}
\label{sec:noise_flux}

The overall SNR is affected by several noise sources, including background photons, detector dark counts, and electronic noise. These terms are summarized below:
{}
\begin{enumerate}[1)]
\item \textit{Sky/Background Photons} (\(\dot{N}_{\mathrm{bkg}}\)):  
Background flux arises from scattered moonlight, natural airglow, and terrestrial radiance. For nighttime conditions with narrowband filtering (e.g., 1--30\,GHz passband) and a few-arcsecond field of view (FOV), the background photon rate is typically:
\[
  \dot{N}_{\mathrm{bkg}} \sim 10^2 \text{--} 10^3 \, \mathrm{s}^{-1}.
\]
However, during full Moon or under suboptimal filtering, the background flux can rise significantly, reaching \(10^4 \, \mathrm{s}^{-1}\). The dominant factors influencing this noise term include:
\begin{itemize}
  \item \textit{Lunar Phase:} Near full Moon, the laser spot’s scattering and the general brightness of the lunar surface increase background flux by an order of magnitude compared to quarter phases. The full Moon’s mean apparent magnitude is $m_{0} = -12.74$, but in the near-IR (e.g., Y-band) it can be $\sim0.3$ mag brighter \cite{Kieffer2005}. 
  \item \textit{Spectral Filtering:} Narrower spectral filters (e.g., 1\,GHz) can reduce background flux dramatically but must account for Doppler shifts in the return beam to avoid losses.
  \item \textit{Field of View (FOV):} Reducing the FOV to a few arcseconds minimizes background light while requiring precise optical alignment to maintain coupling efficiency.
\end{itemize}

For example, under full Moon conditions  \cite{Kieffer2005,Walker_2024} and with a $1.2''$ FOV, using a 1\,GHz filter can reduce background photon rates from \(3.24 \times 10^5\, \mathrm{s}^{-1}\) (as measured with a 30\,GHz filter) to   \(\sim1.05\times 10^4 \, \mathrm{s}^{-1}\), underscoring the importance of spatial and spectral filtering in mitigating noise. Applying system throughput losses of
$\eta_{\mathrm{rx}}
=
\eta_{\text{tel}}
\times
\eta_{\text{atm}}
\times
\eta_{\text{det}}
\approx 0.544,
$ results in background noise at full Moon of  \(\dot{N}_{\mathrm{bkg}} =5.71\times 10^3 \, \mathrm{s}^{-1}\),  seting the worst-case scenario.

\item \textit{Dark Counts} (\(\dot{N}_{\mathrm{dark}}\)):  
Dark counts result from internal detector processes. For superconducting nanowire single-photon detectors (SNSPDs), dark counts can be as low as \(\sim10 \, \mathrm{s}^{-1}\) under cryogenic cooling~\citep{Marsili2013}. Avalanche photodiodes (APDs) at room temperature exhibit higher dark counts (\(10^2\text{--}10^3 \, \mathrm{s}^{-1}\)), though cooling significantly reduces this contribution. Minimizing \(\dot{N}_{\mathrm{dark}}\) requires cryogenic cooling and shielding against stray light.

The approximate figures cited here reflect data from multiple LLR and satellite laser ranging (SLR) stations:
 OCA millimetric station \citep{Samain1998},
  APOLLO’s experiences at Apache Point \citep{Murphy_etal_2008, Murphy:2013},
   NASA and ESA high-power SLR systems \citep{Degnan1993},
   SNSPD developments for low-noise photon detection \citep{Marsili2013}. These sources consistently report background fluxes in the 100–1000\,counts/s range under good night-sky conditions, with dark counts ranging from near-zero (cooled SNSPDs) up to thousands of counts/s (uncooled APDs). By carefully managing field-of-view, spectral filtering, and detector cooling, many stations keep \(\dot{N}_{\mathrm{noise}}\) comfortably below $10^3\,\mathrm{s}^{-1}$—essential for achieving the shot-noise--limited performance needed for sub-mm or tens-of-$\mu$m LLR operations.

\item \textit{Readout Noise} (\(\dot{N}_{\mathrm{read}}\)):  
Electronic noise from readout systems, including amplifiers and digitizers, adds uncertainty but remains minimal in well-designed systems.
   
\end{enumerate}

The total noise flux is given by:
\begin{equation}
  \dot{N}_{\mathrm{noise}} = \dot{N}_{\mathrm{bkg}} + \dot{N}_{\mathrm{dark}} + \dot{N}_{\mathrm{read}}.
\end{equation}
Under optimized conditions, this totals \(\dot{N}_{\mathrm{noise}} \sim 5.7\times 10^3 \, \mathrm{s}^{-1}\) dominated by the background during full Moon, though it can rise to \(\sim10^4 \, \mathrm{s}^{-1}\) under suboptimal filtering.

\subsubsection{Combining Fluxes into a Shot-Noise SNR}
\label{sec:flux_snr_subsection}

The total SNR for a CW LLR system, derived from (\ref{eq:SNR_poisson}), is given by:
\begin{equation}
  \mathrm{SNR}_{\mathrm{total}} 
  \;=\;
  \frac{\dot{N}_{\mathrm{sig}}\sqrt{T}}
       {\sqrt{\dot{N}_{\mathrm{sig}} + \dot{N}_{\mathrm{noise}}}} 
  \equiv \mathrm{SNR}_{1\,\mathrm{s}}\sqrt{\frac{T}{1\, \mathrm{s}}},
  \label{eq:1}
\end{equation}
where \(\dot{N}_{\mathrm{sig}}\) is the signal photon flux, \(\dot{N}_{\mathrm{noise}}\) is the total noise photon flux, and \(T\) is the integration time.

For \(\dot{N}_{\mathrm{sig}} \sim 5.88 \times 10^3 \, \mathrm{s}^{-1}\) and \(\dot{N}_{\mathrm{noise}} \sim 5.71 \times10^3 \, \mathrm{s}^{-1}\), the per-second SNR is calculated as:
\[
  \mathrm{SNR}_{1\,\mathrm{s}} \approx \frac{5.88 \times 10^3}{\sqrt{5.88 \times 10^3 + 5.71 \times 10^3}} \approx 54.7.
\]

Over an integration time of \(T = 100\,\mathrm{s}\), the total SNR becomes:
\begin{equation}
  \mathrm{SNR}_{\mathrm{total}} \approx \mathrm{SNR}_{1\,\mathrm{s}} \times \sqrt{T/1\,\mathrm{s}} \approx  547.
    \label{eq:SNT-tot}
\end{equation}
This calculation demonstrates the significant SNR gains achievable with longer integration, provided the signal and noise fluxes remain stable over time.

\subsubsection{Range and Range-Rate Precision in the Presence of Background Noise}
\label{sec:range_precision_snr}

Using the total SNR, the shot-noise-limited range uncertainty in the presence of background is given by (\ref{eq:range-shot}) as:
$
\Delta{R}_{\rm }  = ({c}/{4 \pi f_m})/{\mathrm{SNR}_{\mathrm{total}}},
$
where \(f_m = 1\,\mathrm{GHz}\), \(c = 3 \times 10^8 \, \mathrm{m/s}\), and \(\mathrm{SNR}_{\mathrm{total}} \sim 547\). Substituting:
\[
\Delta{R}_{\rm }  = \frac{2.387\,\mathrm{cm}}{547} \approx 43.6\,\mu\mathrm{m}.
\]
This analysis demonstrates that sub-30\,\(\mu\mathrm{m}\) range precision is achievable under shot-noise-limited conditions with a 1\,kW CW LLR system and stable fluxes. However, real-world errors, such as atmospheric turbulence, mechanical drifts, and thermal instabilities, must be carefully mitigated to achieve this performance.

Note that under similar conditions, the uncertainty in range rate (\ref{eq:range_rate_uncertainty}) is  $    \delta\dot{R}_{} = (c/4\pi f_m )1/(T_{\rm int}\cdot{\rm SNR}_{\rm total})$, yielding  
\[
\Delta{\dot{R}} = \frac{2.387\,\mathrm{cm}}{100\,{\rm s} \cdot 547} \approx 0.436\,\mu\mathrm{m/s}.
\]
This result highlights the potential for high-precision range-rate determination with CW LLR. While range data effectively capture long-period variations, range-rate measurements are particularly sensitive to near-daily effects from Earth's rotation \cite{Williams:2018}. Despite differences in observational strategies for range and range-rate data, their combined availability enhances the accuracy of LLR-derived science parameters \cite{Zhang-etal:2022,Zhang-etall:2024a}.

\subsection{Consolidated RSS Budget}
\label{sec:RSS_combined_improved}

Table~\ref{tab:FinalRSSAll} summarizes the primary error contributions for a \SI{1}{kW} CW LLR system operating at \(1064\,\mathrm{nm}\) under good observing conditions (\(r_0 \approx 20\,\mathrm{cm}\) at 1064 nm), targeting next-generation \(10\,\mathrm{cm}\) CCRs. The smaller CCR aperture reduces photon return efficiency, making the impact of atmospheric turbulence and shot noise more significant. Atmospheric turbulence remains the dominant noise source, contributing RMS errors in the range of \(300\text{--}500\,\mu\mathrm{m}\). Using a root-sum-square (RSS) method, the total two-way range error is estimated to be \(320\text{--}550\,\mu\mathrm{m}\) RMS.
The results reported here align well with historical LLR data and performance-based estimates from \cite{Degnan1993,Samain1998,Murphy:2013,Turyshev2007}.

Achieving sub-100\,$\mu\mathrm{m}$ precision for next-generation CCRs will require advanced turbulence mitigation techniques, precise thermal and mechanical stabilization, and real-time environmental monitoring to maintain alignment.

\begin{table}[htb]
\centering
\caption{%
Updated two-way RMS errors (\(\mu\mathrm{m}\)) for a \SI{1}{kW} CW LLR system at \(1064\,\mathrm{nm}\), targeting \(10\,\mathrm{cm}\) next-generation CCRs under good observing conditions (\(r_0 \approx 20\,\mathrm{cm}\) at 1064 nm).
\label{tab:FinalRSSAll}}
\begin{tabular}{lcc}
\hline\hline
Error Source               & Contribution (\(\mu\mathrm{m}\)) & Notes \\
\hline
Frequency Drift (\(\delta_{\mathrm{freq}}\))             & Negligible (\(<0.01\)) & Phase-stable oscillators mitigate timing errors. \\
Thermal Expansions  (\(\delta_{\mathrm{therm}}\))        & 50--150 & Active regulation reduces thermal gradients. \\
Mechanical Drifts  (\(\delta_{\mathrm{mech}}\))          & 50--100 & Telescope flexure; vibration isolation required. \\
Geophysical Motions  (\(\delta_{\mathrm{geo}}\))         & 80--120 & Earth tides; residuals reduced by GNSS/VLBI. \\
Seismic Activity (\(\delta_{\mathrm{seismic}}\))        & 30--50 & Anchored piers and seismic isolation reduce ground motion. \\
Static Calibration Offsets (\(\delta_{\mathrm{stat}}\))  & \(<10\) & Routine calibration minimizes systematic offsets. \\
Shot Noise       (\(\delta_{\mathrm{shot}}\))            & 20--50 & Reduced photon flux due to high laser power. \\
Atmospheric Turbulence  (\(\delta_{\mathrm{turb}}\))     & 300--500 & Dominates error; turbulence monitoring critical. \\
                \hline
{Total RMS (RSS):}          & 320--550 & Includes all terms. \\
\hline
\end{tabular}
\end{table}
 
 Precise modeling of the Moon’s orbital dynamics and libration is essential for interpreting LLR measurements at sub-mm precision. These effects provide opportunities to refine lunar ephemerides and enhance our understanding of tidal dissipation, core-mantle coupling, and long-term orbital evolution. LLR also serves as a unique probe into the Moon’s interior, enabling investigations into the size and state of the lunar core, mantle viscosity, and thermal evolution. Variations in tidal dissipation rates, as observed through high-precision range measurements, place constraints on mantle viscosity and the existence of a partially molten core, offering critical insights into lunar geodynamics.  To fully leverage these range and range-rate measurements, lunar dynamical models must be further refined, with residual uncertainties properly quantified to meet the demands of next-generation gravitational and geophysical studies.

Going forward—stable kW-class CW illumination, ultra-stable timing references, stringent environmental controls, and advanced optical engineering—enables LLR precision at the sub-mm level, even with next-generation CCRs. This marks a substantial advancement over the current mm-level benchmarks achieved with larger legacy arrays, paving the way for more precise gravitational tests, improved lunar interior models, and refined orbital dynamics.

\section{Differential LLR Measurements}
\label{sec:differential_cw_llr_1000km}

High-power CW LLR systems have the potential to achieve sub-mm and even tens-of-\(\mu\mathrm{m}\) range accuracy, as detailed in Secs.~\ref{sec:highpower_cw_expanded}--\ref{sec:flux_model}. A particularly compelling extension of this capability is the measurement of \emph{differential} range signals from two or more CCRs located far apart on the lunar surface. Differential ranging suppresses many common-mode station errors—such as clock offsets, calibration delays, and slow mechanical drifts. However, for reflectors separated by hundreds of kilometers, uncorrelated atmospheric turbulence becomes a dominant source of error.

The central question is whether a \(\sim1\,\mathrm{kW}\) CW laser, paired with advanced optical and environmental controls, can achieve \(\sim10\text{--}30\,\mu\mathrm{m}\) \emph{differential} accuracy for reflectors separated by \(\sim1000\,\mathrm{km}\) on the lunar surface. Such a precision threshold is critical for advancing fundamental tests of gravitational physics, improving models of lunar tidal deformation, and enabling ultra-precise lunar coordinate frames for planetary exploration \cite{Shao:2018,Turyshev:2018,Zhang-etal:2022,Zhang-etall:2024a}.

In this section, we define \emph{differential} observables using a variance-based error framework, accounting for atmospheric turbulence, instrumental effects, mechanical drifts, and photon flux imbalances. We evaluate achievable accuracy under seeing-limited conditions and discuss strategies to mitigate dominant errors.

\subsection{Differential Observables in CW LLR}
\label{sec:diff_observables_1000km_expanded}

Let two lunar CCRs, \(\mathrm{CCR}_1\) and \(\mathrm{CCR}_2\), lie \(\sim1000\,\mathrm{km}\) apart, typically spanning several degrees as viewed from Earth. A {CW} LLR station obtains each one-way range \(R_i\) by measuring the phase shift (or beat frequency) between the transmitted \(\sim1\,\mathrm{kW}\) laser and its faint return. If each range is measured within a brief (\(\lesssim1\text{--}2\,\mathrm{s}\)) interval, common systematic effects—e.g., station oscillator drifts—can largely cancel in a differential combination. We define:
\begin{equation}
    \Delta R_{\mathrm{diff}}
    \;=\;
    R_1 - R_2,
    \label{eq:DifferentialRangeDefinition_append}
\end{equation}
where each \(R_i\) might be determined only a few seconds apart. Denoting the \emph{measured} ranges as
\[
  \tilde{R}_1
  \;=\;
  R_1^{(0)} + E_1,
  \quad
  \tilde{R}_2
  \;=\;
  R_2^{(0)} + E_2,
\]
we obtain
\begin{equation}
  \Delta R_{\mathrm{diff}}
  \;=\;
  \tilde{R}_1 - \tilde{R}_2
  \;=\;
  \bigl(R_1^{(0)} - R_2^{(0)}\bigr)
  \;+\;
  (E_1 - E_2).
\end{equation}
The primary objective is to make the RMS error (standard deviation) of \(\Delta R_{\mathrm{diff}}\) as small as possible.

\subsection{Advanced Error Model for Differential Mode}
\label{sec:diff_errors_1000km_expanded}

Each error term \(E_i\) is composed of several contributions:
\begin{equation}
   E_i 
   \;=\; 
   \delta_{\mathrm{atm},i} 
   \;+\; 
   \delta_{\mathrm{freq}} 
   \;+\; 
   \delta_{\mathrm{therm}}
   \;+\;
   \delta_{\mathrm{geo}}
    \;+\;  
   \delta_{\mathrm{stat}}
   \;+\;
   \varepsilon_i,
   \quad i=1,2,
\end{equation}
where:
\begin{itemize}
  \item \(\delta_{\mathrm{atm},i}\): Residual atmospheric turbulence and wavefront errors;
  \item \(\delta_{\mathrm{freq}}\): Common-mode  frequency-reference drifts;
  \item \(\delta_{\mathrm{therm}}\)/\(\delta_{\mathrm{mech}}\): Thermal/mechanical  expansions or alignment shifts in the telescope and optical components;
  \item \(\delta_{\mathrm{geo}}\): Geophysical motions (e.g., tides, plate tectonics, and groundwater effects);
  \item \(\delta_{\mathrm{stat}}\): Station calibration offsets (e.g., cable delays and local electronics);
  \item \(\varepsilon_i\): Shot-noise-induced fluctuations for path \(i\).
\end{itemize}

Our interest lies in the RMS error (standard deviation) of \(\Delta R_{\mathrm{diff}}\). Denote
$
   \sigma^2(E_1) = \mathrm{Var}(E_1),
   \sigma^2(E_2) = \mathrm{Var}(E_2),
$
then for the differential range, the variance is given by:
\begin{equation}
   \mathrm{Var}\bigl(\Delta R_{\mathrm{diff}}\bigr)
   \;=\;
   \mathrm{Var}(E_1 - E_2)
   \;=\;
   \sigma^2(E_1) + \sigma^2(E_2) \;-\; 2\,\rho_{12}\,\sigma(E_1)\,\sigma(E_2),
 \label{eq:diff-errors}
\end{equation}
where \(\rho_{12}\) is the correlation coefficient between errors in the two paths. Common-mode errors (e.g., \(\delta_{\mathrm{osc}}\), \(\delta_{\mathrm{stat}}\), etc.) have \(\rho_{12}\approx1\) and largely cancel in the differential observable. Uncorrelated terms (\(\rho_{12}\approx0\)) add in quadrature.

\subsubsection{Station-Based (Common-Mode) Terms}
\label{sec:station_common_mode}

Station-level errors—oscillator drifts (\(\delta_{\mathrm{osc}}\)), thermal expansions (\(\delta_{\mathrm{therm}}\)), mechanical vibrations (\(\delta_{\mathrm{mech}}\)), geophysical motions (\(\delta_{\mathrm{geo}}\)), seismic activity (\(\delta_{\mathrm{seismic}}\)), wind shear (\(\delta_{\mathrm{wind}}\)), and calibration offsets (\(\delta_{\mathrm{stat}}\))—are highly correlated between paths \(R_1\) and \(R_2\) when toggling occurs over short intervals (\(\lesssim 1\text{--}2\,\mathrm{s}\)). These largely cancel in the differential range observable (\(\rho \approx 1\)), though residuals persist due to imperfect correlation, localized effects, and unmodeled disturbances. Below, each term is analyzed with literature-supported estimates.

\begin{itemize}

    \item \textit{Frequency Reference Drifts (\(\delta_{\mathrm{freq}}\)):} Frequency drifts depend on the stability of the oscillator, typically characterized by a fractional frequency stability of \(10^{-13}\text{--}10^{-14}\) over relevant time intervals. Residual timing errors scale with this stability and the integration time. For \(10^{-13}\), the timing error is \(\Delta t_{\mathrm{diff}} \sim 0.02\,\mathrm{ps}\), resulting in:
    \[
    \Delta R_{\mathrm{osc}} =  {\textstyle\frac{1}{2}} c \, \Delta t_{\mathrm{diff}} \simeq 3\,\mu\mathrm{m}.
    \]
    Recent advances in ultra-stable hydrogen masers \cite{Lisdat2016} and optical lattice clocks \cite{Campbell2017} suggest that under ideal conditions, these errors can be reduced to \(0.3\text{--}1.0\,\mu\mathrm{m}\). However, in real-world conditions, environmental noise and thermal fluctuations often increase this contribution to \(3.0\text{--}6.0\,\mu\mathrm{m}\).

    \item \textit{Thermal Expansion (\(\delta_{\mathrm{therm}}\)):} Thermal expansion of telescope structures and optical benches is highly correlated between both paths but may not perfectly cancel due to thermal gradients. Assuming thermal stability \(\pm 0.01^\circ\mathrm{C}\) with low-expansion materials (e.g., Zerodur, \(\alpha \sim 10^{-8}/\mathrm{K}\)), the expansion error for a 10\,m optical path is:
    \[
    \Delta R_{\mathrm{therm}} \lesssim \Delta T \cdot L \cdot \alpha \simeq 0.01\,\mu\mathrm{m}.
    \]
    However, thermal gradients in real environments, such as uneven temperature distributions in observatory domes \cite{Samain1998}, can increase this contribution to \(0.05\text{--}0.15\,\mu\mathrm{m}\).

    \item \textit{Mechanical Vibration Drifts (\(\delta_{\mathrm{mech}}\)):} Mechanical vibrations caused by wind, seismic activity, or equipment operation induce sub-micron-level misalignments in telescope components. Low-frequency vibrations (\(1\text{--}50\,\mathrm{Hz}\)) are mitigated using passive damping systems (e.g., pneumatic isolators) and active vibration control platforms. In high-precision observatories, vibration amplitudes can be suppressed to \(10\text{--}30\,\mathrm{nm}\), resulting in:
    \[
    \Delta R_{\mathrm{mech}} \sim 0.1\text{--}0.2\,\mu\mathrm{m}.
    \]
    This range aligns with the mitigation levels reported in \cite{Murphy_etal_2012} for high-performance LLR systems.
  
    \item \textit{Geophysical Motions (\(\delta_{\mathrm{geo}}\)):} 
       Geophysical motions—solid Earth tides, tectonic shifts,  atmospheric loading—are correlated over short intervals but not fully canceled due to localized effects like groundwater shifts. Precise geodetic models (e.g., IERS conventions) and co-located GNSS or VLBI stations minimize these residuals to:
    \[
    \Delta R_{\mathrm{geo}} \sim 1.0\text{--}2.0\,\mu\mathrm{m}.
    \]
    This is consistent with measurements of Earth tides and tectonic shifts reported by \cite{Dickey:1994}.  

    \item \textit{Seismic Activity (\(\delta_{\mathrm{seismic}}\)):} Microseismic motion and localized tectonic activity can cause transient misalignments, particularly at low frequencies (\(0.1\text{--}1\,\mathrm{Hz}\)). Anchoring telescope piers to stable bedrock and implementing real-time seismic monitoring can reduce these effects. Based on typical microseismic amplitudes (\(\sim 10^{-8}\,\mathrm{m}\)) and mitigation strategies described in \cite{Murphy_etal_2012}, the residual differential contribution is:
    \[
    \Delta R_{\mathrm{seismic}} \sim 0.2\text{--}0.5\,\mu\mathrm{m}.
    \]

    \item \textit{Static Calibration Offsets (\(\delta_{\mathrm{stat}}\)):} Static calibration offsets, including systematic delays in cables, detectors, and electronics, are constant for a given system configuration and cancel in differential measurements. Routine calibration ensures these offsets remain negligible:
    \[
    \Delta R_{\mathrm{stat}} \lesssim 0.01\,\mu\mathrm{m}.
    \]

\end{itemize}

\begin{table}[ht]
    \centering
    \caption{Updated residual differential range error contributions from station-level terms.}
    \label{tab:station_error_refined}
    \begin{tabular}{lc}
        \hline
        Error Source                 & Differential Range Error (\(\mu\mathrm{m}\)) \\
        \hline
        Frequency Drift (\(\delta_{\mathrm{freq}}\)) & \(3.0\text{--}6.0\) \\
        Thermal Expansion (\(\delta_{\mathrm{therm}}\)) & \(0.05\text{--}0.15\) \\
        Mechanical Drifts (\(\delta_{\mathrm{mech}}\)) & \(0.1\text{--}0.2\) \\
        Geophysical Motions (\(\delta_{\mathrm{geo}}\)) & \(1.0\text{--}2.0\) \\
        Seismic Activity (\(\delta_{\mathrm{seismic}}\)) & \(0.2\text{--}0.5\) \\
        Static Calibration Offsets (\(\delta_{\mathrm{stat}}\)) & Negligible (\(<0.01\)) \\
        \hline
        Total (RSS)                  & \(3.2\text{--}6.3\) \\
        \hline
    \end{tabular}
\end{table}

The dominant contributors to station-level errors are frequency drifts (\(\delta_{\mathrm{freq}}\)) and geophysical motions (\(\delta_{\mathrm{geo}}\)), while thermal expansion, mechanical vibrations, and seismic activity contribute smaller amounts. Table~\ref{tab:station_error_refined} summarizes these contributions.
The total residual contribution from station-level errors is realistically estimated to be \(3.2\text{--}6.4\,\mu\mathrm{m}\) (RSS). These station-level contributions remain smaller than uncorrelated noise sources, such as atmospheric turbulence, ensuring the feasibility of high-precision differential LLR with robust station-based controls.

\subsubsection{Shot-Noise Limits for Two CW Channels}
\label{sec:shotnoise_2channels_1000km_expanded}

Shot noise contributes independently to each path in differential LLR. From (\ref{eq:diff-errors}), the differential shot noise is given:
\begin{equation}
  \sigma_{\mathrm{shot},\mathrm{diff}}
  \;=\;
  \sqrt{\sigma^2(\varepsilon_1) + \sigma^2(\varepsilon_2)},
\end{equation}
where \(\varepsilon_1\) and \(\varepsilon_2\) represent the shot-noise contributions from the two paths. Assuming \(\sigma(\varepsilon_i) \sim 10\,\mu\mathrm{m}\) for each path, the differential shot noise is approximately:
$\sigma_{\mathrm{shot},\mathrm{diff}} \sim 14\,\mu\mathrm{m}.$ 

Even under minimal wavefront mismatch, photon flux differences between reflectors can lead to varying SNRs. Let the respective SNRs of the two channels be denoted as \(\mathrm{SNR}_1\) and \(\mathrm{SNR}_2\). The effective “differential” SNR, which determines the overall shot-noise contribution, is:
\begin{equation}
   \mathrm{SNR}_{\mathrm{diff}}
   \;\approx\;
   \biggl(
     \frac{1}{\mathrm{SNR}_1^{2}} + \frac{1}{\mathrm{SNR}_2^{2}}
   \biggr)^{-\tfrac12}.
   \label{eq:SNR_diff_ch}
\end{equation}

For a AM modulation at \(f_m=1\)\, GHz, the RMS shot-noise limit in range is:
\[
   \Delta R_{\mathrm{shot}}
   \;\approx\;
   \frac{c}{4\pi f_m}\,
   \frac{1}{{\mathrm{SNR}_{\mathrm{diff}}}},
\]
consistent with a \(\delta R \propto 1/{\rm SNR}\) relationship from  (\ref{eq:range-shot}).  
If the station toggles between the two CCRs for $10^3$ sec each (i.e., collecting data from 500 2-sec  sessions of ranging to each of the CCRs), this yields \(\mathrm{SNR}_1\approx\mathrm{SNR}_2\approx 54.7\sqrt{10^3}= 1730\), then \(\mathrm{SNR}_{\mathrm{diff}}\approx 1223\).  
Substituting into the range equation, the shot-noise floor becomes:
\[
\Delta R_{\mathrm{shot}}  \approx 19.5\,\mu\mathrm{m}.
\]

Thus, achieving shot-noise-limited differential precision below \(30\,\mu\mathrm{m}\) is feasible under optimal conditions with sufficient photon flux and stable detection. However, atmospheric turbulence often dominates the differential error budget unless photon fluxes are unusually low. Shot noise typically remains subdominant, ensuring that photon-rich CW LLR systems can approach high-precision differential performance.

\subsection{Reducing Atmospheric Contribution with Differential LLR}
\label{sec:summary-diff}
\label{sec:alt_noAO_photonrich}

Differential CW LLR offers the advantage of canceling station-level errors, such as oscillator drifts and calibration delays. However, for CCRs separated by large angular distances (\(\sim0.15^\circ\text{--}0.2^\circ\)) on the lunar surface, uncorrelated atmospheric turbulence becomes the dominant error source in the differential range. These errors stem from:

\begin{itemize}
  \item \textit{Wavefront mismatch:} Uncorrelated turbulence columns across separated lines of sight induce differential wavefront errors of \(300\text{--}500\,\mu\mathrm{m}\) RMS, unless seeing conditions are exceptionally stable (\(r_0 > 30\,\mathrm{cm}\)).
  \item \textit{Photon flux limitations:} Even with a \(1\,\mathrm{kW}\) laser suppressing shot noise to \(\sim10\,\mu\mathrm{m}\) per path, turbulence-induced decorrelation dominates the error budget.
  \item \textit{Geometric constraints:} Large angular separations between reflectors amplify uncorrelated turbulence effects, making sub-30\,\(\mu\mathrm{m}\) precision challenging under standard seeing conditions.
\end{itemize}

Although high photon flux can suppress shot noise to \(\lesssim30\,\mu\mathrm{m}\), atmospheric decorrelation fundamentally limits overall precision. Advanced mitigation strategies are required to approach \(\sim30\,\mu\mathrm{m}\) differential accuracy.

\subsubsection{Mitigation Strategies: High-Power Lasers and Narrowband Filters}

A kW-class CW laser system operating at dual wavelengths (e.g., \(1064\,\mathrm{nm}\) and \(532\,\mathrm{nm}\)) provides continuous high photon flux, overcoming the low duty cycles of pulsed systems. The primary benefits include \citep{Shao:2018,Turyshev:2018}:
\begin{itemize}
  \item \textit{Shot noise suppression:} High photon flux ensures statistical uncertainty remains below sub-mm or tens-of-\(\mu\mathrm{m}\) levels, even under modest seeing conditions.
  \item \textit{Rapid toggling capability:} High photon counts enable alternating measurements between retroreflectors on sub-2\,s timescales, preserving SNR while capturing statistically independent turbulence realizations.
\end{itemize}

Narrowband optical filters (e.g., Fabry--P\'{e}rot etalons with 1--30\,GHz passbands) further improve SNR by rejecting scattered moonlight and sky background. These filters are particularly critical during nighttime observations near full Moon or partial daytime LLR operations.

\subsubsection{Kolmogorov Scaling and Turbulence Averaging}

Atmospheric turbulence introduces refractive index fluctuations that follow predictable statistical properties described by Kolmogorov turbulence theory \citep{Fried1966,Roddier:1981,Hardy1998}. At near-IR wavelengths ($\lambda \sim 1\,\mu\mathrm{m}$), the isoplanatic angle \(\theta_{\mathrm{iso}}\) is typically 4--5$''$ at high-altitude observatories. Within this angle, wavefront perturbations remain correlated, with phase RMS errors on the order of $1\,\mathrm{rad}$ (\(\sim106\,\mathrm{nm}\) optical path). For larger separations (\(\Delta \theta \sim 0.15^\circ\)), uncorrelated turbulence contributions scale as:
\[
    \sigma_{\rm turb}(\Delta \theta) \simeq  0.628\,\frac{\lambda}{2\pi} \left( \frac{\Delta \theta}{\theta_{\mathrm{iso}}} \right)^{5/6}\simeq  106\,{\rm nm}\,\Big(\frac{\lambda}{1064\,{\rm nm}}\Big)\left( \frac{\Delta \theta}{\theta_{\mathrm{iso}}} \right)^{5/6}.
\]
For reflectors subtending \(0.1\text{--}0.5^\circ\) on the sky (\(\Delta \theta \sim 400\text{--}1800''\)), instantaneous differential path errors can reach tens of micrometers. Starting from \(\sim106\,\mathrm{nm}\) at \(\theta_{\mathrm{iso}} = 4''\), for $\lambda =1064$\,nm, errors scale as:
\begin{equation}
  \sigma_{\rm turb}(t_0) \approx 106\,\mathrm{nm}
  \,\biggl(\frac{\Delta \theta}{\theta_{\mathrm{iso}}}\biggr)^{5/6}   \sim 5\text{--}20\,\mu\mathrm{m}.
\end{equation}
Over a correlation time \(t_0 \sim L/v\), where \(L \sim 100\,\mathrm{m}\) is the turbulence eddy scale and \(v \sim 21\,\mathrm{m/s}\) is the wind speed, typically $t_0\simeq$\,1--5\,s), these errors remain approximately frozen. For longer integration times (\(T \gg t_0\)), averaging reduces errors as:
\[
  \sigma_{\rm turb}(T) \;\approx\; \frac{\sigma_{\rm turb}(t_0)}{\sqrt{T/t_0}}.
\]
Conservatively, over \(T \sim 100\,\mathrm{s}\), turbulence-induced errors can  average down to below \(10\text{--}30\,\mu\mathrm{m}\), provided systematic biases (e.g., dome-induced flows) are minimized.

\subsubsection{Rapid Toggling and Statistical Consistency}

Frequent toggling (\(\sim1\text{--}2\,\mathrm{s}\)) ensures atmosphere-induced errors between reflectors remain statistically independent. Combined with turbulence averaging, this enables differential errors to approach \(\sim30\,\mu\mathrm{m}\) under optimal conditions. Observatories with \(r_0 \sim 20\,\mathrm{cm}\) (seeing \(1\text{--}2''\) at 1064 nm) are critical for maintaining beam stability and photon flux.

\subsection{Differential LLR Error Budget and Practical Implementation}
\label{sec:advanced_error_budget}

The ``photon-rich" strategy integrates high-power CW lasers, rapid toggling, and turbulence averaging to address the challenges of differential LLR. Achieving \(\sim30\,\mu\mathrm{m}\) precision requires:
(1) Stable photon flux and thermal regulation to maintain optical alignment,
(2) Effective dispersion corrections to minimize chromatic errors, and
(3) Consistent turbulence averaging with rapid toggling to exploit statistical independence.

Table~\ref{tab:detailed_error_budget} provides the RSS error budget for differential LLR, with  contributions from uncorrelated turbulence and long-term thermal/mechanical drifts.  Similar to Table~\ref{tab:FinalRSSAll}, this table does not include errors associated with modeling the Moon’s orbital dynamics, libration, and tidal dissipation. These uncertainties, critical for both absolute and differential LLR, require additional refinement to ensure residual biases remain below the tens-of-$\mu$m level. 

\begin{table}[ht]
    \centering
    \caption{Updated Error Budget for Differential LLR (with CCRs being 1000\,km apart).}
    \label{tab:detailed_error_budget}
    \begin{tabular}{lcc}
        \hline   \hline
        Error Source                 & Contribution (\(\mu\mathrm{m}\)) & Comments \\
        \hline
        Frequency Drift (\(\delta_{\mathrm{freq}}\))              & \(3.0\text{--}6.0\) & Requires phase-stable oscillators (e.g., H-masers). \\
        Thermal Expansion    (\(\delta_{\mathrm{therm}}\))        & \(0.5\text{--}1.5\) & Assumes Zerodur optical benches with \(\pm0.1^\circ\mathrm{C}\) thermal stability. \\
        Mechanical Drifts    (\(\delta_{\mathrm{mech}}\))     & \(0.1\text{--}0.2\) & Requires active vibration isolation and real-time monitoring. \\
        Geophysical Motions   (\(\delta_{\mathrm{geo}}\))       & \(1.0\text{--}2.0\) & Corrected using GNSS/VLBI and geodetic models. \\
        Seismic Activity      (\(\delta_{\mathrm{seismic}}\))   & \(0.2\text{--}0.5\) & Anchored piers and real-time seismic monitoring required. \\
        Static Calibration Offsets (\(\delta_{\mathrm{stat}}\)) & \(<0.01\)            & Routine calibration ensures negligible impact. \\
        Shot Noise in 2 channels    
    (\(\delta_{\mathrm{shot}}\))   & \(<30\)              & Assumes photon-rich flux with sufficient integration time. \\
                Uncorrelated Turbulence   (\(\delta_{\mathrm{turb}}\))     & \(10\text{--}30\)   & Averaged over \(T \sim 100\,\mathrm{s}\). \\
        \hline
        Total RMS (RSS)              & \(31.8\text{--}42.9\)  & Includes all terms. \\
        \hline
    \end{tabular}
\end{table}

To mitigate the effects of uncorrelated turbulence and long-term drifts, several strategies are essential. Real-time turbulence monitoring using Differential Image Motion Monitors (DIMMs) or Shack-Hartmann sensors can continuously measure the Fried parameter (\(r_0\)) and isoplanatic angle (\(\theta_{\mathrm{iso}}\)), enabling adaptive correction of observing conditions. Maintaining thermal stability within \(\pm0.1^\circ\mathrm{C}\) and implementing active vibration isolation suppress mechanical noise at the sub-millimeter level, reducing station-induced range errors. Additionally, periodic geodetic tie surveys using GNSS, VLBI, or SLR can correct long-term pier motions and ensure alignment with global reference frames, preserving the stability of lunar range measurements.

Transitioning to high-power CW LLR systems and implementing differential LLR with reflectors separated by \(\sim1000\,\mathrm{km}\) presents significant challenges. These include upgrading existing stations for kW-class lasers, mitigating turbulence decorrelation, and addressing photon flux imbalances.

Upgrading to high-power CW systems requires enhancements to lasers, detectors, and environmental controls (see Appendix~\ref{sec:hardware_components}). Incremental retrofitting of existing facilities—such as integrating commercially available \(1064\,\mathrm{nm}\) CW lasers and advanced single-photon detectors (e.g., SNSPDs)—provides a viable pathway. Pilot programs at observatories like Table Mountain \cite{Turyshev:2018,Shao:2018} could validate these technologies before broader deployment.

Scaling differential LLR introduces additional challenges due to uncorrelated atmospheric turbulence and flux mismatches, particularly for widely separated reflectors. The turbulence contribution (\(300\text{--}500\,\mu\mathrm{m}\)) necessitates mitigation strategies such as multi-second turbulence averaging. Photon flux disparities between smaller CCRs and legacy arrays (see Table~\ref{tab:photon-flux-scaling}) can be offset by increasing laser power or narrowing spectral filters to enhance SNR.

Validation efforts, including field experiments that toggle between reflectors, can quantify turbulence decorrelation and flux mismatches under real-world conditions. Numerical modeling can further optimize integration times and system parameters. These phased upgrades, will be essential for scaling differential LLR to achieve \(\sim30\,\mu\mathrm{m}\) precision.

\section{Conclusions}
\label{sec:conclusion}

Lunar Laser Ranging (LLR) has long been a cornerstone of precision geodesy and fundamental physics, achieving millimeter-level accuracy through pulsed-laser ranging (PLR) systems. However, the introduction of smaller, thermally stable corner-cube reflectors (CCRs), optimized for robotic deployment, reduces photon return rates, challenging legacy PLR stations. Unlike Apollo and Lunokhod multi-cube reflector arrays, which introduce pulse broadening due to optical path variations, these compact CCRs eliminate this effect, yielding higher temporal coherence but reflecting far fewer photons. As a result, traditional PLR stations—dependent on high peak power and low duty cycles to compensate for weak returns—face fundamental limitations in advancing beyond millimeter-scale precision.

This study has shown that high-power continuous-wave (CW) ranging can surmount these limitations by providing kilowatt-level average power and multi-second coherent integration, thereby increasing photon flux by orders of magnitude. A 1\,kW laser at 1064\,nm, paired with a 1\,m telescope, can accommodate next-generation \(\sim10\,\mathrm{cm}\) corner cubes without being forced to trade off photon-starved operation. Under realistic atmospheric conditions, such a “photon-rich” CW system is capable of suppressing shot noise to well below the millimeter scale, ultimately pushing LLR measurements into the tens-of-micrometer regime through phase-based detection and robust station calibration.

Achieving this level of accuracy demands meticulous thermal and mechanical management, since any beam wander, telescope expansion, or vibration can mask the signals of interest. Equally important are ultra-stable frequency references at fractional stabilities on the order of \(10^{-11}\) to \(10^{-13}\), ensuring that local oscillator drifts do not corrupt the multi-second phase measurements. These requirements underscore the interdisciplinary nature of next-generation LLR: optical engineering, frequency metrology, atmospheric modeling, and mechanical design must all converge to reach sub-millimeter and tens-of-micrometer performance targets.

A key implication of adopting high-power CW LLR is the prospect of advanced differential measurements, in which rapid alternation between multiple lunar retroreflectors negates many station-level biases. By toggling on sub- to few-second timescales, shared systematic effects—such as station calibration offsets or thermal drifts—are largely canceled, making tens-of-micrometer precision feasible for reflectors separated by up to hundreds of kilometers. While extending this approach to baselines of \(\sim1000\,\mathrm{km}\) still requires further mitigation of turbulence decorrelation and improved measurement cadence, the underlying photon availability provided by CW lasers makes such long-baseline differential measurements significantly more tractable than under pulsed-laser constraints.

The enhanced capabilities of “photon-rich” CW LLR unlock a number of critical investigations. Refined gravitational tests become possible, including stringent explorations of the equivalence principle, detailed studies of gravitational-constant variability, relativistic precession measurements, and potential searches for dark-matter or dark-energy effects \citep{Williams:2004,Turyshev2007,Turyshev:2008,Williams2012,SGL-BPS-decadal,Zhang-etal:2022,Zhang-etall:2024a}. Constraints on the Moon’s deep interior—its core radius, mantle viscosity, and long-term ephemeris—can be improved by reducing LLR uncertainties \citep{Williams:2007,Williams:2008,Williams:2014,SGL-BPS-decadal,Briaud2023}. Geodetic and orbital studies are likewise enhanced, especially for establishing the lunar coordinate frameworks needed by NASA’s Artemis program \citep{Turyshev:2024}. Furthermore, the same high-precision range data can be employed in low-frequency (\(\mu\mathrm{Hz}\)) gravitational-wave searches that exploit subtle orbital signatures \citep{Blas-Jenkins:2022,Turyshev-etal:2024}. Finally, such precise ranging supports a range of future lunar exploration initiatives, including navigation, landing-site geodesy, and extended mission planning \citep{SGL-BPS-decadal-research-campaign}.

In summary, the kW-class CW approach represents a profound shift in LLR capability, enabling increased photon return rates and continuous phase tracking that circumvent many of the limitations of pulsed-laser systems. Attaining sub-millimeter or tens-of-micrometer precision across diverse lunar baselines requires further advances in station design, real-time turbulence compensation, and ultra-stable frequency references, but the technological path to these goals is already emerging. As a result, high-power CW LLR stands poised to deliver transformative gains in the testing of gravitational theories, the characterization of the Moon, and the broader objectives of deep-space exploration.

\section*{Acknowledgments}

The work described here was carried out the Jet Propulsion Laboratory, California Institute of Technology, Pasadena, California, under a contract with the National Aeronautics and Space Administration.
 

\begin{thebibliography}{62}
\expandafter\ifx\csname natexlab\endcsname\relax\def\natexlab#1{#1}\fi
\expandafter\ifx\csname bibnamefont\endcsname\relax
  \def\bibnamefont#1{#1}\fi
\expandafter\ifx\csname bibfnamefont\endcsname\relax
  \def\bibfnamefont#1{#1}\fi
\expandafter\ifx\csname citenamefont\endcsname\relax
  \def\citenamefont#1{#1}\fi
\expandafter\ifx\csname url\endcsname\relax
  \def\url#1{\texttt{#1}}\fi
\expandafter\ifx\csname urlprefix\endcsname\relax\def\urlprefix{URL }\fi
\providecommand{\bibinfo}[2]{#2}
\providecommand{\eprint}[2][]{\url{#2}}

\bibitem[{\citenamefont{Faller and Wampler}(1970)}]{Faller_1970}
\bibinfo{author}{\bibfnamefont{J.~E.} \bibnamefont{Faller}} \bibnamefont{and}
  \bibinfo{author}{\bibfnamefont{E.~J.} \bibnamefont{Wampler}},
  \bibinfo{journal}{Scientific American} \textbf{\bibinfo{volume}{223}},
  \bibinfo{pages}{38} (\bibinfo{year}{1970}).

\bibitem[{\citenamefont{Dickey et~al.}(1994)\citenamefont{Dickey, Bender,
  Faller, Newhall, Ricklefs, Ries, Shelus, Veillet, Whipple, Wiant
  et~al.}}]{Dickey:1994}
\bibinfo{author}{\bibfnamefont{J.~O.} \bibnamefont{Dickey}},
  \bibinfo{author}{\bibfnamefont{P.~L.} \bibnamefont{Bender}},
  \bibinfo{author}{\bibfnamefont{J.~E.} \bibnamefont{Faller}},
  \bibinfo{author}{\bibfnamefont{X.~X.} \bibnamefont{Newhall}},
  \bibinfo{author}{\bibfnamefont{R.}~\bibnamefont{Ricklefs}},
  \bibinfo{author}{\bibfnamefont{J.~G.} \bibnamefont{Ries}},
  \bibinfo{author}{\bibfnamefont{P.~J.} \bibnamefont{Shelus}},
  \bibinfo{author}{\bibfnamefont{C.}~\bibnamefont{Veillet}},
  \bibinfo{author}{\bibfnamefont{A.~L.} \bibnamefont{Whipple}},
  \bibinfo{author}{\bibfnamefont{J.~R.} \bibnamefont{Wiant}},
  \bibnamefont{et~al.}, \bibinfo{journal}{Science}
  \textbf{\bibinfo{volume}{265}}, \bibinfo{pages}{482} (\bibinfo{year}{1994}).

\bibitem[{\citenamefont{{Williams} et~al.}(2004)\citenamefont{{Williams},
  {Turyshev}, and {Boggs}}}]{Williams:2004}
\bibinfo{author}{\bibfnamefont{J.~G.} \bibnamefont{{Williams}}},
  \bibinfo{author}{\bibfnamefont{S.~G.} \bibnamefont{{Turyshev}}},
  \bibnamefont{and} \bibinfo{author}{\bibfnamefont{D.~H.}
  \bibnamefont{{Boggs}}}, \bibinfo{journal}{Phys. Rev. Lett.}
  \textbf{\bibinfo{volume}{93}}, \bibinfo{eid}{261101} (\bibinfo{year}{2004}).

\bibitem[{\citenamefont{{Williams} et~al.}(2009)\citenamefont{{Williams},
  {Turyshev}, and {Boggs}}}]{Williams:2009}
\bibinfo{author}{\bibfnamefont{J.~G.} \bibnamefont{{Williams}}},
  \bibinfo{author}{\bibfnamefont{S.~G.} \bibnamefont{{Turyshev}}},
  \bibnamefont{and} \bibinfo{author}{\bibfnamefont{D.~H.}
  \bibnamefont{{Boggs}}}, \bibinfo{journal}{IJMPD}
  \textbf{\bibinfo{volume}{18}}, \bibinfo{pages}{1129} (\bibinfo{year}{2009}).

\bibitem[{\citenamefont{{Williams} et~al.}(2012)\citenamefont{{Williams},
  {Turyshev}, and {Boggs}}}]{Williams2012}
\bibinfo{author}{\bibfnamefont{J.~G.} \bibnamefont{{Williams}}},
  \bibinfo{author}{\bibfnamefont{S.~G.} \bibnamefont{{Turyshev}}},
  \bibnamefont{and} \bibinfo{author}{\bibfnamefont{D.~H.}
  \bibnamefont{{Boggs}}}, \bibinfo{journal}{CQG} \textbf{\bibinfo{volume}{29}},
  \bibinfo{eid}{184004} (\bibinfo{year}{2012}).

\bibitem[{\citenamefont{{Williams}}(2008)}]{Williams:2008}
\bibinfo{author}{\bibfnamefont{J.~G.} \bibnamefont{{Williams}}}, in
  \emph{\bibinfo{booktitle}{16th International Workshop on Laser Ranging}}
  (\bibinfo{year}{2008}), p.~\bibinfo{pages}{17}.

\bibitem[{\citenamefont{{Williams} et~al.}(2014)\citenamefont{{Williams},
  {Konopliv}, {Boggs}, {Park}, {Yuan}, {Lemoine}, {Goossens}, {Mazarico},
  {Nimmo}, {Weber} et~al.}}]{Williams:2014}
\bibinfo{author}{\bibfnamefont{J.~G.} \bibnamefont{{Williams}}},
  \bibinfo{author}{\bibfnamefont{A.~S.} \bibnamefont{{Konopliv}}},
  \bibinfo{author}{\bibfnamefont{D.~H.} \bibnamefont{{Boggs}}},
  \bibinfo{author}{\bibfnamefont{R.~S.} \bibnamefont{{Park}}},
  \bibinfo{author}{\bibfnamefont{D.-N.} \bibnamefont{{Yuan}}},
  \bibinfo{author}{\bibfnamefont{F.~G.} \bibnamefont{{Lemoine}}},
  \bibinfo{author}{\bibfnamefont{S.}~\bibnamefont{{Goossens}}},
  \bibinfo{author}{\bibfnamefont{E.}~\bibnamefont{{Mazarico}}},
  \bibinfo{author}{\bibfnamefont{F.}~\bibnamefont{{Nimmo}}},
  \bibinfo{author}{\bibfnamefont{R.~C.} \bibnamefont{{Weber}}},
  \bibnamefont{et~al.}, \bibinfo{journal}{JGR (Planets)}
  \textbf{\bibinfo{volume}{119}}, \bibinfo{pages}{1546} (\bibinfo{year}{2014}).

\bibitem[{\citenamefont{Briaud et~al.}(2023)\citenamefont{Briaud, Ganino,
  Fienga, Mémin, and Rambaux}}]{Briaud2023}
\bibinfo{author}{\bibfnamefont{A.}~\bibnamefont{Briaud}},
  \bibinfo{author}{\bibfnamefont{C.}~\bibnamefont{Ganino}},
  \bibinfo{author}{\bibfnamefont{A.}~\bibnamefont{Fienga}},
  \bibinfo{author}{\bibfnamefont{A.}~\bibnamefont{Mémin}}, \bibnamefont{and}
  \bibinfo{author}{\bibfnamefont{N.}~\bibnamefont{Rambaux}},
  \bibinfo{journal}{Nature} \textbf{\bibinfo{volume}{617}},
  \bibinfo{pages}{743} (\bibinfo{year}{2023}).

\bibitem[{\citenamefont{{Turyshev}}(2008)}]{Turyshev:2008}
\bibinfo{author}{\bibfnamefont{S.~G.} \bibnamefont{{Turyshev}}},
  \bibinfo{journal}{Ann. Rev. Nucl. Part. Sci.} \textbf{\bibinfo{volume}{58}},
  \bibinfo{pages}{207} (\bibinfo{year}{2008}).

\bibitem[{\citenamefont{{Zhang} et~al.}(2022)\citenamefont{{Zhang},
  {M{\"u}ller}, {Biskupek}, and {Singh}}}]{Zhang-etal:2022}
\bibinfo{author}{\bibfnamefont{M.}~\bibnamefont{{Zhang}}},
  \bibinfo{author}{\bibfnamefont{J.}~\bibnamefont{{M{\"u}ller}}},
  \bibinfo{author}{\bibfnamefont{L.}~\bibnamefont{{Biskupek}}},
  \bibnamefont{and} \bibinfo{author}{\bibfnamefont{V.~V.}
  \bibnamefont{{Singh}}}, \bibinfo{journal}{Astron. Asrophys.}
  \textbf{\bibinfo{volume}{659}}, \bibinfo{eid}{A148} (\bibinfo{year}{2022}).

\bibitem[{\citenamefont{{Konopliv} et~al.}(2013)\citenamefont{{Konopliv},
  {Park}, {Yuan}, {Asmar}, {Watkins}, {Williams}, {Fahnestock}, {Kruizinga},
  {Paik}, {Strekalov} et~al.}}]{Konopliv-etal:2013}
\bibinfo{author}{\bibfnamefont{A.~S.} \bibnamefont{{Konopliv}}},
  \bibinfo{author}{\bibfnamefont{R.~S.} \bibnamefont{{Park}}},
  \bibinfo{author}{\bibfnamefont{D.-N.} \bibnamefont{{Yuan}}},
  \bibinfo{author}{\bibfnamefont{S.~W.} \bibnamefont{{Asmar}}},
  \bibinfo{author}{\bibfnamefont{M.~M.} \bibnamefont{{Watkins}}},
  \bibinfo{author}{\bibfnamefont{J.~G.} \bibnamefont{{Williams}}},
  \bibinfo{author}{\bibfnamefont{E.}~\bibnamefont{{Fahnestock}}},
  \bibinfo{author}{\bibfnamefont{G.}~\bibnamefont{{Kruizinga}}},
  \bibinfo{author}{\bibfnamefont{M.}~\bibnamefont{{Paik}}},
  \bibinfo{author}{\bibfnamefont{D.}~\bibnamefont{{Strekalov}}},
  \bibnamefont{et~al.}, \bibinfo{journal}{J. Geophys. Res. (Planets)}
  \textbf{\bibinfo{volume}{118}}, \bibinfo{pages}{1415} (\bibinfo{year}{2013}).

\bibitem[{\citenamefont{{Zuber} et~al.}(2013)\citenamefont{{Zuber}, {Smith},
  {Watkins}, {Asmar}, {Konopliv}, {Lemoine}, {Melosh}, {Neumann}, {Phillips},
  {Solomon} et~al.}}]{Zuber-etal:2013}
\bibinfo{author}{\bibfnamefont{M.~T.} \bibnamefont{{Zuber}}},
  \bibinfo{author}{\bibfnamefont{D.~E.} \bibnamefont{{Smith}}},
  \bibinfo{author}{\bibfnamefont{M.~M.} \bibnamefont{{Watkins}}},
  \bibinfo{author}{\bibfnamefont{S.~W.} \bibnamefont{{Asmar}}},
  \bibinfo{author}{\bibfnamefont{A.~S.} \bibnamefont{{Konopliv}}},
  \bibinfo{author}{\bibfnamefont{F.~G.} \bibnamefont{{Lemoine}}},
  \bibinfo{author}{\bibfnamefont{H.~J.} \bibnamefont{{Melosh}}},
  \bibinfo{author}{\bibfnamefont{G.~A.} \bibnamefont{{Neumann}}},
  \bibinfo{author}{\bibfnamefont{R.~J.} \bibnamefont{{Phillips}}},
  \bibinfo{author}{\bibfnamefont{S.~C.} \bibnamefont{{Solomon}}},
  \bibnamefont{et~al.}, \bibinfo{journal}{Science}
  \textbf{\bibinfo{volume}{339}}, \bibinfo{pages}{668} (\bibinfo{year}{2013}).

\bibitem[{\citenamefont{{Blas} and {Jenkins}}(2022)}]{Blas-Jenkins:2022}
\bibinfo{author}{\bibfnamefont{D.}~\bibnamefont{{Blas}}} \bibnamefont{and}
  \bibinfo{author}{\bibfnamefont{A.~C.} \bibnamefont{{Jenkins}}},
  \bibinfo{journal}{Phys. Rev. Lett.} \textbf{\bibinfo{volume}{128}},
  \bibinfo{eid}{101103} (\bibinfo{year}{2022}).

\bibitem[{\citenamefont{{Turyshev}
  et~al.}(2024{\natexlab{a}})\citenamefont{{Turyshev}, {Shao}, {Hahn}, {Blas},
  and {Jenkins}}}]{Turyshev-etal:2024}
\bibinfo{author}{\bibfnamefont{S.~G.} \bibnamefont{{Turyshev}}},
  \bibinfo{author}{\bibfnamefont{M.}~\bibnamefont{{Shao}}},
  \bibinfo{author}{\bibfnamefont{I.}~\bibnamefont{{Hahn}}},
  \bibinfo{author}{\bibfnamefont{D.}~\bibnamefont{{Blas}}}, \bibnamefont{and}
  \bibinfo{author}{\bibfnamefont{A.~C.} \bibnamefont{{Jenkins}}}, in
  \emph{\bibinfo{booktitle}{LPI Contributions}}
  (\bibinfo{year}{2024}{\natexlab{a}}), vol. \bibinfo{volume}{3065}, p.
  \bibinfo{pages}{5001}.

\bibitem[{\citenamefont{{Shao} et~al.}(2018)\citenamefont{{Shao}, {Turyshev},
  {Hanh}, and {Trahan}}}]{Shao:2018}
\bibinfo{author}{\bibfnamefont{M.}~\bibnamefont{{Shao}}},
  \bibinfo{author}{\bibfnamefont{S.~G.} \bibnamefont{{Turyshev}}},
  \bibinfo{author}{\bibfnamefont{I.}~\bibnamefont{{Hanh}}}, \bibnamefont{and}
  \bibinfo{author}{\bibfnamefont{R.}~\bibnamefont{{Trahan}}}, in
  \emph{\bibinfo{booktitle}{Proceedings of the 21st International Workshop on
  Laser Ranging, Canberra, Australia, 5-9 November 2018}}
  (\bibinfo{organization}{IILRS}, \bibinfo{year}{2018}).

\bibitem[{\citenamefont{{Turyshev} et~al.}(2018)\citenamefont{{Turyshev},
  {Shao}, {Hanh}, {Williams}, and {Trahan}}}]{Turyshev:2018}
\bibinfo{author}{\bibfnamefont{S.~G.} \bibnamefont{{Turyshev}}},
  \bibinfo{author}{\bibfnamefont{M.}~\bibnamefont{{Shao}}},
  \bibinfo{author}{\bibfnamefont{I.}~\bibnamefont{{Hanh}}},
  \bibinfo{author}{\bibfnamefont{J.~G.} \bibnamefont{{Williams}}},
  \bibnamefont{and} \bibinfo{author}{\bibfnamefont{R.}~\bibnamefont{{Trahan}}},
  in \emph{\bibinfo{booktitle}{Proceedings of the 21st International Workshop
  on Laser Ranging, Canberra, Australia, 5-9 November 2018}}
  (\bibinfo{organization}{IILRS}, \bibinfo{year}{2018}).

\bibitem[{\citenamefont{{Murphy} et~al.}(2008)\citenamefont{{Murphy},
  {Adelberger}, {Battat}, {Carey}, {Hoyle}, {LeBlanc}, {Michelsen},
  {Nordtvedt}, {Orin}, {Strasburg} et~al.}}]{Murphy_etal_2008}
\bibinfo{author}{\bibfnamefont{T.~W.} \bibnamefont{{Murphy}},
  \bibfnamefont{Jr.}}, \bibinfo{author}{\bibfnamefont{E.~G.}
  \bibnamefont{{Adelberger}}}, \bibinfo{author}{\bibfnamefont{J.~B.~R.}
  \bibnamefont{{Battat}}}, \bibinfo{author}{\bibfnamefont{L.~N.}
  \bibnamefont{{Carey}}}, \bibinfo{author}{\bibfnamefont{C.~D.}
  \bibnamefont{{Hoyle}}},
  \bibinfo{author}{\bibfnamefont{P.}~\bibnamefont{{LeBlanc}}},
  \bibinfo{author}{\bibfnamefont{E.~L.} \bibnamefont{{Michelsen}}},
  \bibinfo{author}{\bibfnamefont{K.}~\bibnamefont{{Nordtvedt}}},
  \bibinfo{author}{\bibfnamefont{A.~E.} \bibnamefont{{Orin}}},
  \bibinfo{author}{\bibfnamefont{J.~D.} \bibnamefont{{Strasburg}}},
  \bibnamefont{et~al.}, \bibinfo{journal}{PASP} \textbf{\bibinfo{volume}{120}},
  \bibinfo{pages}{20} (\bibinfo{year}{2008}).

\bibitem[{\citenamefont{Samain et~al.}(1998)\citenamefont{Samain, Veillet,
  Fridelance, and {et al.}}}]{Samain1998}
\bibinfo{author}{\bibfnamefont{E.}~\bibnamefont{Samain}},
  \bibinfo{author}{\bibfnamefont{P.}~\bibnamefont{Veillet}},
  \bibinfo{author}{\bibfnamefont{C.}~\bibnamefont{Fridelance}},
  \bibnamefont{and} \bibinfo{author}{\bibnamefont{{et al.}}},
  \bibinfo{journal}{Astron. Astrophys.} \textbf{\bibinfo{volume}{336}},
  \bibinfo{pages}{L17} (\bibinfo{year}{1998}).

\bibitem[{\citenamefont{{Murphy}}(2013)}]{Murphy:2013}
\bibinfo{author}{\bibfnamefont{T.~W.} \bibnamefont{{Murphy}},
  \bibfnamefont{Jr.}}, \bibinfo{journal}{Rep. Progr. Phys.}
  \textbf{\bibinfo{volume}{76}}, \bibinfo{eid}{076901} (\bibinfo{year}{2013}).

\bibitem[{\citenamefont{Degnan}(1993)}]{Degnan1993}
\bibinfo{author}{\bibfnamefont{J.~J.} \bibnamefont{Degnan}},
  \emph{\bibinfo{title}{Millimeter Accuracy Satellite Laser Ranging: a Review}}
  (\bibinfo{publisher}{AGU Geodynamics Series}, \bibinfo{year}{1993}),
  vol.~\bibinfo{volume}{25}, pp. \bibinfo{pages}{133--162}.

\bibitem[{\citenamefont{{Murphy} et~al.}(2010)\citenamefont{{Murphy},
  {Adelberger}, {Battat}, {Hoyle}, {McMillan}, {Michelsen}, {Samad}, {Stubbs},
  and {Swanson}}}]{Murphy_etal_2010}
\bibinfo{author}{\bibfnamefont{T.~W.} \bibnamefont{{Murphy}},
  \bibfnamefont{Jr.}}, \bibinfo{author}{\bibfnamefont{E.~G.}
  \bibnamefont{{Adelberger}}}, \bibinfo{author}{\bibfnamefont{J.~B.~R.}
  \bibnamefont{{Battat}}}, \bibinfo{author}{\bibfnamefont{C.~D.}
  \bibnamefont{{Hoyle}}}, \bibinfo{author}{\bibfnamefont{R.~J.}
  \bibnamefont{{McMillan}}}, \bibinfo{author}{\bibfnamefont{E.~L.}
  \bibnamefont{{Michelsen}}}, \bibinfo{author}{\bibfnamefont{R.~L.}
  \bibnamefont{{Samad}}}, \bibinfo{author}{\bibfnamefont{C.~W.}
  \bibnamefont{{Stubbs}}}, \bibnamefont{and}
  \bibinfo{author}{\bibfnamefont{H.~E.} \bibnamefont{{Swanson}}},
  \bibinfo{journal}{Icarus} \textbf{\bibinfo{volume}{208}}, \bibinfo{pages}{31}
  (\bibinfo{year}{2010}).

\bibitem[{\citenamefont{{Battat} et~al.}(2009)\citenamefont{{Battat}, {Murphy},
  {Adelberger}, {Gillespie}, {Hoyle}, {McMillan}, {Michelsen}, {Nordtvedt},
  {Orin}, {Stubbs} et~al.}}]{Battat_etal_2009}
\bibinfo{author}{\bibfnamefont{J.~B.~R.} \bibnamefont{{Battat}}},
  \bibinfo{author}{\bibfnamefont{T.~W.} \bibnamefont{{Murphy}}},
  \bibinfo{author}{\bibfnamefont{E.~G.} \bibnamefont{{Adelberger}}},
  \bibinfo{author}{\bibfnamefont{B.}~\bibnamefont{{Gillespie}}},
  \bibinfo{author}{\bibfnamefont{C.~D.} \bibnamefont{{Hoyle}}},
  \bibinfo{author}{\bibfnamefont{R.~J.} \bibnamefont{{McMillan}}},
  \bibinfo{author}{\bibfnamefont{E.~L.} \bibnamefont{{Michelsen}}},
  \bibinfo{author}{\bibfnamefont{K.}~\bibnamefont{{Nordtvedt}}},
  \bibinfo{author}{\bibfnamefont{A.~E.} \bibnamefont{{Orin}}},
  \bibinfo{author}{\bibfnamefont{C.~W.} \bibnamefont{{Stubbs}}},
  \bibnamefont{et~al.}, \bibinfo{journal}{PASP} \textbf{\bibinfo{volume}{121}},
  \bibinfo{pages}{29} (\bibinfo{year}{2009}).

\bibitem[{\citenamefont{Lisdat et~al.}(2016)\citenamefont{Lisdat, Grosche,
  Quintin, Shi, Lipphardt, Tamm, Al-Masoudi, Fitzek, B{\"u}ckle, Jeon
  et~al.}}]{Lisdat2016}
\bibinfo{author}{\bibfnamefont{C.}~\bibnamefont{Lisdat}},
  \bibinfo{author}{\bibfnamefont{G.}~\bibnamefont{Grosche}},
  \bibinfo{author}{\bibfnamefont{N.}~\bibnamefont{Quintin}},
  \bibinfo{author}{\bibfnamefont{C.}~\bibnamefont{Shi}},
  \bibinfo{author}{\bibfnamefont{B.}~\bibnamefont{Lipphardt}},
  \bibinfo{author}{\bibfnamefont{C.}~\bibnamefont{Tamm}},
  \bibinfo{author}{\bibfnamefont{A.}~\bibnamefont{Al-Masoudi}},
  \bibinfo{author}{\bibfnamefont{F.}~\bibnamefont{Fitzek}},
  \bibinfo{author}{\bibfnamefont{T.}~\bibnamefont{B{\"u}ckle}},
  \bibinfo{author}{\bibfnamefont{J.}~\bibnamefont{Jeon}}, \bibnamefont{et~al.},
  \bibinfo{journal}{Nature Communications} \textbf{\bibinfo{volume}{7}},
  \bibinfo{pages}{12443} (\bibinfo{year}{2016}).

\bibitem[{\citenamefont{{Park} et~al.}(2021)\citenamefont{{Park}, {Folkner},
  {Williams}, and {Boggs}}}]{Park-etal:2021}
\bibinfo{author}{\bibfnamefont{R.~S.} \bibnamefont{{Park}}},
  \bibinfo{author}{\bibfnamefont{W.~M.} \bibnamefont{{Folkner}}},
  \bibinfo{author}{\bibfnamefont{J.~G.} \bibnamefont{{Williams}}},
  \bibnamefont{and} \bibinfo{author}{\bibfnamefont{D.~H.}
  \bibnamefont{{Boggs}}}, \bibinfo{journal}{Astron. J.}
  \textbf{\bibinfo{volume}{161}}, \bibinfo{eid}{105} (\bibinfo{year}{2021}).

\bibitem[{\citenamefont{Williams et~al.}(2014)\citenamefont{Williams, Boggs,
  and Ratcliff}}]{Williams_etal_2014}
\bibinfo{author}{\bibfnamefont{J.~G.} \bibnamefont{Williams}},
  \bibinfo{author}{\bibfnamefont{D.~H.} \bibnamefont{Boggs}}, \bibnamefont{and}
  \bibinfo{author}{\bibfnamefont{J.~T.} \bibnamefont{Ratcliff}},
  \bibinfo{journal}{JGR (Planets)} \textbf{\bibinfo{volume}{119}},
  \bibinfo{pages}{1546} (\bibinfo{year}{2014}).

\bibitem[{\citenamefont{{Murphy} et~al.}(2012)\citenamefont{{Murphy},
  {Adelberger}, {Battat}, {Hoyle}, {Johnson}, {McMillan}, {Stubbs}, and
  {Swanson}}}]{Murphy_etal_2012}
\bibinfo{author}{\bibfnamefont{T.~W.} \bibnamefont{{Murphy}},
  \bibfnamefont{Jr.}}, \bibinfo{author}{\bibfnamefont{E.~G.}
  \bibnamefont{{Adelberger}}}, \bibinfo{author}{\bibfnamefont{J.~B.~R.}
  \bibnamefont{{Battat}}}, \bibinfo{author}{\bibfnamefont{C.~D.}
  \bibnamefont{{Hoyle}}}, \bibinfo{author}{\bibfnamefont{N.~H.}
  \bibnamefont{{Johnson}}}, \bibinfo{author}{\bibfnamefont{R.~J.}
  \bibnamefont{{McMillan}}}, \bibinfo{author}{\bibfnamefont{C.~W.}
  \bibnamefont{{Stubbs}}}, \bibnamefont{and}
  \bibinfo{author}{\bibfnamefont{H.~E.} \bibnamefont{{Swanson}}},
  \bibinfo{journal}{CQG} \textbf{\bibinfo{volume}{29}}, \bibinfo{eid}{184005}
  (\bibinfo{year}{2012}).

\bibitem[{\citenamefont{Mendes and Pavlis}(2004)}]{Mendes2004}
\bibinfo{author}{\bibfnamefont{V.}~\bibnamefont{Mendes}} \bibnamefont{and}
  \bibinfo{author}{\bibfnamefont{E.}~\bibnamefont{Pavlis}},
  \bibinfo{journal}{Geophysical Research Letters}
  \textbf{\bibinfo{volume}{31}}, \bibinfo{pages}{L14602}
  (\bibinfo{year}{2004}).

\bibitem[{\citenamefont{{Turyshev} et~al.}(2013)\citenamefont{{Turyshev},
  {Williams}, {Folkner}, {Gutt}, {Baran}, {Hein}, {Somawardhana}, {Lipa}, and
  {Wang}}}]{Turyshev-etal:2013}
\bibinfo{author}{\bibfnamefont{S.~G.} \bibnamefont{{Turyshev}}},
  \bibinfo{author}{\bibfnamefont{J.~G.} \bibnamefont{{Williams}}},
  \bibinfo{author}{\bibfnamefont{W.~M.} \bibnamefont{{Folkner}}},
  \bibinfo{author}{\bibfnamefont{G.~M.} \bibnamefont{{Gutt}}},
  \bibinfo{author}{\bibfnamefont{R.~T.} \bibnamefont{{Baran}}},
  \bibinfo{author}{\bibfnamefont{R.~C.} \bibnamefont{{Hein}}},
  \bibinfo{author}{\bibfnamefont{R.~P.} \bibnamefont{{Somawardhana}}},
  \bibinfo{author}{\bibfnamefont{J.~A.} \bibnamefont{{Lipa}}},
  \bibnamefont{and} \bibinfo{author}{\bibfnamefont{S.}~\bibnamefont{{Wang}}},
  \bibinfo{journal}{Experimental Astronomy} \textbf{\bibinfo{volume}{36}},
  \bibinfo{pages}{105} (\bibinfo{year}{2013}).

\bibitem[{\citenamefont{{Lyard} et~al.}(2021)\citenamefont{{Lyard}, {Allain},
  {Cancet}, {Carr\`ere}, and {Picot}}}]{fes2014}
\bibinfo{author}{\bibfnamefont{F.~H.} \bibnamefont{{Lyard}}},
  \bibinfo{author}{\bibfnamefont{D.~J.} \bibnamefont{{Allain}}},
  \bibinfo{author}{\bibfnamefont{M.}~\bibnamefont{{Cancet}}},
  \bibinfo{author}{\bibfnamefont{L.}~\bibnamefont{{Carr\`ere}}},
  \bibnamefont{and} \bibinfo{author}{\bibfnamefont{N.}~\bibnamefont{{Picot}}},
  \bibinfo{journal}{Ocean Science} \textbf{\bibinfo{volume}{17}},
  \bibinfo{pages}{615} (\bibinfo{year}{2021}).

\bibitem[{\citenamefont{Vokrouhlick{\'y}
  et~al.}(2019)\citenamefont{Vokrouhlick{\'y}, M{\"u}ller, Dirkx, and {et
  al.}}}]{Vokrouhlicky_etal_2019}
\bibinfo{author}{\bibfnamefont{D.}~\bibnamefont{Vokrouhlick{\'y}}},
  \bibinfo{author}{\bibfnamefont{J.}~\bibnamefont{M{\"u}ller}},
  \bibinfo{author}{\bibfnamefont{D.}~\bibnamefont{Dirkx}}, \bibnamefont{and}
  \bibinfo{author}{\bibnamefont{{et al.}}}, \bibinfo{journal}{Space Science
  Reviews} \textbf{\bibinfo{volume}{215}}, \bibinfo{pages}{16}
  (\bibinfo{year}{2019}).

\bibitem[{\citenamefont{Altamimi et~al.}(2016)\citenamefont{Altamimi,
  Rebischung, M{\'e}tivier, and Collilieux}}]{altamimi2016}
\bibinfo{author}{\bibfnamefont{Z.}~\bibnamefont{Altamimi}},
  \bibinfo{author}{\bibfnamefont{P.}~\bibnamefont{Rebischung}},
  \bibinfo{author}{\bibfnamefont{L.}~\bibnamefont{M{\'e}tivier}},
  \bibnamefont{and}
  \bibinfo{author}{\bibfnamefont{X.}~\bibnamefont{Collilieux}},
  \bibinfo{journal}{JGR: Solid Earth} \textbf{\bibinfo{volume}{121}},
  \bibinfo{pages}{6109} (\bibinfo{year}{2016}).

\bibitem[{\citenamefont{Sakamoto et~al.}(2014)\citenamefont{Sakamoto, Sato, and
  Imamura}}]{Sakamoto2014}
\bibinfo{author}{\bibfnamefont{M.}~\bibnamefont{Sakamoto}},
  \bibinfo{author}{\bibfnamefont{Y.}~\bibnamefont{Sato}}, \bibnamefont{and}
  \bibinfo{author}{\bibfnamefont{Y.}~\bibnamefont{Imamura}},
  \bibinfo{journal}{Optics Letters} \textbf{\bibinfo{volume}{39}},
  \bibinfo{pages}{1581} (\bibinfo{year}{2014}).

\bibitem[{\citenamefont{Marsili et~al.}(2013)\citenamefont{Marsili, Verma,
  Stern, and {et al.}}}]{Marsili2013}
\bibinfo{author}{\bibfnamefont{F.}~\bibnamefont{Marsili}},
  \bibinfo{author}{\bibfnamefont{V.~B.} \bibnamefont{Verma}},
  \bibinfo{author}{\bibfnamefont{J.~A.} \bibnamefont{Stern}}, \bibnamefont{and}
  \bibinfo{author}{\bibnamefont{{et al.}}}, \bibinfo{journal}{Nature Photonics}
  \textbf{\bibinfo{volume}{7}}, \bibinfo{pages}{210} (\bibinfo{year}{2013}).

\bibitem[{\citenamefont{{Fienga} et~al.}(2021)\citenamefont{{Fienga}, {Deram},
  {Di Ruscio}, {Viswanathan}, {Camargo}, {Bernus}, {Gastineau}, and
  {Laskar}}}]{Fienga_etal_2021}
\bibinfo{author}{\bibfnamefont{A.}~\bibnamefont{{Fienga}}},
  \bibinfo{author}{\bibfnamefont{P.}~\bibnamefont{{Deram}}},
  \bibinfo{author}{\bibfnamefont{A.}~\bibnamefont{{Di Ruscio}}},
  \bibinfo{author}{\bibfnamefont{V.}~\bibnamefont{{Viswanathan}}},
  \bibinfo{author}{\bibfnamefont{J.~I.~B.} \bibnamefont{{Camargo}}},
  \bibinfo{author}{\bibfnamefont{L.}~\bibnamefont{{Bernus}}},
  \bibinfo{author}{\bibfnamefont{M.}~\bibnamefont{{Gastineau}}},
  \bibnamefont{and} \bibinfo{author}{\bibfnamefont{J.}~\bibnamefont{{Laskar}}},
  \bibinfo{journal}{Notes Scientifiques et Techniques de l'Institut de
  Mecanique Celeste} \textbf{\bibinfo{volume}{110}} (\bibinfo{year}{2021}).

\bibitem[{\citenamefont{{Turyshev} and {Williams}}(2007)}]{Turyshev2007}
\bibinfo{author}{\bibfnamefont{S.~G.} \bibnamefont{{Turyshev}}}
  \bibnamefont{and} \bibinfo{author}{\bibfnamefont{J.~G.}
  \bibnamefont{{Williams}}}, \bibinfo{journal}{IJMPD}
  \textbf{\bibinfo{volume}{16}}, \bibinfo{pages}{2165} (\bibinfo{year}{2007}).

\bibitem[{\citenamefont{{Currie} et~al.}(2011)\citenamefont{{Currie},
  {Dell'Agnello}, and {Delle Monache}}}]{Currie_etal_2011}
\bibinfo{author}{\bibfnamefont{D.}~\bibnamefont{{Currie}}},
  \bibinfo{author}{\bibfnamefont{S.}~\bibnamefont{{Dell'Agnello}}},
  \bibnamefont{and} \bibinfo{author}{\bibfnamefont{G.}~\bibnamefont{{Delle
  Monache}}}, \bibinfo{journal}{Acta Astronautica}
  \textbf{\bibinfo{volume}{68}}, \bibinfo{pages}{667} (\bibinfo{year}{2011}).

\bibitem[{\citenamefont{Samain et~al.}(2009)\citenamefont{Samain, Guillemot,
  Courde, and Torre}}]{Samain2009}
\bibinfo{author}{\bibfnamefont{E.}~\bibnamefont{Samain}},
  \bibinfo{author}{\bibfnamefont{P.}~\bibnamefont{Guillemot}},
  \bibinfo{author}{\bibfnamefont{C.}~\bibnamefont{Courde}}, \bibnamefont{and}
  \bibinfo{author}{\bibfnamefont{L.}~\bibnamefont{Torre}}, in
  \emph{\bibinfo{booktitle}{16th International Workshop on Laser Ranging}}
  (\bibinfo{year}{2009}),
  \urlprefix\url{https://ilrs.gsfc.nasa.gov/lw16/docs/papers/new_4_Samain_p.pdf}.

\bibitem[{\citenamefont{Courde et~al.}(2017)\citenamefont{Courde, Torre,
  Samain, Martinot-Lagarde, Aimar, Albanese, Exertier, Fienga, Mariey, Metris
  et~al.}}]{Courde-etal:2017}
\bibinfo{author}{\bibfnamefont{C.}~\bibnamefont{Courde}},
  \bibinfo{author}{\bibfnamefont{J.~M.} \bibnamefont{Torre}},
  \bibinfo{author}{\bibfnamefont{E.}~\bibnamefont{Samain}},
  \bibinfo{author}{\bibfnamefont{G.}~\bibnamefont{Martinot-Lagarde}},
  \bibinfo{author}{\bibfnamefont{M.}~\bibnamefont{Aimar}},
  \bibinfo{author}{\bibfnamefont{D.}~\bibnamefont{Albanese}},
  \bibinfo{author}{\bibfnamefont{P.}~\bibnamefont{Exertier}},
  \bibinfo{author}{\bibfnamefont{A.}~\bibnamefont{Fienga}},
  \bibinfo{author}{\bibfnamefont{H.}~\bibnamefont{Mariey}},
  \bibinfo{author}{\bibfnamefont{G.}~\bibnamefont{Metris}},
  \bibnamefont{et~al.}, \bibinfo{journal}{Astron. Astrophys.}
  \textbf{\bibinfo{volume}{602}}, \bibinfo{pages}{A90} (\bibinfo{year}{2017}).

\bibitem[{\citenamefont{Campbell et~al.}(2017)\citenamefont{Campbell, Hutson,
  Marti, Goban, Darkwah~Oppong, McNally, Sonderhouse, Robinson, Zhang, Bloom
  et~al.}}]{Campbell2017}
\bibinfo{author}{\bibfnamefont{S.~L.} \bibnamefont{Campbell}},
  \bibinfo{author}{\bibfnamefont{R.~B.} \bibnamefont{Hutson}},
  \bibinfo{author}{\bibfnamefont{G.~E.} \bibnamefont{Marti}},
  \bibinfo{author}{\bibfnamefont{A.}~\bibnamefont{Goban}},
  \bibinfo{author}{\bibfnamefont{N.}~\bibnamefont{Darkwah~Oppong}},
  \bibinfo{author}{\bibfnamefont{R.~L.} \bibnamefont{McNally}},
  \bibinfo{author}{\bibfnamefont{L.}~\bibnamefont{Sonderhouse}},
  \bibinfo{author}{\bibfnamefont{J.~M.} \bibnamefont{Robinson}},
  \bibinfo{author}{\bibfnamefont{W.}~\bibnamefont{Zhang}},
  \bibinfo{author}{\bibfnamefont{B.~J.} \bibnamefont{Bloom}},
  \bibnamefont{et~al.}, \bibinfo{journal}{Science}
  \textbf{\bibinfo{volume}{358}}, \bibinfo{pages}{90} (\bibinfo{year}{2017}).

\bibitem[{\citenamefont{{Turyshev} et~al.}(2010)\citenamefont{{Turyshev},
  {Farr}, {Folkner}, {Girerd}, {Hemmati}, {Murphy}, {Williams}, and
  {Degnan}}}]{Turyshev-etal:2010}
\bibinfo{author}{\bibfnamefont{S.~G.} \bibnamefont{{Turyshev}}},
  \bibinfo{author}{\bibfnamefont{W.}~\bibnamefont{{Farr}}},
  \bibinfo{author}{\bibfnamefont{W.~M.} \bibnamefont{{Folkner}}},
  \bibinfo{author}{\bibfnamefont{A.~R.} \bibnamefont{{Girerd}}},
  \bibinfo{author}{\bibfnamefont{H.}~\bibnamefont{{Hemmati}}},
  \bibinfo{author}{\bibfnamefont{T.~W.} \bibnamefont{{Murphy}}},
  \bibinfo{author}{\bibfnamefont{J.~G.} \bibnamefont{{Williams}}},
  \bibnamefont{and} \bibinfo{author}{\bibfnamefont{J.~J.}
  \bibnamefont{{Degnan}}}, \bibinfo{journal}{Experimental Astronomy}
  \textbf{\bibinfo{volume}{28}}, \bibinfo{pages}{209} (\bibinfo{year}{2010}).

\bibitem[{\citenamefont{{Williams}}(2018)}]{Williams:2018}
\bibinfo{author}{\bibfnamefont{J.~G.} \bibnamefont{{Williams}}},
  \bibinfo{journal}{Cel. Mech. Dyn. Astron.} \textbf{\bibinfo{volume}{130}},
  \bibinfo{eid}{63} (\bibinfo{year}{2018}).

\bibitem[{\citenamefont{Rife and Boorstyn}(1974)}]{Rife1974}
\bibinfo{author}{\bibfnamefont{D.~C.} \bibnamefont{Rife}} \bibnamefont{and}
  \bibinfo{author}{\bibfnamefont{R.~R.} \bibnamefont{Boorstyn}},
  \bibinfo{journal}{IEEE Transactions on Information Theory}
  \textbf{\bibinfo{volume}{20}}, \bibinfo{pages}{591} (\bibinfo{year}{1974}).

\bibitem[{\citenamefont{{Kay}}(1993)}]{kay1993}
\bibinfo{author}{\bibfnamefont{S.~M.} \bibnamefont{{Kay}}},
  \emph{\bibinfo{title}{Fundamentals of Statistical Signal Processing:
  Estimation Theory}} (\bibinfo{publisher}{Prentice Hall},
  \bibinfo{year}{1993}).

\bibitem[{\citenamefont{Van~Trees}(2001)}]{vantrees2001}
\bibinfo{author}{\bibfnamefont{H.~L.} \bibnamefont{Van~Trees}},
  \emph{\bibinfo{title}{Detection, Estimation, and Modulation Theory, Part I}}
  (\bibinfo{publisher}{Wiley-Interscience}, \bibinfo{year}{2001}).

\bibitem[{\citenamefont{{Murphy, Jr.} et~al.}(2006)\citenamefont{{Murphy, Jr.},
  {Adelberger}, {Battat}, {Hoyle}, {Michelsen}, {Stubbs}, and
  {Swanson}}}]{Murphy:2006}
\bibinfo{author}{\bibfnamefont{T.~W.} \bibnamefont{{Murphy, Jr.}}},
  \bibinfo{author}{\bibfnamefont{E.~G.} \bibnamefont{{Adelberger}}},
  \bibinfo{author}{\bibfnamefont{J.~B.} \bibnamefont{{Battat}}},
  \bibinfo{author}{\bibfnamefont{C.~D.} \bibnamefont{{Hoyle}}},
  \bibinfo{author}{\bibfnamefont{E.~L.} \bibnamefont{{Michelsen}}},
  \bibinfo{author}{\bibfnamefont{C.~W.} \bibnamefont{{Stubbs}}},
  \bibnamefont{and} \bibinfo{author}{\bibfnamefont{H.~E.}
  \bibnamefont{{Swanson}}}, in \emph{\bibinfo{booktitle}{Proceedings of the
  15st International Workshop on Laser Ranging, Canberra, Australia, 15-20
  October 2006}} (\bibinfo{organization}{IILRS}, \bibinfo{year}{2006}).

\bibitem[{\citenamefont{Murphy~Jr. et~al.}(2010)\citenamefont{Murphy~Jr.,
  Adelberger, Battat, Hoyle, McMillan, Michelsen, Samad, Stubbs, and
  Swanson}}]{Murphy2010}
\bibinfo{author}{\bibfnamefont{T.~W.} \bibnamefont{Murphy~Jr.}},
  \bibinfo{author}{\bibfnamefont{E.~G.} \bibnamefont{Adelberger}},
  \bibinfo{author}{\bibfnamefont{J.~B.~R.} \bibnamefont{Battat}},
  \bibinfo{author}{\bibfnamefont{C.~D.} \bibnamefont{Hoyle}},
  \bibinfo{author}{\bibfnamefont{R.~J.} \bibnamefont{McMillan}},
  \bibinfo{author}{\bibfnamefont{E.~L.} \bibnamefont{Michelsen}},
  \bibinfo{author}{\bibfnamefont{R.~L.} \bibnamefont{Samad}},
  \bibinfo{author}{\bibfnamefont{C.~W.} \bibnamefont{Stubbs}},
  \bibnamefont{and} \bibinfo{author}{\bibfnamefont{H.~E.}
  \bibnamefont{Swanson}}, \bibinfo{journal}{Icarus}
  \textbf{\bibinfo{volume}{208}}, \bibinfo{pages}{31} (\bibinfo{year}{2010}).

\bibitem[{\citenamefont{Telle et~al.}(1999)\citenamefont{Telle, Lipphardt, and
  Stenger}}]{Telle1999}
\bibinfo{author}{\bibfnamefont{H.~R.} \bibnamefont{Telle}},
  \bibinfo{author}{\bibfnamefont{B.}~\bibnamefont{Lipphardt}},
  \bibnamefont{and} \bibinfo{author}{\bibfnamefont{J.}~\bibnamefont{Stenger}},
  \bibinfo{journal}{Applied Physics B} \textbf{\bibinfo{volume}{69}},
  \bibinfo{pages}{327} (\bibinfo{year}{1999}).

\bibitem[{\citenamefont{Beland}(1993)}]{Beland1993}
\bibinfo{author}{\bibfnamefont{R.~R.} \bibnamefont{Beland}}, in
  \emph{\bibinfo{booktitle}{The Infrared \& Electro-Optical Systems Handbook}},
  edited by \bibinfo{editor}{\bibfnamefont{F.~G.} \bibnamefont{Smith}}
  (\bibinfo{publisher}{SPIE Press}, \bibinfo{year}{1993}),
  vol.~\bibinfo{volume}{2}, pp. \bibinfo{pages}{157--232}.

\bibitem[{\citenamefont{{Dro{\.z}d{\.z}ewski} and
  {So{\'s}nica}}(2021)}]{Drozdzewski:2021}
\bibinfo{author}{\bibfnamefont{M.}~\bibnamefont{{Dro{\.z}d{\.z}ewski}}}
  \bibnamefont{and}
  \bibinfo{author}{\bibfnamefont{K.}~\bibnamefont{{So{\'s}nica}}},
  \bibinfo{journal}{J. Geodesy} \textbf{\bibinfo{volume}{95}},
  \bibinfo{eid}{100} (\bibinfo{year}{2021}).

\bibitem[{\citenamefont{{Arnold} et~al.}(2022)\citenamefont{{Arnold},
  {Couhert}, {Montenbruck}, {Kobel}, {Saquet}, {Peter}, {Mercier}, and
  {J{\"a}ggi}}}]{Arnold-etal:2022}
\bibinfo{author}{\bibfnamefont{D.}~\bibnamefont{{Arnold}}},
  \bibinfo{author}{\bibfnamefont{A.}~\bibnamefont{{Couhert}}},
  \bibinfo{author}{\bibfnamefont{O.}~\bibnamefont{{Montenbruck}}},
  \bibinfo{author}{\bibfnamefont{C.}~\bibnamefont{{Kobel}}},
  \bibinfo{author}{\bibfnamefont{E.}~\bibnamefont{{Saquet}}},
  \bibinfo{author}{\bibfnamefont{H.}~\bibnamefont{{Peter}}},
  \bibinfo{author}{\bibfnamefont{F.}~\bibnamefont{{Mercier}}},
  \bibnamefont{and}
  \bibinfo{author}{\bibfnamefont{A.}~\bibnamefont{{J{\"a}ggi}}}, in
  \emph{\bibinfo{booktitle}{44th COSPAR Scientific Assembly. Held 16-24 July}}
  (\bibinfo{year}{2022}), vol.~\bibinfo{volume}{44}, p. \bibinfo{pages}{3396}.

\bibitem[{\citenamefont{Petit and Luzum}(2010)}]{iers2010}
\bibinfo{author}{\bibfnamefont{G.}~\bibnamefont{Petit}} \bibnamefont{and}
  \bibinfo{author}{\bibfnamefont{B.}~\bibnamefont{Luzum}},
  \emph{\bibinfo{title}{{IERS} Conventions (2010)}}, no.~\bibinfo{number}{36}
  in \bibinfo{series}{IERS Technical Note} (\bibinfo{publisher}{International
  Earth Rotation and Reference Systems Service (IERS)},
  \bibinfo{address}{Frankfurt am Main, Germany}, \bibinfo{year}{2010}).

\bibitem[{\citenamefont{{Pavlis} et~al.}(2018)\citenamefont{{Pavlis}, {Luceri},
  {Kuzmicz-Cieslak}, {Pirri}, {Evans}, and {Bianco}}}]{Pavlis-etal:2018}
\bibinfo{author}{\bibfnamefont{E.~C.} \bibnamefont{{Pavlis}}},
  \bibinfo{author}{\bibfnamefont{V.}~\bibnamefont{{Luceri}}},
  \bibinfo{author}{\bibfnamefont{M.}~\bibnamefont{{Kuzmicz-Cieslak}}},
  \bibinfo{author}{\bibfnamefont{M.}~\bibnamefont{{Pirri}}},
  \bibinfo{author}{\bibfnamefont{K.}~\bibnamefont{{Evans}}}, \bibnamefont{and}
  \bibinfo{author}{\bibfnamefont{G.}~\bibnamefont{{Bianco}}}, in
  \emph{\bibinfo{booktitle}{EGU General Assembly Conference Abstracts}}
  (\bibinfo{year}{2018}), EGU General Assembly Conference Abstracts, p.
  \bibinfo{pages}{9216}.

\bibitem[{\citenamefont{Kieffer and Stone}(2005)}]{Kieffer2005}
\bibinfo{author}{\bibfnamefont{H.~H.} \bibnamefont{Kieffer}} \bibnamefont{and}
  \bibinfo{author}{\bibfnamefont{T.~C.} \bibnamefont{Stone}},
  \bibinfo{journal}{Astron. J.} \textbf{\bibinfo{volume}{129}},
  \bibinfo{pages}{2887} (\bibinfo{year}{2005}).

\bibitem[{\citenamefont{Walker et~al.}(2024)\citenamefont{Walker, Barker,
  Mazarico, Sun, Neumann, Smith, Head, and Zuber}}]{Walker_2024}
\bibinfo{author}{\bibfnamefont{R.~T.} \bibnamefont{Walker}},
  \bibinfo{author}{\bibfnamefont{M.~K.} \bibnamefont{Barker}},
  \bibinfo{author}{\bibfnamefont{E.}~\bibnamefont{Mazarico}},
  \bibinfo{author}{\bibfnamefont{X.}~\bibnamefont{Sun}},
  \bibinfo{author}{\bibfnamefont{G.~A.} \bibnamefont{Neumann}},
  \bibinfo{author}{\bibfnamefont{D.~E.} \bibnamefont{Smith}},
  \bibinfo{author}{\bibfnamefont{J.~W.} \bibnamefont{Head}}, \bibnamefont{and}
  \bibinfo{author}{\bibfnamefont{M.~T.} \bibnamefont{Zuber}},
  \bibinfo{journal}{The Planetary Science Journal} \textbf{\bibinfo{volume}{5}}
  (\bibinfo{year}{2024}).

\bibitem[{\citenamefont{{Zhang} et~al.}(2024)\citenamefont{{Zhang},
  {M{\"u}ller}, and {Biskupek}}}]{Zhang-etall:2024a}
\bibinfo{author}{\bibfnamefont{M.}~\bibnamefont{{Zhang}}},
  \bibinfo{author}{\bibfnamefont{J.}~\bibnamefont{{M{\"u}ller}}},
  \bibnamefont{and}
  \bibinfo{author}{\bibfnamefont{L.}~\bibnamefont{{Biskupek}}},
  \bibinfo{journal}{Astron. Asrophys.} \textbf{\bibinfo{volume}{681}},
  \bibinfo{eid}{A5} (\bibinfo{year}{2024}).

\bibitem[{\citenamefont{Fried}(1966)}]{Fried1966}
\bibinfo{author}{\bibfnamefont{D.~L.} \bibnamefont{Fried}},
  \bibinfo{journal}{J. Opt. Soc. Am.} \textbf{\bibinfo{volume}{56}},
  \bibinfo{pages}{1372} (\bibinfo{year}{1966}).

\bibitem[{\citenamefont{{Roddier}}(1981)}]{Roddier:1981}
\bibinfo{author}{\bibfnamefont{F.}~\bibnamefont{{Roddier}}},
  \bibinfo{journal}{Progress in Optics} \textbf{\bibinfo{volume}{19}},
  \bibinfo{pages}{281} (\bibinfo{year}{1981}).

\bibitem[{\citenamefont{Hardy}(1998)}]{Hardy1998}
\bibinfo{author}{\bibfnamefont{J.~W.} \bibnamefont{Hardy}},
  \emph{\bibinfo{title}{Adaptive Optics for Astronomical Telescopes}}
  (\bibinfo{publisher}{Oxford University Press}, \bibinfo{address}{New York},
  \bibinfo{year}{1998}).

\bibitem[{\citenamefont{Turyshev
  et~al.}(2021{\natexlab{a}})\citenamefont{Turyshev, Shao, and
  Hahn}}]{SGL-BPS-decadal}
\bibinfo{author}{\bibfnamefont{S.~G.} \bibnamefont{Turyshev}},
  \bibinfo{author}{\bibfnamefont{M.}~\bibnamefont{Shao}}, \bibnamefont{and}
  \bibinfo{author}{\bibfnamefont{I.}~\bibnamefont{Hahn}},
  \emph{\bibinfo{title}{Fundamental physics and lunar science investigations
  with advanced lunar laser ranging}} (\bibinfo{year}{2021}{\natexlab{a}}),
  \bibinfo{note}{{NAS Decadal Survey on Biological and Physical Sciences (BPS)
  Research in Space 2023-2032}}.

\bibitem[{\citenamefont{{Williams}}(2007)}]{Williams:2007}
\bibinfo{author}{\bibfnamefont{J.~G.} \bibnamefont{{Williams}}},
  \bibinfo{journal}{Geophys. Res. Let.} \textbf{\bibinfo{volume}{34}},
  \bibinfo{eid}{L03202} (\bibinfo{year}{2007}).

\bibitem[{\citenamefont{{Turyshev}
  et~al.}(2024{\natexlab{b}})\citenamefont{{Turyshev}, {Williams}, {Boggs}, and
  {Park}}}]{Turyshev:2024}
\bibinfo{author}{\bibfnamefont{S.~G.} \bibnamefont{{Turyshev}}},
  \bibinfo{author}{\bibfnamefont{J.~G.} \bibnamefont{{Williams}}},
  \bibinfo{author}{\bibfnamefont{D.~H.} \bibnamefont{{Boggs}}},
  \bibnamefont{and} \bibinfo{author}{\bibfnamefont{R.~S.} \bibnamefont{{Park}}}
  (\bibinfo{year}{2024}{\natexlab{b}}), \bibinfo{note}{arXiv:2406.16147
  [astro-ph.EP]}.

\bibitem[{\citenamefont{Turyshev
  et~al.}(2021{\natexlab{b}})\citenamefont{Turyshev, Shao, and
  Hahn}}]{SGL-BPS-decadal-research-campaign}
\bibinfo{author}{\bibfnamefont{S.~G.} \bibnamefont{Turyshev}},
  \bibinfo{author}{\bibfnamefont{M.}~\bibnamefont{Shao}}, \bibnamefont{and}
  \bibinfo{author}{\bibfnamefont{I.}~\bibnamefont{Hahn}},
  \emph{\bibinfo{title}{Night-time power delivery to lunar outposts with a
  high-power ground-based laser array}} (\bibinfo{year}{2021}{\natexlab{b}}),
  \bibinfo{note}{{NAS Decadal Survey on Biological and Physical Sciences (BPS)
  Research in Space 2023-2032}}.

\end{thebibliography}

\appendix

\section{Calibration and Maintenance}
\label{sec:calibration_maintenance}

High-precision LLR systems require that mechanical and thermal drifts be controlled to better than tens of $\mu$ms, as even small variations can obscure the sub-mm signals obtained during multi-second to multi-minute integrations. This section outlines strategies for calibration, external referencing, and maintaining required stability during observations.

\subsection{Daily and Continuous Calibration}

LLR facilities typically implement \emph{daily calibration} routines, supplemented by \emph{continuous checkouts} at hourly or sub-hourly intervals. Key methods include:
\begin{enumerate}
  \item \textit{Short-Distance Reference Target:}  
  A stable local retroreflector (e.g., a CCR or precision cavity) positioned at a known baseline (10--100\,m) from the main telescope serves as a zero-point reference. Regular ranging to this target—performed every 1--2 hours under variable conditions, or once per session in stable environments—enables tracking of drifts on the order of a few \(\mu\mathrm{m}/\)hr \citep{Murphy_etal_2008}.
  
  \item \textit{GPS or Frequency-Comb Tie:}  
  The station's local oscillator, which determines the CW modulation frequency \(f_m\) or chirp rate \(\beta\), is phase-locked to an external reference (such as a GPS-disciplined clock or optical frequency comb). Since drifts in \(f_m\) can lead to cm- to mm-scale range biases (see Sec.~\ref{sec:freq_drift}), using a high-stability standard (fractional stabilities of \(10^{-12}\) to \(10^{-14}\) over 100--1000\,s) is essential.
  
  \item \textit{Telescope Alignment Tracking:}  
  Internal metrology beams or star-based auto-collimators periodically verify alignment and focus, preventing sub-arcsec misalignments from degrading photon return or introducing timing errors.
  \end{enumerate}
Calibration re-baselines the system, keeping range offsets below 1\,mm---and even to tens of $\mu$m with precise surveying.

\subsection{Maintaining Sub-\texorpdfstring{\(\mu\)}{\textmu}m/s Stability}

Daily calibration may not suffice when environmental variations are rapid. To achieve sub-\(\mu\mathrm{m}/\mathrm{s}\) stability during live observations, it is necessary to:
\begin{itemize}
  \item \textit{Thermal Control:}  
  Maintain the optical assembly (optical bench, relay optics, detector housing) to \(\pm0.1^\circ\mathrm{C}\). Low-expansion materials (e.g., Zerodur, Invar) with thermal coefficients \(\lesssim10^{-7}\,\mathrm{K}^{-1}\) can limit path-length variations to \(\sim1\,\mu\mathrm{m}/\)hr. Active cooling loops around key components further mitigate thermal lensing and beam wander.
  
  \item \textit{Vibration Isolation:}  
  Use high-stiffness telescope piers, passive damping, or active isolation platforms to reduce seismic and mechanical noise in the 0.1--10\,Hz range. Such measures can restrict mirror shifts to below \(10\,\mu\mathrm{m}\), thereby preventing multi-\(\mu\mathrm{m}\) range errors.
  
  \item \textit{Rapid In-Situ Checks:}  
  A partial reflection from an internal beamsplitter or local retroreflector inserted into the signal path every few minutes verifies the stability of the optical train. Any deviations exceeding a preset threshold (e.g., \(5\text{--}10\,\mu\mathrm{m}\)) trigger recalibration or compensation.
\end{itemize}
Collectively, these measures can confine drifts to below \(\sim10\,\mu\mathrm{m}/\)hr, ensuring stable sub-mm LLR performance.

\subsection{External Ties to Metrological Networks}

Anchoring LLR data to global geodetic frames requires station positions accurate to within \(\lesssim1\,\mathrm{mm}\)  in every axis:
\begin{itemize}
  \item \textit{Local Surveys:}  
  Instruments such as GNSS antennas, satellite laser ranging (SLR) systems, or VLBI radio telescopes provide precise station coordinates in the International Terrestrial Reference Frame (ITRF) \citep{Vokrouhlicky_etal_2019}. Annual tie surveys detect tectonic shifts or pier settling at the sub-millimeter level.
  
  \item \textit{Clock Synchronization:}  
  Cross-referencing multiple timing references (e.g., hydrogen masers, optical clocks) with picosecond precision prevents internal drifts from masquerading as lunar range offsets. Short baselines (10--50\,m) between reference clocks and the ranging system help isolate local instabilities from global geophysical motions.
\end{itemize}

\subsection{Calibration Frequency and Adaptation}

Environmental conditions dictate the optimal calibration frequency:
\begin{itemize}
  \item \textit{Rapid Weather Changes:}  
  Under unstable conditions (e.g., large temperature swings or high winds), calibrating every 1--2 hours is recommended.
  
  \item \textit{Stable Conditions:}  
  During consistently calm nighttime periods with minimal dome seeing, once-daily calibration and pre/post-session checks are typically sufficient.
\end{itemize}

Rigorous calibration, continuous monitoring, and advanced metrological ties provide the foundation for achieving tens-of-\(\mu\mathrm{m}\) precision in high-power CW LLR, enabling next-gen lunar geodesy and gravitational physics experiments.

\section{Required Hardware Components}
\label{sec:hardware_components}

Sub-mm precision in high-power CW LLR requires more than just a laser; the observatory must manage high power, maximize photon returns, and mitigate turbulence and mechanical instabilities. An integrated approach optimizing all subsystems is crucial for long-term stability and precision. Below, we summarize the key hardware components.

\subsection{High-Power Laser System}
\label{sec:hardware_laser}

The laser system is the cornerstone of CW LLR, providing the high photon flux required to suppress shot noise and achieve sub-mm precision:
\begin{itemize}
  \item \textit{Wavelength (1064\,nm):} Widely used in Nd:YAG and fiber lasers, this wavelength offers high atmospheric transmission and excellent retroreflection efficiency from lunar corner cubes \citep{Currie_etal_2011}.
  \item \textit{Output Power (\(\sim1\,\mathrm{kW}\)):} 
  High average power boosts photon return by  orders of magnitude over pulsed systems (\(\lesssim10\,\mathrm{W}\) average), while amplitude stability (\(\lesssim0.5\%\)) is essential to suppress phase noise in coherent integration.
    \item \textit{Beam Quality (\(M^2 \sim 1.1\)):} Near-diffraction-limited output ensures optimal power density on lunar CCRs.
  \item \textit{Modulation Bandwidth (to 10\,GHz):} Electro- or acousto-optic modulators must tolerate kW-level beams and support multi-frequency or chirped modulation for resolving \(2\pi\)-phase ambiguities (Secs.~\ref{sec:AM_observable}, \ref{sec:FM_observable}).
  \item \textit{Thermal Management:} Gain media and optics must be stabilized within \(\pm0.1^\circ\mathrm{C}\) to prevent thermal lensing and beam wander. Active cooling loops around amplifier heads are essential.
  \item \textit{Pointing Stability:} Active beam steering or rigid optical mounts are required to maintain alignment within \(\pm5\,\mu\mathrm{rad}/\mathrm{hr}\), minimizing photon losses and systematic timing offsets \citep{Murphy_etal_2012}.
\end{itemize}

\subsection{Telescope System}
\label{sec:hardware_telescope}

The telescope must both transmit a high-power beam and efficiently collect the faint returns:
\begin{itemize}
  \item \textit{Aperture (1--2\,m):} A 1\,m Ritchey--Chr\'etien design strikes a balance between photon collection efficiency, mechanical complexity, and cost \citep{Samain1998}. Larger apertures boost photon flux but require tighter thermal control.
  \item \textit{Optical Coatings:} High-reflectivity (\(>99.5\%\)) and anti-reflection coatings minimize link losses \citep{Degnan1993}.
  \item \textit{Precision Tracking:} Direct-drive alt-azimuth mounts with sub-arcsecond RMS pointing errors ensure consistent illumination of lunar corner cubes \citep{Murphy_etal_2008}.
\end{itemize}

\subsection{Beam Expander}
\label{sec:hardware_expander}

Beam expanders are used to reduce near-field intensity and improve the far-field spot size:
\begin{itemize}
  \item \textit{Expansion Ratio (10--20$\times$):} Adjustable beam diameters (10--20\,cm) help mitigate optic damage at high power levels and counteract atmospheric seeing effects.
  \item \textit{High-Damage-Threshold Coatings:} Coatings must withstand \(\gtrsim1\,\mathrm{kW}\) power, and active cooling prevents wavefront distortions or lensing \citep{Samain1998}.
\end{itemize}

\subsection{Detector System}
\label{sec:hardware_detector}

The detector system determines the ultimate timing precision and shot-noise limit:
\begin{itemize}
  \item \textit{Superconducting Nanowire SPDs (SNSPDs):} These detectors achieve \(>80\%\) quantum efficiency at 1064\,nm, with dark counts below \(10\,\mathrm{s}^{-1}\) and timing jitter of 20--50\,ps \citep{Marsili2013}.
  \item \textit{Avalanche Photodiodes (APDs):} APDs are easier to operate at moderate temperatures (\(-30^\circ\mathrm{C}\)) but typically have higher dark counts and lower quantum efficiency compared to SNSPDs.
  \item \textit{Spectral Filters:} Narrowband filters (e.g., 0.1--0.2\,nm) centered at 1064\,nm suppress sky and lunar background light. Combined with small FOVs (\(<2\,\mathrm{arcsec}\)), these filters reduce background noise to \(10^2\text{--}10^3\,\mathrm{s}^{-1}\) \citep{Battat_etal_2009}.
\end{itemize}

\subsection{Timing and Control Electronics}
\label{sec:hardware_electronics}

Ultra-stable timing electronics are essential for resolving small phase shifts over multi-second integrations:
\begin{itemize}
  \item \textit{Time-to-Digital Converters (TDCs):} Sub-10\,ps bin resolution minimizes timing error. Multi-second averaging enhances precision while suppressing random noise \citep{Battat_etal_2009}.
  \item \textit{Ultra-Stable References:} Hydrogen masers or GPS-locked oscillators with stabilities of \(10^{-13}\text{--}10^{-14}\) over $10^2-10^3$\,s ensure accurate frequency control (Sec.~\ref{sec:freq_drift}).
  \item \textit{Phase-Locked Loop (PLL) and FPGA Control:} Real-time demodulation, Doppler compensation, and multi-frequency integration rely on high-speed FPGA pipelines \citep{Murphy_etal_2012}.
\end{itemize}

\subsection{Environmental Control and Safety Systems}
\label{sec:hardware_environment}

A high-power CW LLR system requires precise environmental controls and robust safety mechanisms:
\begin{itemize}
  \item \textit{Thermal Regulation:} Dome interiors and optical components must be maintained within \(\pm0.01^\circ\mathrm{C}\) to limit thermal expansion to sub-\(\mu\mathrm{m}\) levels \citep{Samain1998}.
  \item \textit{Vibration Isolation:} High-stiffness piers and passive or active damping systems minimize mechanical noise \citep{Murphy_etal_2008}.
  \item \textit{Safety Protocols:} Aircraft and satellite detection systems ensure the laser shuts down during intrusions, maintaining compliance with safety regulations \citep{Samain1998}.
\end{itemize}

\subsection{Control Software}
\label{sec:hardware_software}

Advanced control software is essential for real-time operation and diagnostics:
\begin{itemize}
  \item \textit{Dynamic Scheduling:} Automated scripts integrate real-time ephemerides, atmospheric data, and calibration results to optimize observing sequences. Adaptive modes or pauses protect the system under poor conditions.
  \item \textit{Data Logging and Diagnostics:} Real-time logging of phase measurements, timing offsets, and environmental parameters allows for post-session diagnostics and precise range modeling.
  \item \textit{Anomaly Detection:} Automated monitoring of laser power, detector performance, alignment, and environmental conditions flags anomalies for operator intervention or maintenance.
\end{itemize}

\textit{In summary}, the integration of these hardware components—spanning high-power lasers, precision optics, ultra-stable timing, environmental controls, and sophisticated software—is essential for achieving and maintaining the sub-mm precision required for next-generation lunar laser ranging. Each subsystem is designed to complement the others, ensuring that both systematic and random errors are minimized over extended observation periods.

\end{document}